\pdfoutput=1
\documentclass[11pt]{article}
\pdfoutput=1
\usepackage{jheppub}

\usepackage{amsmath} 
\usepackage{multirow} 
\usepackage{amsfonts}
\usepackage{amssymb,graphics,psfrag}
\usepackage{array,epsfig,multirow,graphicx,mathtools}
\usepackage{hyperref}
\def\CH{{\cal H}}

\newcommand{\be}{\begin{equation}}
\newcommand{\ee}{\end{equation}}
\newcommand{\bea}{\begin{eqnarray}}
\newcommand{\eea}{\end{eqnarray}}
\def\lam{{\lambda}}

\title{Partition functions of web diagrams with an O7$^-$-plane}

\author[a, b]{Hirotaka Hayashi,}
\author[b]{Gianluca Zoccarato}

\affiliation[a]{Tokai University, 4-1-1 Kitakaname, Hiratsuka, Kanagawa 259-1292, Japan}
\affiliation[b]{Departamento de F\'isica Te\'orica and Instituto de F\'isica Te\'orica UAM/CSIC,\\ Universidad Aut\'onoma de Madrid, Cantoblanco, 28049 Madrid, Spain}

\emailAdd{h.hayashi@tokai.ac.jp}
\emailAdd{gianluca.zoccarato@csic.es}

\abstract{We consider the computation of the topological string partition function for 5-brane web diagrams with
an O7$^-$-plane. Since upon quantum resolution of the orientifold plane these diagrams become non-toric web diagrams without the orientifold we are able to apply the topological vertex to obtain the Nekrasov partition function of the corresponding 5d theory.  We apply this procedure to the case of 5d $SU(N)$ theories with one hypermultiplet in the antisymmetric representation and to the case of 5d pure $USp(2N)$ theories.  For these cases we discuss the 
dictionary between parameters and moduli of the 5d gauge theory and lengths of 5-branes in the web diagram and moreover we
perform comparison of the results obtained via application
of the topological vertex and the one obtained via localisation techniques, finding in all instances we consider perfect agreement.}

\begin{document}

\makeatletter
\let\old@fpheader\@fpheader
\renewcommand{\@fpheader}{\old@fpheader\hfill
IFT-UAM/CSIC-16-085}
\makeatother

\maketitle


%
%

\section{Introduction}

Recently, there has been much progress in the understanding of supersymmetric quantum field theories in five-dimensions. The existence of various ultraviolet (UV) complete five-dimensional (5d) $\mathcal{N}=1$ supersymmetric gauge theories was originally studied in \cite{Seiberg:1996bd, Morrison:1996xf, Douglas:1996xp, Intriligator:1997pq} and shows interesting features such as an enhancement of flavour symmetries. 
Recently, a larger class of new UV complete 5d $\mathcal{N}=1$ supersymmetric gauge theories has been conjectured to exist by 
the existence of 5-brane webs in type IIB string theory \cite{Bergman:2013aca, Zafrir:2014ywa, Bergman:2014kza, Hayashi:2015fsa, Bergman:2015dpa, Zafrir:2015rga, Hayashi:2015zka, Ohmori:2015tka, Hayashi:2015vhy, Zafrir:2015ftn, Zafrir:2016jpu}\footnote{It is also possible to give an evidence for the existence of the new UV complete 5d gauge theories by a field theory method using the instanton operator analysis \cite{Tachikawa:2015mha, Zafrir:2015uaa, Yonekura:2015ksa, Gaiotto:2015una}.} whose worldvolume theories yield the corresponding 5d gauge theories at low energies \cite{Kol:1997fv, Aharony:1997ju, Aharony:1997bh, Benini:2009gi}. If a 5-brane web has an $S^1$ direction or infinitely rotating structure, the 5-brane web can even realise a 5d gauge theory which has a UV completion as a 6d $\mathcal{N}=(1,0)$ superconformal field theory (SCFT) \cite{Hayashi:2015fsa, Zafrir:2015rga, Hayashi:2015zka, Ohmori:2015tka, Hayashi:2015vhy}.  

The 5-brane web diagram is useful for exploring the landscape of UV complete 5d gauge theories as just described, but it also has another important feature. Namely, it is possible to compute the exact partition function of the 5d theory realised on a 5-brane web by using the topological vertex. This is based on a duality between a 5-brane web in type IIB string theory and a Calabi-Yau threefold in M-theory \cite{Leung:1997tw}. The duality maps the BPS states of the 5d theory to M2-branes wrapping two-cycles in the dual Calabi-Yau threefold, whose contribution can be captured by the topological string partition function. Therefore, the topological string partition function for the dual Calabi-Yau threefold exactly gives the partition function of a 5d gauge theory on the 5-brane web, see for example \cite{Iqbal:2003ix, Iqbal:2003zz, Eguchi:2003sj, Hollowood:2003cv}. 

Although the technique of the topological vertex was originally developed for computing the topological string partition function for a toric Calabi-Yau threefold, further studies revealed that the topological vertex can be also employed for the computation of the topological string partition function for a certain class of non-toric Calabi-Yau threefolds \cite{Hayashi:2013qwa,  Hayashi:2014wfa, Kim:2015jba, Hayashi:2015xla, Hayashi:2016abm}. These threefolds may be obtained by a topology changing transition from a toric Calabi-Yau threefold. This transition has a nice interpretation in terms of physics of the 5d theory obtained from the toric Calabi-Yau
threefold: initially the theory is in a Coulomb branch and the transition is triggered by going in a particular region of the Coulomb branch as well as the parameter space where some hypermultiplets become massless. Having massless hypermultiplets then allows entrance in a Higgs branch which in turns yields a different theory at low energies. The renormalisation group (RG) flow triggered by the Higgs vacuum expectation value (vev) of the UV 5d theory to the infrared (IR) 5d theory corresponds to the topology changing transition of the toric Calabi-Yau threefold to a non-toric Calabi-Yau threefold. The partition function of the UV 5d theory can be easily computed by using the topological vertex since it is characterised by a toric Calabi-Yau threefold. Ref.~\cite{Hayashi:2013qwa, Hayashi:2014wfa, Hayashi:2015xla} have applied the Higgsing prescription to the topological vertex formalism and found out that essentially the same topological vertex can be used for computing the topological string partition function for such a type of non-toric Calabi-Yau threefolds. The upshot is that  it is possible to apply the topological vertex to a 5-brane web even if it is dual to a non-toric Calabi-Yau threefold.

In this paper we will apply this technique and compute the Nekrasov partition function of two classes of 5d $\mathcal{N}=1$ supersymmetric gauge theories:
we will consider $SU(N)$ gauge theories with a hypermultiplet in the antisymmetric representation and pure $USp(2N)$ gauge theories.
For both class of theories the corresponding 5-brane diagrams were proposed in in \cite{Bergman:2015dpa} and have the common feature of having 
an O7$^-$-plane in the web diagram.
 However it has not been checked whether the topological string partition function for the dual non-toric Calabi-Yau threefold reproduces the corresponding Nekrasov partition function. Although the gauge theory content is manifestly seen in the presence of an O7$^-$-plane, we need to resolve the O7$^-$-plane in order to apply the topological vertex formalism. After decomposing the O7$^-$-plane into two 7-branes by quantum effects \cite{Sen:1996vd}, the resulting 5-brane web does not manifestly show the gauge theory content of the 5d $SU(N)$ gauge theory with antisymmetric matter or the 5d $USp(2N)$ gauge theory. However, we will see that the topological string partition function remarkably recovers the Nekrasov partition function of the 5d theories. Also, we will find a way to identify the gauge theory parameters from the 5-brane web diagrams, which gives a systematic method for computing the Nekrasov partition functions from the 5-brane webs.

The organisation of the paper is as follows. In section \ref{sec:SU2NA}, we compute the Nekrasov partition function of a 5d $SU(2N)$ gauge theory with a hypermultiplet in the antisymmetric representation by using the topological vertex. We explicitly compare the perturbative part and the one-instanton part of the partition function with the known field theory result from the localisation for the cases of 5d $SU(4), SU(6)$ gauge theories finding  agreement between the two results. In section \ref{sec:USp}, we turn to the computation of the Nekrasov partition function of the 5d $USp(2N)$ gauge theory. 
We start with the observation that the web diagram for an $USp(2N)$ gauge theory may be obtained by a Higgsing of the web diagram of an $SU(2N)$ gauge theory with a hypermultiplet in the antisymmetric representation. This allows us to obtain the result for the partition function of a pure $USp(2N)$ gauge 
theory by applying the Higgsing to the partition function obtained in section \ref{sec:SU2NA}. Also in this case we compare the result obtained via
topological string computation with localisation techniques for the cases of pure $USp(4)$ and $USp(6)$ gauge theories finding perfect agreement.
Lastly, we compute the partition function of a 5d $SU(2N-1)$ gauge theory with one antisymmetric hypermultiplet in section \ref{sec:SU2N-1A}. We discuss 
how going to an extreme region in Coulomb branch moduli space it is possible to obtain the diagram realising a $SU(2N-1)$ gauge theory with a 
hypermultiplet in the antisymmetric representation from the diagram realising a $SU(2N)$ theory with one antisymmetric hypermultiplet. Therefore 
using the results obtained in section \ref{sec:SU2NA} we easily obtain the partition function for the case of an $SU(5)$ gauge theory with one antisymmetric
hypermultiplet. Again we compare the result obtained via topological string computation with the one obtained by applying localisation techniques 
finding perfect agreement. Some technical details employed in the computations throughout the paper are reviewed in the appendices.
Appendix \ref{sec:top} summarises the formalism of the topological vertex including its application to non-toric Calabi-Yau threefolds. Appendix \ref{sec:Nek} collects the result of the Nekrasov partition function of a 5d $SU(N)$ gauge theory with a hypermultiplet in the antisymmetric representation and of a 5d $USp(2N)$ gauge theory.

\section{5d $SU(2N)$ gauge theory with antisymmetric matter}
\label{sec:SU2NA}

In this section, we analyse a 5d $SU(2N)$ gauge theory with a hypermultiplet in the antisymmetric representation and the Chern-Simons (CS) level $\kappa$
and their realisation via 5-brane webs. We always assume that $N$ is greater than $1$. In the case of $N=1$, the antisymmetric representation 
is trivial. 
In order to have a 5d UV fixed point, the value of the Chern-Simons level is restricted to $|\kappa| \leq N+3$ \cite{Yonekura:2015ksa, Hayashi:2015zka}. We will employ the notation $G_{\kappa}$ where $G$ stands for a gauge group and $\kappa$ is the CS level. 
Note also that the quantisation condition of the CS level is given by
\be
\kappa + \frac{2N-4}{2} = \kappa + N-2 \in \mathbb{Z}. \label{CSquantization}
\ee
Therefore, the CS level $\kappa$ is always integer in this case, which will be also seen from the corresponding 5-brane web. 

\subsection{The 5-brane web}
A 5d $SU(2N)_{\kappa}$ gauge theory with a hypermultiplet in the antisymmetric representation can be realised on a 5-brane web with an O7$^-$-plane \cite{Bergman:2015dpa}. The 5d theory is conjectured to have a 5d UV fixed point \cite{Intriligator:1997pq, Yonekura:2015ksa,Bergman:2015dpa, Hayashi:2015zka} and its perturbative flavour symmetry is $U(1) \times U(1)_I$ where the latter $U(1)_I$ is associated to the instanton flavour current. 
A the 5d UV fixed point there is a possibility of further enhancement of
the flavour symmetry depending on the specific value of the CS level. 
In the particular case of $N=2$ since the antisymmetric representation is real
the perturbative flavour symmetry is actually $USp(2) \times U(1)_I$.

\begin{figure}
\centering
\includegraphics[width=13cm]{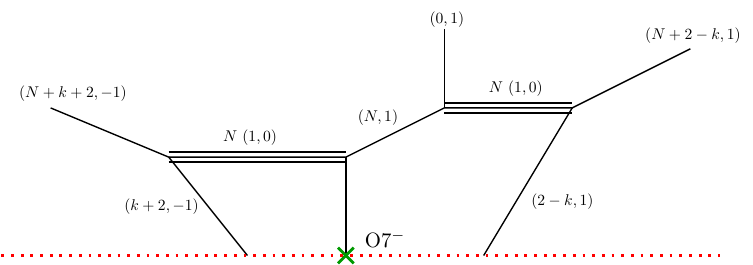}
\caption{The web diagram realising an $SU(2N)_k$ gauge theory with antisymmetric matter in the presence of an O7$^-$-plane.}
\label{fig:SU2Nantisym}
\end{figure}

The 5-brane web for the 5d $SU(2N)_{k}$ gauge theory with a hypermultiplet in the antisymmetric representation is depicted in Figure \ref{fig:SU2Nantisym}.
The directions in which 5-branes extend are summarised in Table \ref{tb:brane}.
\begin{table}[t]
\begin{center}
\begin{tabular}{c|c c c c c|c c|c c c}
\hline
&0&1&2&3&4&5&6&7&8&9\\
\hline
D5&$\times$&$\times$&$\times$&$\times$&$\times$&$\times$&&&&\\

NS5&$\times$&$\times$&$\times$&$\times$&$\times$&&$\times$&&&\\
$(p, q)$ 5-brane & $\times$& $\times$& $\times$ & $\times$ &$\times$& \multicolumn{2}{c|}{\text{angle}}&&& \\
7-brane & $\times$& $\times$&$\times$ & $\times$ &$\times$ &  &  &$\times$ &$\times$  &$\times$\\
\hline
\end{tabular}
\end{center}
\caption{Type IIB brane configuration for a 5-brane web diagram. The slope of the $(p, q)$ 5-brane is $\frac{q}{p}$ in the $(x_5, x_6)$-plane.}
\label{tb:brane}
\end{table}
Since the whole structure of the web diagram may be analysed by projection in the $(x_5, x_6)$-plane we shall always restrict all figures in the paper to
this particular plane choosing the orientation so that $x_5$ corresponds to the horizontal direction and $x_6$ to the vertical one. Therefore in all figures 
a horizontal line corresponds to a D5--brane and a vertical one to an NS5--brane.
In Figure \ref{fig:SU2Nantisym}, fundamental strings which cross the middle NS5-brane gives a hypermultiplet in the antisymmetric representation of the 
$SU(2N)$ gauge group. Furthermore, $k$ is integer and this reflects the fact that the CS level is integer \eqref{CSquantization}.

\begin{figure}
\centering
\includegraphics[width=11cm]{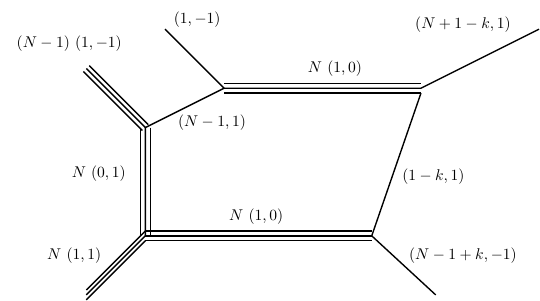}
\caption{The web diagram realising an $SU(2N)_k$ theory with antisymmetric matter after the quantum resolution of the orientifold plane and the extraction of the 7-branes to infinity.}
\label{fig:SU2Nantisym1}
\end{figure}

While the 5-brane web in the presence of an O7$^-$-plane is well suited for reading off the gauge group and the matter fields
we cannot directly use it for computation of the 5d Nekrasov partition function for there is no prescription for the application of the topological vertex
in the presence of orientifold planes. In the case of an O7$^-$-plane it is still possible to find a web-diagram to which it is possible to apply the topological 
vertex for an O7$^-$-plane splits into a pair of 7--branes when non--perturbative effects in the string coupling $g_s$ are taken into account \cite{Sen:1996vd}. In the case at 
hand since the middle NS5-brane is attached to the O7$^-$-plane, the O7$^-$-plane should split into a $(0, -1)$ 7-brane and a $(2, 1)$ 7-brane and the NS5-brane ends on the $(0, -1)$ 7--brane \cite{Bergman:2015dpa, Hayashi:2015zka}. The resolution at the same time creates a 5-brane loop. Pulling off the 7-branes from the 5-brane loop and taking into account the various Hanany--Witten
transitions \cite{Hanany:1996ie} due to the branch cuts of the 7-branes we finally obtain an equivalent 5-brane web diagram given by Figure \ref{fig:SU2Nantisym1}.\footnote{In fact, we performed the overall $T^{-1}$-transformation compared to the web in Figure \ref{fig:SU2Nantisym}. The charge of the $(p, q)$ 5-branes becomes $(p-q, q)$ by the $T^{-1}$-transformation.}

The resulting 5-brane web diagram contains configurations of some 5-branes jumping over other 5-branes, and hence it is not dual to a toric Calabi--Yau threefold in M-theory. Although the original topological vertex formalism in \cite{Iqbal:2002we, Aganagic:2003db} was intended for the application to toric Calabi--Yau threefolds, Ref.~\cite{Hayashi:2015xla} has shown that it is possible to apply the topological vertex directly to such a 5-brane web diagram involving the configurations of 5-brane jumps, which is dual to a certain non-toric Calabi-Yau threefold. Therefore, we can obtain the 5d Nekrasov partition function of the 5d $SU(2N)$ gauge theory with a hypermultiplet in the antisymmetric representation from computing the topological string partition function obtained by applying the topological vertex to the 5-brane web in Figure \ref{fig:SU2Nantisym1}. Although the presence of the hypermultiplet in the antisymmetric representation is not manifest after splitting the O7$^-$-plane, we will see that the topological string computation precisely recovers its contribution from the 5-brane web diagram in Figure \ref{fig:SU2Nantisym1}. 

Let us also point out that the 5-brane web depicted in Figure \ref{fig:SU2Nantisym1} is valid only when the Chern-Simons level $k$ lies in the interval $-N \leq k \leq N+2$. In the other cases within $|k| \leq N+3$, we need to further manipulate the 5-brane web so that all the 7-branes attached to external 5-branes are put at infinity. We will restrict ourselves to the cases where the 5-brane web diagram is in the form given by Figure \ref{fig:SU2Nantisym1}, namely the CS leve is $-N \leq k \leq N+2$, for computing the topological string partition function, but the computation for the other cases can be done in a similar manner in principle.

\begin{figure}
\centering
\includegraphics[width=13cm]{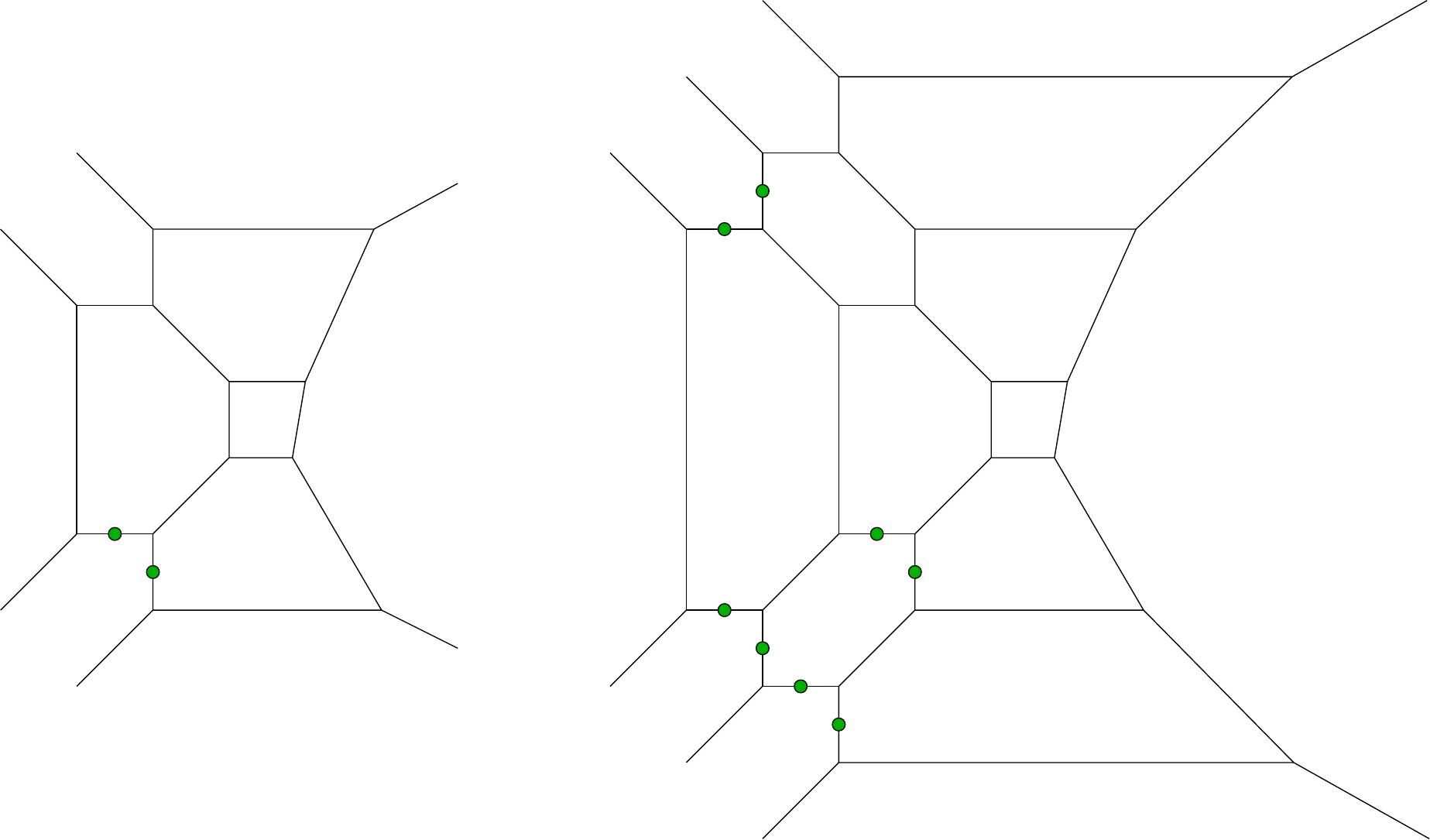}
\caption{Examples of $SU(2N)_k$ with antisymmetric matter as a Higgsing of a linear quiver. On the left the case of $N=2$, on the right the case of $N=3$. The 5-branes with the green dot get shrunken for the Higgsing.}
\label{fig:Higgsantisym}
\end{figure}

Finally we would like to note that 
the web diagram of the 5d $SU(2N)$ gauge theory with a hypermultiplet in the antisymmetric representation may be also obtained by Higgsing a 5-brane web diagram yielding a 5d linear quiver theory of the following type
\be
SU(2) - SU(4) - SU(6) - \cdots - SU(2N-2) - SU(2N)
\ee
where we have $N$ gauge nodes. The examples of the Higgsing in the case of $N=2, 3$ are depicted in Figure \ref{fig:Higgsantisym}.
This Higgsing in fact cannot be seen in a perturbative description of the 5d linear quiver theory for it
uses a vev of hypermultiplet charged under the instanton flavour current. Those hypermultiplets do not appear in a perturbative regime but appear non-perturbatively. The 5-brane web diagram enables us to see such a Higgsing as in Figure \ref{fig:Higgsantisym}.

\subsection{Parametrisation}
\label{sec:SU2Npara}

\begin{figure}
\centering
\includegraphics[width=12cm]{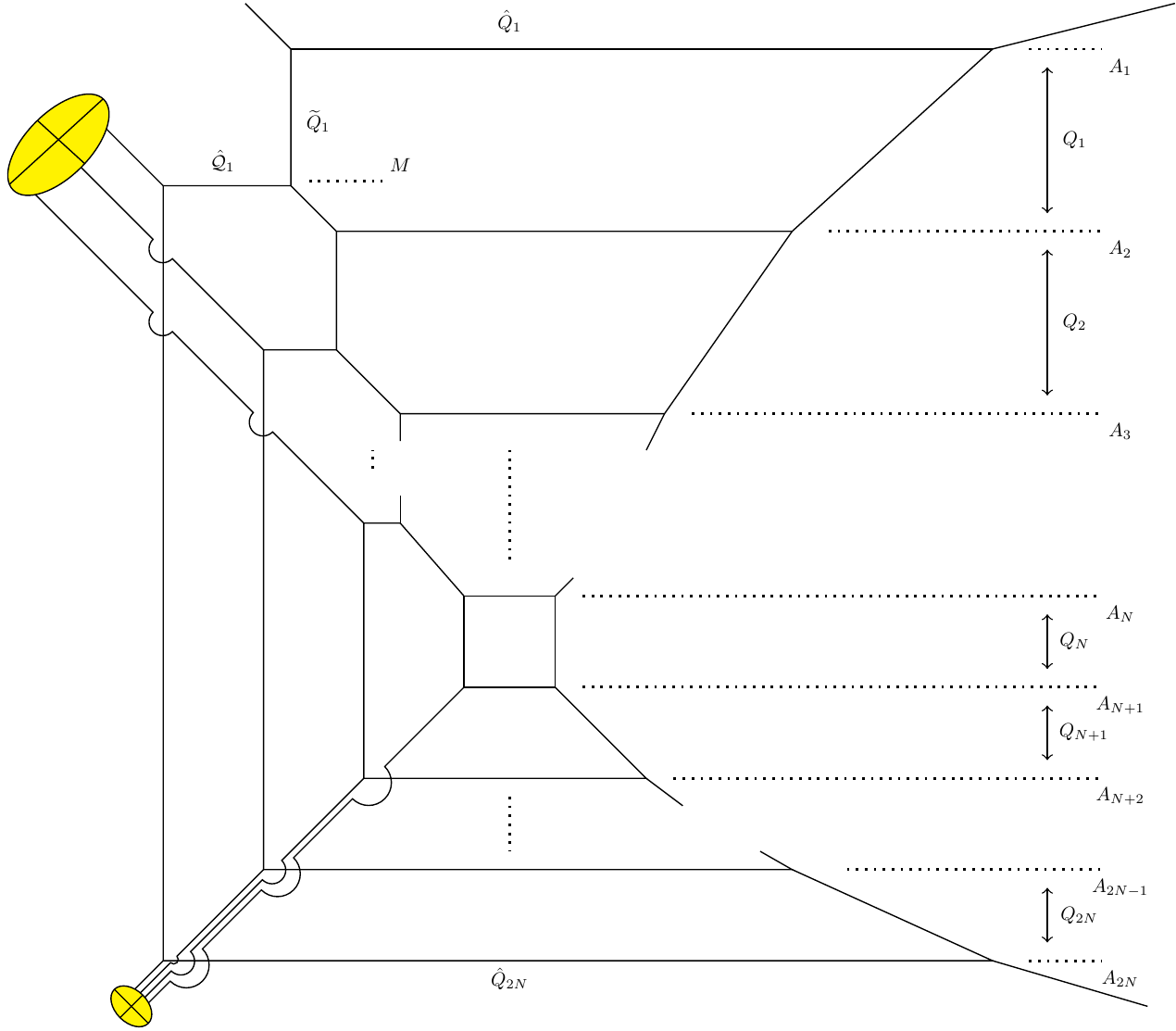}
\caption{The definition of the parameters entering in the brane web of the 5d $SU(2N)_k$ gauge theory with antisymmetric matter.}
\label{fig:SU2Nantisymdiag}
\end{figure}

Before going to the actual computation of the topological string partition function we will first determine how the parameters and the moduli of the 5d $SU(2N)$ gauge theory with a hypermultiplet in the antisymmetric representation are related to the parameters of the 5-brane web yielding the 5d theory. For that we will use a 5-brane web in Figure \ref{fig:SU2Nantisymdiag} instead of the one in Figure \ref{fig:SU2Nantisym1}. The two figures are equivalent as they are related by flop transitions in the dual Calabi-Yau geometry.

The parameters of the 5-brane web are the lengths of the finite size 5-branes. They are dual to the volume of two-cycles in the dual Calabi--Yau threefold in M-theory. 
The relations between the parameters of a 5-brane web and the moduli and the parameters of the corresponding 5d theory are the followings: local deformations which do not move external 5-branes give moduli of the corresponding 5d theory whereas global deformations which move external 5-branes yield parameters of the corresponding 5d theory.

From the matter content of the 5d gauge theory, we should have $2N-1$ Coulomb branch moduli, one gauge coupling, and one mass parameter for the antisymmetric hypermultiplet. The Coulomb branch moduli of the 5d theory can be easily identified with the parameters in the 5-brane web diagram. The 5-brane web in Figure \ref{fig:SU2Nantisymdiag} has $2N$ color D5-branes giving the $SU(2N)$ gauge group. Hence, the length between each color D5-brane corresponds to the $2N-1$ Coulomb branch moduli. More specifically, we parametrise $Q_i$ in Figure \ref{fig:SU2Nantisymdiag} as
\be
Q_i = \exp[-(\alpha_i - \alpha_{i+1})], \label{coulomb.SU2N}
\ee 
for $i=1, \cdots, 2N-1$. $\alpha_i, (i=1, \cdots, 2N)$ with $\sum_{i=1}^{2N}\alpha_i = 0$ are the Coulomb branch moduli of the $SU(2N)$ gauge theory. Here $\alpha_i-\alpha_{i+1}$ is the length between the $i$-th\footnote{The order of the color D5-branes is given from top to bottom.} color D5-brane and the $i+1$-th color D5-brane. The parameter $Q$ associated with a length $L$ is generally defined as $Q = e^{-L}$. In this paper, the length $Q$ always means $L$ of $e^{-L}$.

It is also convenient for later discussion to define 
\be
A_i = \exp[-\alpha_i],
\ee
for $i=1, \cdots, 2N$ and $\prod_{i=1}^{2N}A_i = 1$, which parameterise the height of the color D5-branes as in Figure \ref{fig:SU2Nantisymdiag}. With the $A_i$'s, Eq.~\eqref{coulomb.SU2N} can be rewritten as
\be
Q_i = A_iA_{i+1}^{-1}.\label{QandA}
\ee
for $i=1, \cdots, 2N-1$. Using \eqref{QandA} with $\prod_{i=1}^{2N}A_i = 1$, one can inversely solve \eqref{QandA} and obtains
\be
A_{2N} = Q_1^{-\frac{1}{2N}}Q_2^{-\frac{2}{2N}}\cdots Q_{2N-2}^{-\frac{2N-2}{2N}}Q_{2N-1}^{-\frac{2N-1}{2N}}. \label{A2N}
\ee
The expression of the rest of $A_i, i=1, \cdots, 2N-1$ can be constructed by using \eqref{QandA} and \eqref{A2N}.

Another parameter we need to determine is the mass parameter of the hypermultiplet in the antisymmetric representation. The identification of the mass parameter can be understood by considering a Higgsing to a 5d $USp(2N)$ gauge theory from the 5d $SU(2N)$ gauge theory with the one antisymmetric hypermultiplet. The antisymmetric hypermultiplet can acquire a vacuum expectation value when the mass parameter is zero. In terms of the 5-brane web in Figure \ref{fig:SU2Nantisymdiag}, the Higgsing occurs when we put the one $(1, -1)$ 5-brane in the upper middle part of the diagram on the $(1, -1)$ 7-brane on which $N-1$ 5-branes end. Therefore, the mass parameter $M$ of the hypermultiplet in the antisymmetric representation should satisfy the following condition
\be
-\log\left(\tilde{Q}_1\hat{\mathcal{Q}}_1\right) \propto m, \label{antisymmass1}
\ee
where $\tilde{Q}_1$ and $\hat{\mathcal{Q}}_1$ are depicted in Figure \ref{fig:SU2Nantisymdiag}. From the geometry of the web diagram, one can obtain 
\bea
\tilde{Q}_1 &=& A_1M^{-1}, \label{AMs1}\\
\hat{\mathcal{Q}}_1 &=& M^{-N+1}A_{2N}^{-N}\left(\prod_{i=2}^{2N}A_i\right)\\
&=&M^{-N+1}A_1^{-1}A_{2N}^{-N},\label{AMs2}
\eea
where $M$ is the height of the D5-brane on the upper-left part of the diagram as in Figure \ref{fig:SU2Nantisymdiag}. By using the parameters of \eqref{AMs1}, \eqref{AMs2}, Eq.~\eqref{antisymmass1} becomes
\be
N\log\left(M A_{2N}\right) \propto m.
\ee
Then it is natural to choose $-N$\footnote{The overall negative sign may be a convention of the mass parameter.} for the proportionality constant of \eqref{antisymmass1} and hence \eqref{antisymmass1} becomes
\be
\tilde{Q}_1\hat{\mathcal{Q}}_1 = \exp[Nm],\label{mass.antisym}
\ee
which gives the relation between the mass parameter of the antisymmetric matter and the lengths in the 5-brane web. 

\begin{figure}
\centering
\includegraphics[width=15cm]{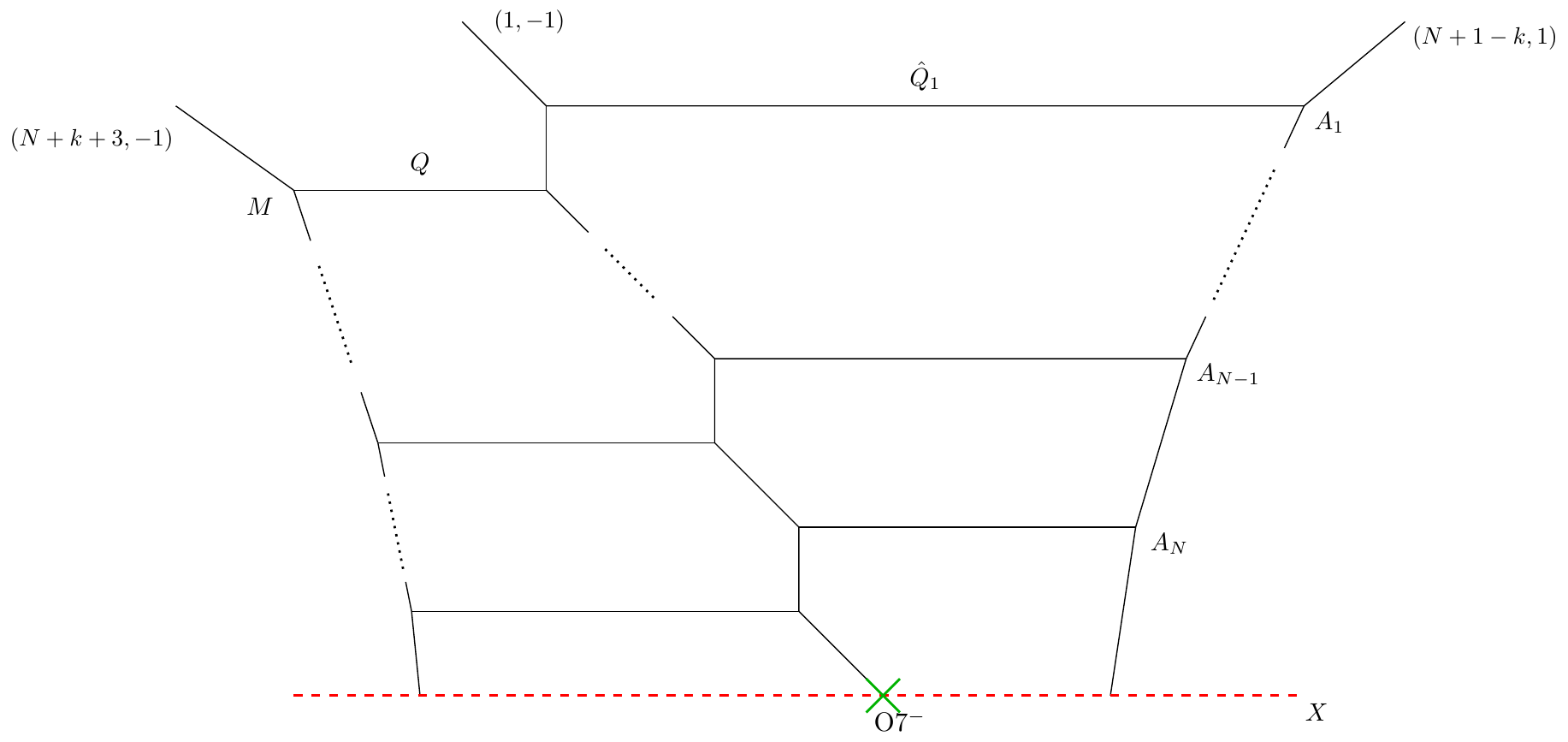}
\caption{The web diagram of Figure \ref{fig:SU2Nantisym} after performing the $T^{-1}$ transformation.}
\label{fig:SU2NantisymT}
\end{figure}

The last parameter is the gauge coupling or the instanton fugacity of the 5d $SU(2N)$ gauge theory. In order to determine the instanton fugacity from the web diagram, we go back to the web diagram with an O7$^-$-plane in Figure \ref{fig:SU2NantisymT}.
Note that the web diagram in Figure \ref{fig:SU2NantisymT} is related by the $T^{-1}$-transformation 
to the original web diagram in Figure \ref{fig:SU2Nantisym}. The $T^{-1}$-transformed diagram in Figure \ref{fig:SU2NantisymT} is directly related to the 5-brane web in Figure \ref{fig:SU2Nantisym1} after resolving the O7$^-$-plane. Some parameters in Figure \ref{fig:SU2NantisymT} are common to the ones in Figure \ref{fig:SU2Nantisymdiag} and hence we use the same parameters. More specifically, $\hat{Q}_1, A_1, \cdots, A_N$ and $M$ in Figure \ref{fig:SU2NantisymT} are the same as the ones in Figure \ref{fig:SU2Nantisymdiag}. However, some parameters are new and we define $Q$ as the exponential of the minus of the length of the D5-brane in the upper-left part. Also we define $X$ as the exponential of the minus of the height of the O7$^-$-plane. We will soon determine $Q$ and $X$ by the parameters in Figure \ref{fig:SU2Nantisymdiag} but let us first see how the instanton fugacity of the $SU(2N)$ gauge theory with one antisymmetric hypermultiplet can be determined from the web diagram in Figure \ref{fig:SU2NantisymT}.

\begin{figure}
\centering
\includegraphics[width=12cm]{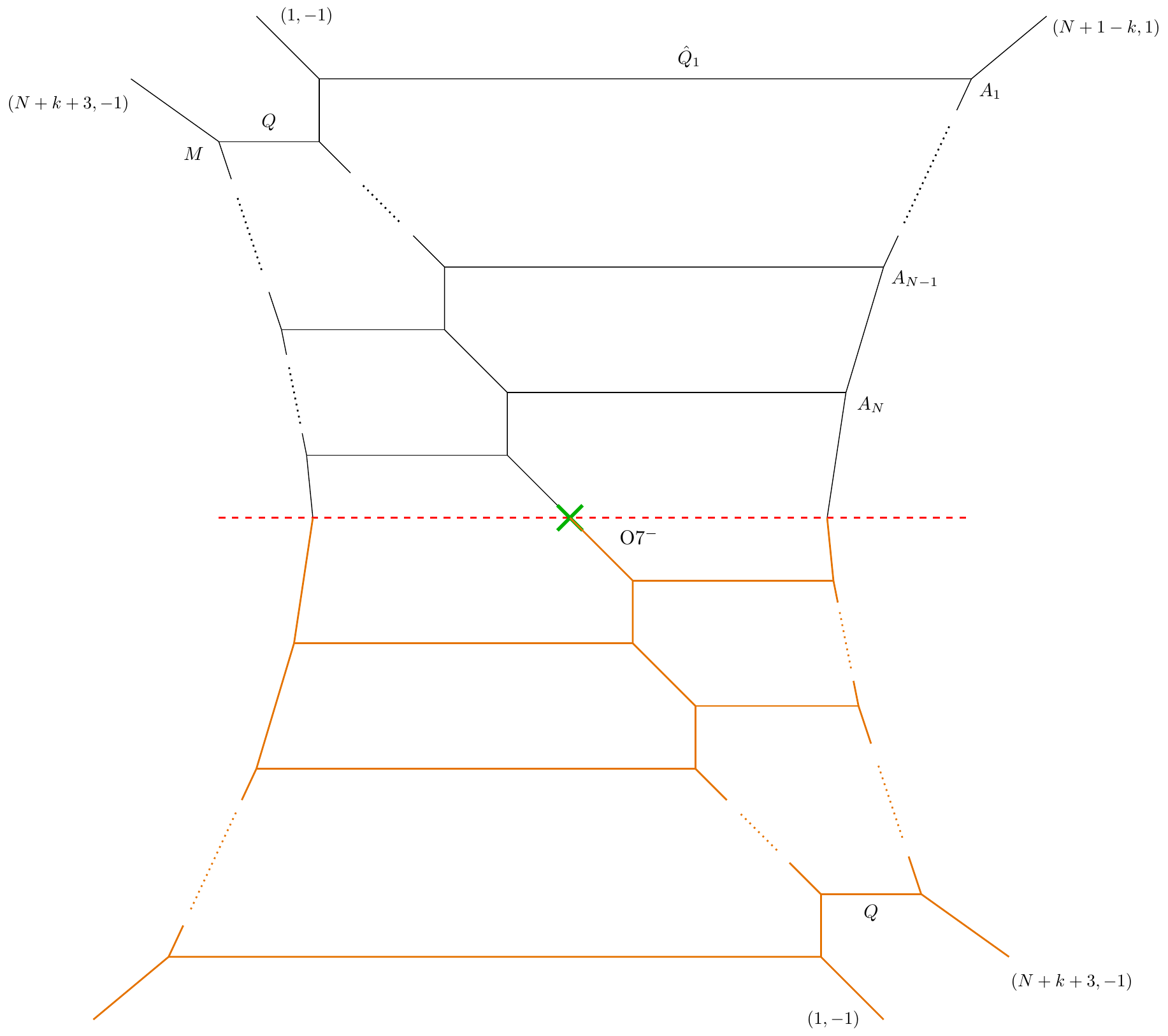}
\caption{The double cover of the diagram appearing in Figure \ref{fig:SU2NantisymT}.}
\label{fig:SU2NantisymTdouble}
\end{figure}

\begin{figure}
\centering
\includegraphics[width=16cm]{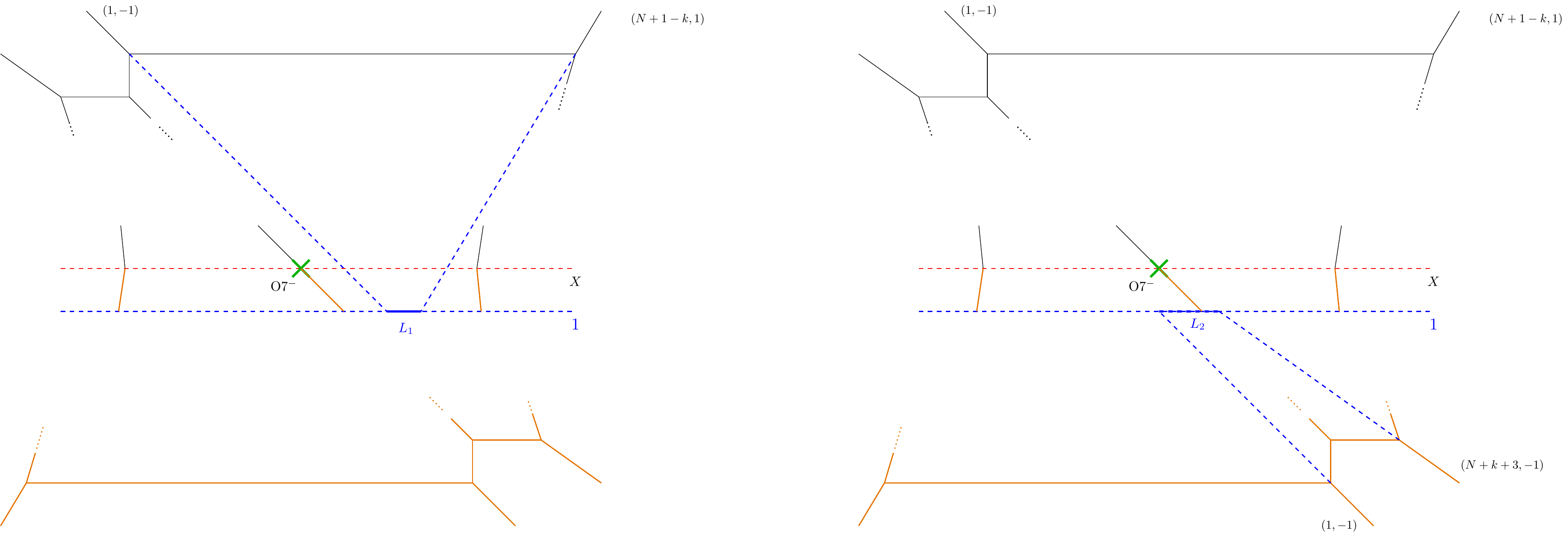}
\caption{Left: The length $L_1$ obtained from the extrapolation of the two upper-right external 5-branes. Right: The length $L_2$ obtained from the extrapolation of the two lower-right external 5-branes.}
\label{fig:gaugecoupling1}
\end{figure}

For that, we consider a double cover of Figure \ref{fig:SU2NantisymT} by adding the mirror image of the upper-half diagram as in Figure \ref{fig:SU2NantisymTdouble}. 
The 5-brane web diagram in orange in Figure \ref{fig:SU2NantisymTdouble} is mirror to the 5-brane web diagram in black in Figure \ref{fig:SU2NantisymTdouble}. From the double cover, one can choose a fundamental region as we like. In order to determine the instanton fugacity, we choose the right-half fundamental region. Then the instanton fugacity or the gauge coupling can be determined by shrinking all the Coulomb branch moduli \cite{Aharony:1997bh}. In practice, the instanton fugacity can be obtained by taking the average of the lengths given by extrapolating upper and lower external 5-branes to the origin \cite{Bao:2011rc}. When we extrapolate the two upper external 5-branes to a horizontal line which is located at the origin in the vertical direction as on the left of Figure \ref{fig:gaugecoupling1}, 
the length between the external 5-branes on the horizontal line is given by
\be
L_1 = \hat{Q}_1\left(A_1\right)^{-1}\left(A_1^{N+1-k}\right)^{-1} = \hat{Q}_1A_1^{-N-2+k}.
\ee
On the other hand, when we extrapolate the two lower external 5-branes to the horizontal line which is located at the origin in the vertical direction as on the right of Figure \ref{fig:gaugecoupling1}, the length between the external 5-branes on the horizontal line becomes
\be
L_2 = Q\left(A_1^{-1}X^{2}\right)^{-1}\left(\left(X^{-2}M\right)^{N+k+3}\right)^{-1}= QA_1M^{-N-k-3}X^{2N+2k+4}.
\ee
Note that the heights of the left-lowest D5-brane and right-lowest D5-brane are $A_1^{-1}X^2$, $X^2M^{-1}$ respectively. Then the square of the instanton fugacity $u$ is $L_1L_2$, namely
\be
u^2_{SU(2N)} = L_1L_2 =  \hat{Q}_1QA_1^{-N-1+k}M^{-N-k-3}X^{2N+2k+4}. \label{instantonpre}
\ee

The last task is to rewrite $X$ and $Q$ in terms of the parameters in Figure \ref{fig:SU2Nantisymdiag}. Hence, we carefully follow what happens when the O7$^-$-plane is resolved into an $[1, -1]$ 7-brane and a $[1, 1]$ 7-brane. 
\begin{figure}
\centering
\includegraphics[width=12cm]{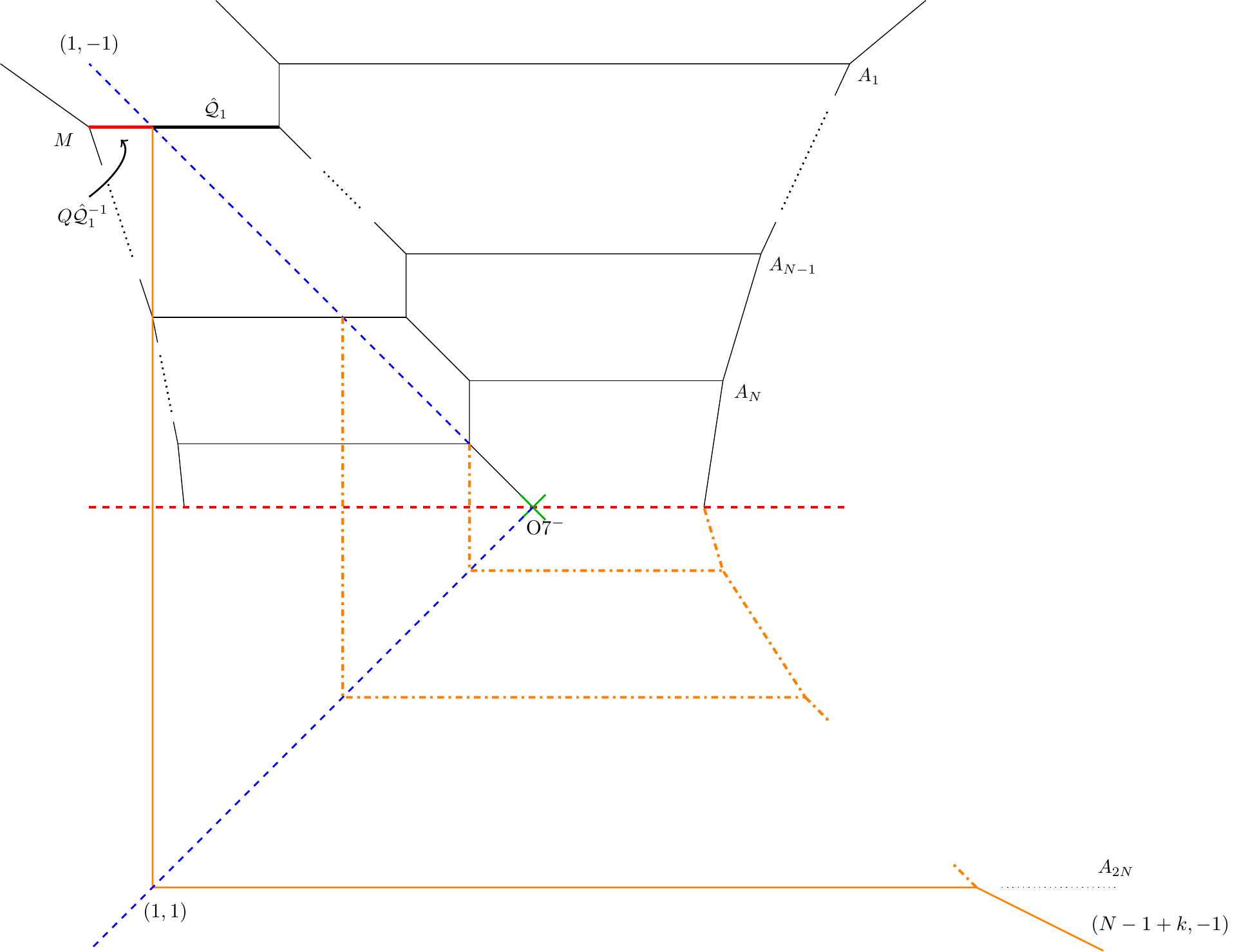}
\caption{Result of Hanany-Witten transitions on the web diagram after resolution of the orientifold plane.}
\label{fig:resolutionO7}
\end{figure}
One can see that the $[1, -1]$ 7-brane which ends on the upper-left external $(1. -1)$ 5-brane originates from the resolution of the O7$^-$-plane. Also, the $[1, 1]$ 7-brane which ends on the lower-left external $(1, 1)$ 5-brane originates from the resolution of the O7$^-$-plane. In fact, the position of the O7$^-$-plane of the 5-brane web in Figure \ref{fig:SU2Nantisymdiag} is identified by the intersection of two lines which are obtained by extrapolating the two external 5-branes on the lefthand side in Figure \ref{fig:SU2Nantisymdiag} \cite{Hayashi:2016abm}. The $(N+k+3, -1)$ external 5-brane in Figure \ref{fig:SU2NantisymT} becomes the lower-right $(N-1+k, -1)$ 5-brane after crossing the branch cut of the O7$^-$-plane. The structure is summarised in Figure \ref{fig:resolutionO7}. From the diagram in Figure \ref{fig:resolutionO7}, the height of the location of the O7$^-$-plane is the middle between the height $M$ and the height $A_{2N}$. Therefore, $X$ can be identified as
\be
X = M^{\frac{1}{2}}A_{2N}^{\frac{1}{2}}.\label{Xis}
\ee
Then the instanton fugacity \eqref{instantonpre} can be written by
\be
u^2_{SU(2N)} = \hat{Q}_1QA_1^{-N-1+k}A_{2N}^{N+k+2}M^{-1}. \label{instantonpre2}
\ee

A part of the 5-brane whose length is $Q$ is affected by the branch cut of the O7$^-$-plane. In $Q$, the $\hat{\mathcal{Q}}_1$ part remains the same as the original one but the rest part which we denote by a red line in Figure \ref{fig:resolutionO7} changes into the orange straight lines in Figure \ref{fig:resolutionO7}. Therefore, 
\be
Q\left(\hat{\mathcal{Q}}_1\right)^{-1} = MA_{2N}^{-1}\hat{Q}_{2N}\,,\ \
\longleftrightarrow \ \ Q = \hat{Q}_{2N}A_1^{-1}M^{-N+2}A_{2N}^{-N-1}. \label{Qis}
\ee
Then, inserting \eqref{Xis} and \eqref{Qis} into the expression of the instanton fugacity \eqref{instantonpre} finally yields
\be
u^2_{SU(2N)} = \hat{Q}_1\hat{Q}_{2N}A_1^{-N-2+k}A_{2N}^{k+1}M^{-N+1}.\label{inst.fugacity}
\ee
It is also possible to rewrite \eqref{inst.fugacity} in terms of $Q_1, \cdots, Q_{2N-1}, \tilde{Q}_1, \hat{Q}_1, \hat{Q}_{2N}$ by using \eqref{QandA}, \eqref{A2N} and \eqref{AMs1}. 

In summary, the relations between the parameters in the 5-brane web and the moduli and the parameters of the 5d $SU(2N)$ gauge theory with antisymmetric matter are \eqref{coulomb.SU2N}, \eqref{mass.antisym} and \eqref{inst.fugacity}.

\subsection{The partition function}

\begin{figure}
\centering
\includegraphics[width=7cm]{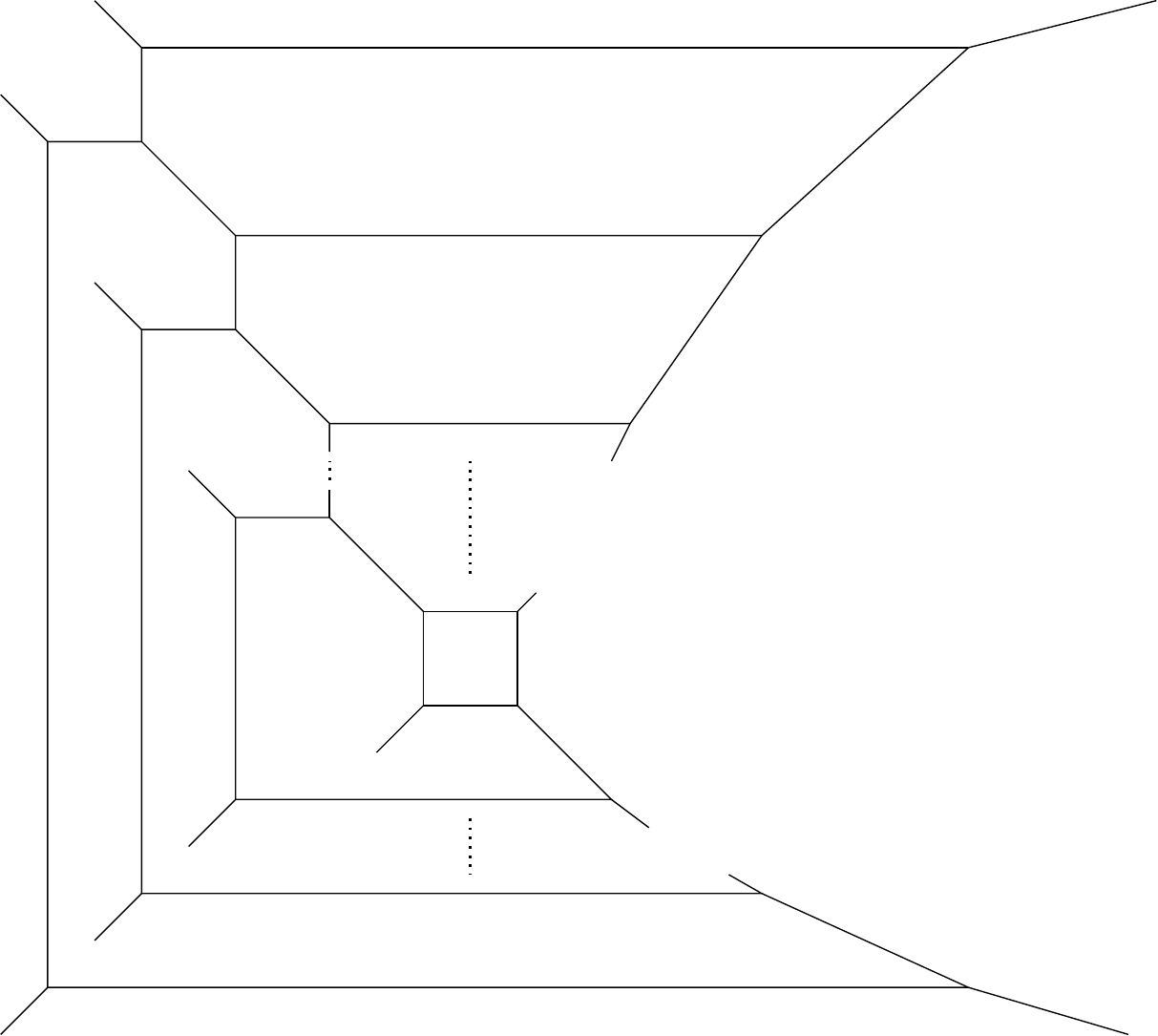}
\caption{Non-toric web diagram giving an $SU(2N)_k$ theory with antisymmetric matter.}
\label{fig:replace}
\end{figure}

Given the parametrisations \eqref{coulomb.SU2N}, \eqref{mass.antisym} and \eqref{inst.fugacity}, we have all the necessary ingredients for the computation of the topological string partition function and its comparison with the corresponding Nekrasov partition function obtained by localisation techniques. The basic tool for computing the topological string partition function is the topological vertex which we briefly review  in Appendix \ref{sec:top}. In Figure \ref{fig:SU2Nantisymdiag}, some external 5-branes jump over other internal 5-branes. For those cases, we can simply replace the configuration with a 5-brane web without any jump as in Figure \ref{fig:replace} \cite{Hayashi:2013qwa, Hayashi:2015xla} by shrinking the external 5-branes until no jumps occur.
 Putting trivial representations for the shrunken 5-branes as well as remaining external 5-branes, we can apply the topological vertex to the 5-brane web given in Figure \ref{fig:replace}.

\begin{figure}
\centering
\includegraphics[width=7cm]{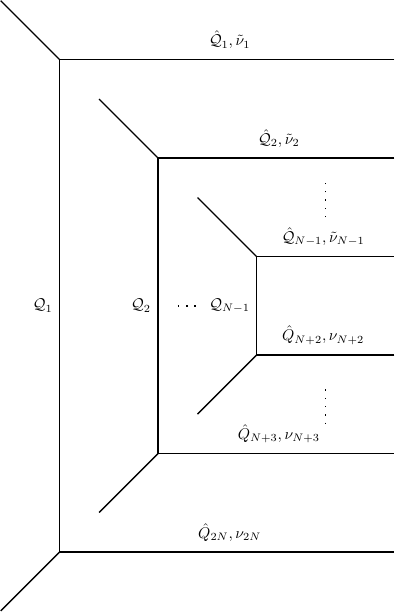}
\caption{The first strip of the diagram appearing in Figure \ref{fig:replace}.}
\label{fig:strip1}
\end{figure}

\begin{figure}
\centering
\includegraphics[width=7cm]{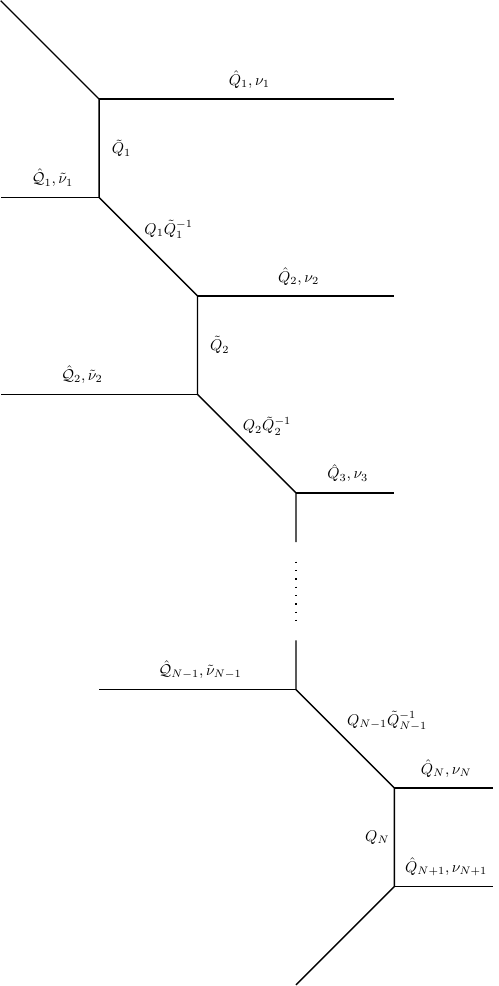}
\caption{The second strip of the diagram appearing in Figure \ref{fig:replace}.}
\label{fig:strip2}
\end{figure}

\begin{figure}
\centering
\includegraphics[width=9cm]{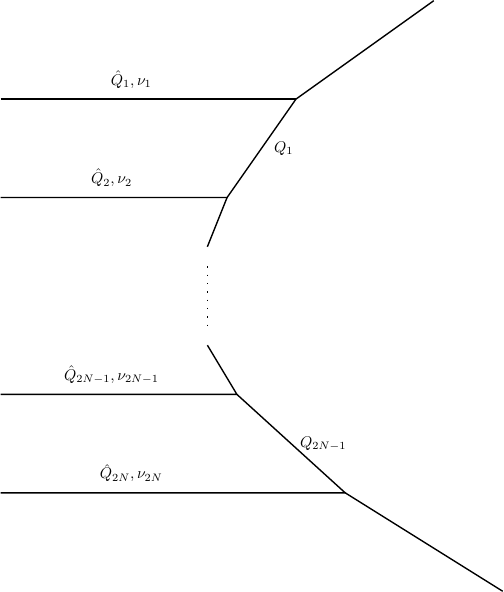}
\caption{The third strip of the diagram appearing in Figure \ref{fig:replace}.}
\label{fig:strip3}
\end{figure}

 We find it convenient for the application of the topological vertex to the 5-brane web in Figure \ref{fig:replace} to first cut the diagram into three vertical strips and then glue the three pieces after computing the partition function of each part. Each strip is depicted in Figure \ref{fig:strip1}, Figure \ref{fig:strip2} and Figure \ref{fig:strip3}.
We denote the topological string partition function from the vertical strip geometry by $Z_1\left(\{\nu_i\}, \{\tilde{\nu}_k\}\right)$ from Figure \ref{fig:strip1}, $Z_2\left(\{\nu_i\}, \{\tilde{\nu}_k\}\right)$ from Figure \ref{fig:strip2} and $Z_3\left(\{\nu_i\}\right)$ from Figure \ref{fig:strip3}. $\nu_i, (i=1, \cdots, 2N)$ and $\tilde{\nu}_k, (k=1, \cdots, N-1)$ are the Young diagrams associated to the horizontal legs in Figure \ref{fig:replace}. The application of the formulae of the topological vertex in appendix \ref{sec:top} is straightforward and we obtain 
\be
Z_1\left(\{\nu_i\}, \{\tilde{\nu}_k\}\right) = \prod_{i=1}^{N-1} \CH ( {\mathcal Q}_i , \tilde \nu_i, \nu_{2N+1-i}^t)^{-1},\\
\ee
\be\begin{split}
Z_2(\nu,\tilde \nu) =& \left[\prod_{i=1}^{N-1} \prod_{j=i}^{N-1} \CH (\tilde  Q_j \prod_{k=i+1}^j Q_{k-1},\nu_i,\tilde \nu_j^t)\right]\left[
\prod_{j=1}^{N-1}\prod_{i=j+1}^{N+1} \CH (\tilde  Q_j^{-1} \prod_{k=j+1}^i Q_{k-1}, \nu_i^t,\tilde \nu_{j})\right]\times \\&
\left[\prod_{i=1}^N \prod_{j=i+1}^{N+1} \CH (\prod_{k= i}^{j-1} Q_k,\nu_i,\nu_j^t)\right]^{-1} \left[\prod_{j=1}^N\prod_{i=j+1}^{N-1}
\mathcal H (\prod_{k=j}^{i-1} Q_{k-1} \tilde Q_{k-1}^{-1} \tilde Q_k ,\tilde \nu_j, \tilde \nu_i^t)  \right]^{-1}\,,
\end{split}\ee
 \be
Z_3\left(\{\nu_i\}\right)  = \prod_{\beta=1}^{2N-1} \prod_{\alpha = \beta}^{2N-1} \CH (\prod_{\kappa = \beta}^\alpha Q_\kappa,\nu_\beta,\nu^t_{\alpha+1})^{-1}\,,
\ee
where we introduced definitions
\be
\CH(Q , \nu_1, \nu_2) :=\prod_{i,j=1}^\infty (1-Q q^{i+j-1-\nu_{1,i}-\nu_{2,j}})\,,
\ee
and 
\be\begin{split}
&\mathcal Q_j := Q_{2N-j}^2 \,\mathcal Q_{j+1}\,, \quad j = 1,\dots,N-2\,,\\
& \mathcal Q_{N-1} := Q_{N+1} Q_{N} Q_{N-1} \tilde Q_{N-1}\,.
\end{split}\ee
Here $q=e^{g_s}$ where $g_s$ is the topological string coupling constant. In terms of the Nekrasov partition function, it becomes the $\Omega$-deformation parameter.\footnote{Note that for comparison between the topological string and the Nekrasov partition function we need to 
work in the case in which the two $\epsilon_i, i=1,2$ parameters of the $\Omega$-background satisfy $\epsilon_1 = -\epsilon_2 = \epsilon$.}

Then, the total topological string partition function is given by gluing the three pieces by summing the Young diagrams $\nu_i$ for $i=1, \cdots, 2N$ and $\tilde{\nu}_k$ for $a=1, \cdots, N-1$ with some weights. The gluing along the horizontal lines parametrised by the Young diagrams $\tilde{\nu}_k, (k=1, \cdots, N-1)$ is 
\be
Z_{1+2}\left(\{\nu_i\}\right) = \sum_{\tilde{\nu}_1, \cdots, \tilde{\nu}_{N-1}} \left[\prod_{k=1}^{N-1}(- \hat {\mathcal Q}_k)^{|\tilde \nu_{k}|}\mathcal K(\tilde{\nu}_k)\right]Z_1\left(\{\nu_i\}, \{\tilde{\nu}_k\}\right)  Z_2\left(\{\nu_i\}, \{\tilde{\nu}_k\}\right),
\ee
where we defined 
\be
\mathcal K (\lambda) = q^{\frac{||\lam||^2+||\lam^t||^2}{2}} \; \tilde Z_\lambda^2(q)\,.
\ee
Further gluing along the horizontal lines parametrised by the Young diagrams $\nu_i, (i=1, \cdots, 2N)$ gives
\be\begin{split}\label{top.part}
Z_{top} = \sum_{\nu_1, \cdots, \nu_{2N}}&\left[\prod_{i=1}^{N}(-\hat Q_i)^{|\nu_i|} f_{\nu_i}(q)^{N+2-\kappa-i}\mathcal K(\nu_i)\right]\left[\prod_{i=N+1}^{2N}(-\hat Q_{i})^{|\nu_{i}|} f_{\nu_{i}}(q)^{-\kappa-(i-N-1)}\mathcal K(\nu_i)\right]\times\\& Z_{1+2}\left(\{\nu_i\}\right) Z_3\left(\{\nu_i\}\right).
\end{split}\ee

Eq.~\eqref{top.part} yields the full topological string partition function by applying the topological vertex to the 5-brane web in Figure \ref{fig:SU2Nantisym1} or equivalently Figure \ref{fig:replace}. However, when comparing this result with the Nekrasov partition function it is necessary to remove the contributions coming from decoupled factors  \cite{Bergman:2013ala, Hayashi:2013qwa, Bao:2013pwa, Bergman:2013aca}. Without the cancellation of these contributions the topological string partition function would otherwise contain contributions of incomplete 5d BPS multiplets.
The contributions can be easily read off from those of strings between parallel external legs in the 5-brane web diagram. This is also true when we consider 5-brane web diagrams which include the configuration of 5-brane jumps \cite{Hayashi:2015xla}. 
It is clear from the 5-brane web in Figure \ref{fig:SU2Nantisym1} that we have a universal decoupled factor
\be
Z_{dec}^{\text{univ}} = \CH \left(\tilde{Q}_1^2\left[\prod_{k=2}^{N-1}\tilde{Q}_k\right]\left[\prod_{k=1}^{N-1}Q_i^{-1}Q_{i+N}\right]\right)^{-1}.
\ee
Here we used an abbreviated notation
\be
\CH(Q) := \CH(Q, \; \emptyset, \; \emptyset).
\ee
For special value of the CS level, there can be an additional decoupled factor. For example, when $\kappa = N+2$, we have 
\be
Z_{dec}^{\kappa=N+2} = \CH(\hat{Q}_1)^{-1}.
\ee
When $\kappa = -N$, we have 
\be
Z_{dec}^{\kappa=-N} = \CH(\hat{Q}_{2N})^{-1}.
\ee
Namely, the total decoupled factor $Z_{dec}$ is $Z_{dec}^{\text{univ}}$ for $-N+1 \leq \kappa \leq N+1$, $Z_{dec}^{\text{univ}}\cdot Z_{dec}^{\kappa = N+2}$ for $\kappa = N+2$ or $Z_{dec}^{\text{univ}} \cdot Z_{dec}^{\kappa = -N}$ for $\kappa = -N$. 

Therefore, we claim that the Nekrasov partition function of the 5d $SU(2N)$ gauge theory with a hypermultiplet in the antisymmetric representation is give by
\be
Z_{Nek}^I\left(\{\alpha_i\}, m, u_{SU(2N)} \right) = \frac{Z_ {top}}{Z_{dec}}. \label{nek1}
\ee
At the moment, the correspondence is obscure. We will see more explicit correspondence by splitting the right-hand side of \eqref{nek1} into the perturbative part and the instanton part of the 5d $SU(2N)$ gauge theory with one antisymmetric hypermultiplet.

\subsubsection{Perturbative part}

We first see the perturbative part of the Nekrasov partition function \eqref{nek1} of the 5d $SU(2N)$ gauge theory with the antisymmetric hypermultiplet. The perturbative part is reproduced by taking the limit of  zero gauge coupling or $u_{SU(2N)} \rightarrow 0$. In terms of the parameters in the 5-brane web in Figure \ref{fig:replace}, we set $\hat{Q}_i = 0$ for $i=1, \cdots, 2N$. This condition reduces all the Young diagram summations with respect to $\nu_i, (i=1, \cdots 2N)$ in \eqref{top.part} and we finally obtain
\be
Z_{pert}^I = \frac{Z_{1+2}\left(\{\nu_i = \emptyset \}\right)Z_3\left(\{\nu_i= \emptyset\}\right)}{Z_{dec}^{\text{univ}}}. \label{pert1}
\ee
Note that this part does not depend on the CS level $k$, which is consistent with the field theory expectation. By using the identity \cite{Nakajima:2003pg},
\be
\prod_{i,j=1}^{\infty}\left(1 - Qq^{i+j-1-\nu_{1, j} - \nu_{2,i}}\right) = \prod_{i,j=1}^{\infty}\left(1 - Qq^{i+j-1}\right)\prod_{s\in \nu_2}\left(1 - Qq^{-a_{\nu_1^t}(s) - l_{\nu_2} - 1}\right)\prod_{s\in\nu_1^t}\left(1 - Qq^{a_{\nu_2(s)}+l_{\nu_1^t}+1}\right),
\ee
where $l_{\nu}(i, j) = \nu_i - j$ and $a_{\nu}(i, j) = \nu^t_j - i$, it is possible to rewrite \eqref{pert1} as
\be\label{pertSU2Nantisym}
Z_{pert}^I = Z_{pert1}^I\frac{Z_{pert2}^I}{Z_{dec}^{\text{univ}}},
\ee
where
\be\begin{split}
Z_{pert1}^I  =& \left[\prod_{i=1}^{2N-1} \prod_{j = i}^{2N-1} \CH (\prod_{k = i}^j Q_k)\right]^{-1} \left[\prod_{i=1}^{N-1} \prod_{j=i-1}^{N-2} \CH (\tilde  Q_i \prod_{k=i}^j Q_{2N-k})\right]\left[
\prod_{j=1}^{N-1}\prod_{i=j}^{N} \CH (Q_j\tilde  Q_j^{-1} \prod_{k=j}^{i-1} Q_{k+1})\right]\times \\&
\left[\prod_{i=1}^N \prod_{j=i+1}^{N+1} \CH (\prod_{k= i}^{j-1} Q_k)\right]^{-1} \left[\prod_{i=1}^{N-2}\prod_{j=i}^{N-2}
\mathcal H (\prod_{k=i}^{j} Q_{2N-k})  \right]^{-1}\left[\prod_{i=1}^{N-1} \CH (\mathcal Q_i)\right]^{-1}\,,
\end{split}\ee
and 
\be
Z_{pert2}^I =\sum_{\tilde{\nu}_1, \cdots, \tilde{\nu}_{N-1}}\prod_{i=1}^{N=1}\left(-\hat{\mathcal{Q}}_i\right)^{|\tilde{\nu}_i|}\mathcal{K}\left(\tilde{\nu}_i\right)\mathcal{A}_i\left(\left\{\tilde{\nu}\right\}\right),
\ee
with
\be
\mathcal A_j (\tilde \nu) = \prod_{s \in \tilde \nu_j} \frac{\prod_{i=1}^j \mathcal L (\tilde Q_i \prod_{k=i+1}^j Q_{k-1},\tilde \nu_j,
\emptyset,2)\prod_{i=j+1}^{N+1} \mathcal L (\tilde Q_j^{-1} \prod_{k = j+1}^i Q_{k-1},\tilde \nu_j,\emptyset,1)}{
\mathcal L (\mathcal Q_j,\tilde \nu_j,\emptyset,1)\prod_{i=j+1}^{N-1}
\mathcal L (\prod_{k=j}^{i-1} Q_{2N-k} ,\tilde \nu_j, \tilde \nu_i,1) \prod_{i=1}^{j-1} \mathcal L (\prod_{k=i}^{j-1} Q_{2N-k},\tilde \nu_j,\tilde
\nu_i,2)}\,.
\ee
Here we introduced the notation 
\be
\mathcal L (Q,\nu_1,\nu_2,k) := (1- Q q^{(-1)^k(l_{\nu_1}(s)+a_{\nu_2}(s)+1)})\,.
\ee
Due to geometric constraints we have $\hat {\mathcal{Q}}_k = Q_{N+1} Q_{N-1}^{-1} \tilde Q_{N-1} \prod_{l=k+1}^{N-1} \tilde Q_l$. 

Since the eq.~\eqref{pertSU2Nantisym} should be equal to the perturbative partition function of a 5d $SU(2N)$ gauge theory with a hypermultiplet in the antisymmetric representation, we conjecture that 
\be\label{eq:1}
Z_{pert2}^I = \frac{\left[\prod_{k=1}^{N-1} \prod_{\alpha = N-1}^{2N-k-1} \CH(\tilde Q_k \prod_{l=k}^\alpha Q_{2N-l})\right]\left[
\prod_{k=1}^{N-1} \prod_{\alpha =N+1}^{2N-k-1} \CH(Q_k \tilde Q_k^{-1} \prod_{l=k}^{\alpha-1} Q_{l+1})\right]\prod_{i=1}^{N-1}\CH(\mathcal Q_i)}{\CH (\tilde Q_1 \hat {\mathcal Q}_1)\prod_{i=1}^{N+1}\prod_{j=1}^{N-1}\CH(Q_{N+1} \prod_{k=i}^{N}Q_{k}\prod_{l=1}^{j-1} Q_{N+1+l})}\,.
\ee
In \eqref{eq:1} we defined $\tilde Q_N \equiv Q_{N+1} Q_{N-1}^{-1} \tilde Q_{N-1}$. Then, the partition function \eqref{pertSU2Nantisym} exactly reproduces the perturbative partition function of a 5d $SU(2N)$ gauge theory with a hypermultiplet in the antisymmetric representation with the contribution from the antisymmetric hypermultiplet
\be
Z_{pert\;hyper}^I  = \left[\prod_{k=1}^{N-1} \prod_{\alpha = k-1}^{2N-k-1} \CH (\tilde Q_k \prod_{l=k}^\alpha Q_{2N-l})\right]\left[ \prod_{k=1}^{N-1} \prod_{\alpha = k}
^{2N-k-1} \CH (Q_k \tilde Q_k^{-1} \prod_{l=k}^{\alpha-1} Q_{l+1})\right]\,.
\ee

We will check the equality \eqref{eq:1} in the examples of $N=2, 3$ and see the partition function \eqref{pertSU2Nantisym} reproduces the perturbative partition function of the 5d $SU(2N)$ gauge theory with a antisymmetric hypermultiplet more explicitly.  

\paragraph{Examples: $N=2,3$} 

Here we concentrate on the cases of $N=2, 3$. 

In the case of $N=2$. the perturbative parts of the partition function is given by 
\be\begin{split}
Z_{pert1}^I =& \,\CH (Q_1)^{-2} \CH(Q_2)^{-2} \CH(Q_1 Q_2)^{-2} \CH(Q_3)^{-1} \CH(Q_1 Q_2 Q_3)^{-1} \CH(Q_2 Q_3)^{-1} \\&\,\CH(Q_m) \CH(Q_m Q_2) \CH (Q_1 Q_m^{-1})
\CH(Q_m Q_2 Q_3)^{-1}\,.
\end{split}\ee
and 
\be\begin{split}\label{pert2SU4antisym}
Z_{pert2}^I &= \sum_{\tilde{\nu}_1}(- Q_3 Q_m^{-1})^{|\tilde{\nu}_1|} \,\mathcal K \left(\tilde{\nu}_1\right)\\&\, \prod_{s \in \tilde{\nu}_1} \frac{(1- Q_1Q_m^{-1} q^{l_{\tilde{\nu}_1}(s) +a_{\emptyset}(s) +1})(1-  Q_m q^{-l_{\tilde{\nu}_1}(s) -a_{\emptyset}(s) -1})(1- Q_2 Q_m q^{-l_{\tilde{\nu}_1}(s) -a_{\emptyset}(s) -1})}{(1-Q_m Q_2 Q_3 q^{-l_{\tilde{\nu}_1}(s)-a_{\emptyset}(s)-1)})}.
\end{split}\ee
In this case, the conjecture \eqref{eq:1} becomes
\be\label{pert2conjSU4antisym}
Z_{pert2}^I  = \frac{\CH (Q_2 Q_3 Q_m)\CH (Q_3 Q_m^{-1}) \CH(Q_1 Q_3 Q_m^{-1}) \CH(Q_1 Q_2 Q_3 Q_m^{-1})}{\CH(Q_1 Q_3 Q_m^{-2}) \CH(Q_3) \CH(Q_2 Q_3) 
\CH(Q_1 Q_2 Q_3)}\,.
\ee
While it is not easy to analytically prove this identity it is possible to check it expanding both sides in series of $Q_3$. This is because the right-hand side of \eqref{pert2SU4antisym} is expanded in $Q_3$ and the argument in each factor on the righthand side of \eqref{pert2conjSU4antisym} contains $Q_3$ with a positive power. Explicit computation shows the agreement up to order $Q_3^7$. 

Then, the perturbative partition function becomes
\be\begin{split}
Z_{pert}^I =& \,\CH (Q_1)^{-2} \CH(Q_2)^{-2} \CH(Q_1 Q_2)^{-2} \CH(Q_3)^{-2} \CH(Q_1 Q_2 Q_3)^{-2} \CH(Q_2 Q_3)^{-2} \\&\,\CH(Q_m) \CH(Q_m Q_2) \CH (Q_1 Q_m^{-1})
\CH(Q_3 Q_m^{-1}) \CH (Q_1 Q_3 Q_m^{-1}) \CH(Q_1 Q_2 Q_3 Q_m^{-1})\,,
\end{split}\ee
By using the parametrisation \eqref{coulomb.SU2N} and \eqref{mass.antisym} or more explicitly, 
\be
Q_1 = e^{-(\alpha_1-\alpha_2)}\,, \quad Q_2 = e^{-(\alpha_2-\alpha_3)}\,, \quad Q_3 = e^{-(\alpha_3-\alpha_4)}\,, \quad Q_m = e^{-m +\alpha_4+\alpha_2}\,,
\ee
we obtain the following perturbative partition function 
\be\begin{split}
Z_{pert}^I =& \frac{\CH(e^{m-\alpha_1-\alpha_4}) \CH(e^{-m+\alpha_2+\alpha_4})
\CH(e^{-m+\alpha_3+\alpha_4})\CH(e^{m-\alpha_2-\alpha_3})\CH(e^{m-\alpha_1-\alpha_3})\CH(e^{m-\alpha_1-\alpha_2}) }{\CH (e^{-(\alpha_1-\alpha_2)})^2\CH (e^{-(\alpha_1-\alpha_3)})^2\CH (e^{-(\alpha_1-\alpha_4)})^2\CH (e^{-(\alpha_2-\alpha_3)})^2\CH (e^{-(\alpha_2-\alpha_4)})^2  
\CH (e^{-(\alpha_3-\alpha_4)})^2}\,,
\end{split}\ee
which exactly agrees with the perturbative partition function of a 5d $SU(4)$ gauge theory with a hypermultiplet in the antisymmetric representation. 

In the case of $N=3$. the perturbative parts of the partition function is given by 
\be\begin{split}
Z_{pert1}^I = \frac{\CH \left(\frac{Q_1 Q_2 }{Q_m Q_5}\right)\CH \left(\frac{Q_1 Q_2}{Q_m}\right) \CH \left(\frac{Q_m Q_5}{Q_2}\right)\CH \left(\frac{Q_2}{Q_m}\right)\CH(Q_m) \CH (Q_m Q_5) \CH (Q_m Q_3) \CH (Q_m Q_3 Q_5)}{\prod_{i = 1}^ 5\left[\prod_{k=i}^3\CH (Q_i \prod_{j=i+1}^k Q_{j})^2 \prod_{k=4}^5\CH (Q_i \prod_{j=i+1}^k Q_{j})
  \right]\CH (Q_m Q_3 Q_4) \CH (Q_m Q_3 Q_4 Q_5^2)}
\end{split}\ee
and 
\be\begin{split}\label{pert2SU6antisym}
Z_{pert2}^I =& \sum_{\tilde{\nu}_1, \tilde{\nu}_2} \left(-\frac{Q_2 Q_4}{Q_m^2}\right)^{|\tilde{\nu}_1|}\left(-\frac{Q_4}{Q_m}\right)^{|\tilde{\nu}_2|}\mathcal K (\tilde{\nu}_1) \mathcal K (\tilde{\nu}_2)\\&
\prod_{s \in \tilde{\nu}_1} \frac{\mathcal L\left(\frac{Q_1 Q_2}{Q_m Q_5},\tilde{\nu}_1,\emptyset,2\right)\mathcal L \left(\frac{Q_5 Q_m}{Q_2},\tilde{\nu}_1,\emptyset,1\right)
\mathcal L (Q_5 Q_m,\tilde{\nu}_1,\emptyset,1) \mathcal L (Q_5 Q_m Q_3 ,\tilde{\nu}_1,\emptyset,1)}{
\mathcal L (Q_4 Q_3 Q_5^2 Q_m,\tilde{\nu}_1,\emptyset,1)\mathcal L (Q_5,\tilde{\nu}_1,\tilde{\nu}_2,1)}\\&
\prod_{s \in \tilde{\nu}_2}\frac{\mathcal L (Q_m, \tilde{\nu}_2, \emptyset,1)\mathcal L \left(\frac{Q_2}{Q_m}, \tilde{\nu}_2, \emptyset,2\right)\mathcal L (Q_3 Q_m, \tilde{\nu}_2, \emptyset,1)\mathcal L \left(\frac{Q_1 Q_2}{ Q_m}, \tilde{\nu}_2, \emptyset,2\right)}{\mathcal L(Q_3 Q_4 Q_m,\tilde{\nu}_2,\emptyset,1)
\mathcal L (Q_5,\tilde{\nu}_2, \tilde{\nu}_1,2)}\,.
\end{split}\ee
The conjecture \eqref{eq:1} becomes
\be\begin{split}\label{pert2conjSU6antisym}
Z_{pert2}^I =&\frac{\CH\left(\frac{Q_4}{Q_m}\right) \CH \left(\frac{Q_2 Q_4}{Q_m}\right) \CH \left(\frac{Q_1 Q_2 Q_4}{Q_m}\right) \CH \left(\frac{Q_2 Q_3 Q_4}{Q_m}\right)
\CH (\frac{Q_1 Q_2 Q_3 Q_4}{Q_m}) \CH(\frac{Q_1 Q_2^2 Q_3 Q_4}{Q_m}) }
{\CH(Q_4) \CH (Q_3 Q_4) \CH (Q_2 Q_3 Q_4) \CH (Q_1 Q_2 Q_3 Q_4) \CH (Q_4 Q_5)}\times\\&
\frac{\CH (Q_3 Q_4 Q_m) \CH (Q_3 Q_4 Q_5 Q_m)\CH (Q_3Q_4 Q_5^2 Q_m)}{\CH (Q_1 Q_2 Q_3 Q_4 Q_5) \CH \left(\frac{Q_1 Q_2^2 Q_4 }{Q_5 Q_m^3}\right) \CH (Q_3 Q_4 Q_5) \CH (Q_2 Q_3 Q_4 Q_5)}\,.
\end{split}\ee
It is possible to check this identity expanding in $Q_4$ both sides since the righthand side of \eqref{pert2SU6antisym} is expanded by $Q_4$ and the argument in each factor on the righthand side of \eqref{pert2conjSU6antisym} contains $Q_4$ with a positive power. The agreement has been found up to order $Q_4^4$. 

By using the parametrisation \eqref{coulomb.SU2N} and \eqref{mass.antisym}, we finally obtain   
\be\begin{split}
Z_{pert}^I =& \frac{\CH(e^{m-\alpha _1-\alpha _2}) \CH(e^{m-\alpha _1-\alpha_3} )\CH(e^{m-\alpha _1-\alpha_4})\CH(e^{m-\alpha_1-\alpha _5}) \CH(e^{m-\alpha _1-\alpha _6})
}{\CH (e^{-(\alpha_1-\alpha_2)})^2\CH (e^{-(\alpha_1-\alpha_3)})^2\CH (e^{-(\alpha_1-\alpha_4)})^2\CH (e^{-(\alpha_1-\alpha_5)})^2\CH (e^{-(\alpha_1-\alpha_6)})^2}\\&
\frac{\CH(e^{m-\alpha _2-\alpha_3})\CH (e^{m-\alpha _2-\alpha_4}) \CH (e^{m-\alpha _2-\alpha _5}) \CH(e^{-m+\alpha _2+\alpha_6})\CH(e^{m-\alpha _3-\alpha_4})
}{\CH (e^{-(\alpha_2-\alpha_3)})^2 \CH (e^{-(\alpha_2-\alpha_4)})^2 \CH (e^{-(\alpha_2-\alpha_5)})^2
\CH (e^{-(\alpha_2-\alpha_6)})^2\CH (e^{-(\alpha_3-\alpha_4)})^2 }\\&
\frac{\CH (e^{-m+\alpha_3+\alpha _5}) \CH (e^{-m+\alpha _3+\alpha_6}) \CH (e^{-m+\alpha _4+\alpha _5}) \CH (e^{-m+\alpha_4+\alpha _6})\CH(e^{-m+\alpha _5+\alpha _6})
}{\CH (e^{-(\alpha_3-\alpha_5)})^2\CH (e^{-(\alpha_3-\alpha_6)})^2 \CH (e^{-(\alpha_4-\alpha_5)})^2\CH (e^{-(\alpha_4-\alpha_6)})^2 \CH (e^{-(\alpha_5-\alpha_6)})^2 },
\end{split}\ee
which exactly agrees with the perturbative partition function of a 5d $SU(6)$ gauge theory with a hypermultiplet in the antisymmetric representation. 

\subsubsection{Instanton part}\label{sec:instsu2n}

The instanton part in the full partition function \eqref{nek1} can be obtained by removing the perturbative partition function \eqref{pertSU2Nantisym}, namely
\be
Z_{inst}^I = \frac{Z_{Nek}^I \left(\left\{\alpha_i\right\}, m , u_{SU(2N)}\right)}{Z_{pert}^I}.\label{instSU2Nantisym}
\ee
The instanton partition function \eqref{instSU2Nantisym} involves the Young diagram summations both of $\nu_i,\, i = 1, \cdots, 2N$ and of $\tilde{\nu}_k,\, k = 1, \cdots, N-1$. The Young diagram summations with respect to $\nu_i, i = 1, \cdots, 2N$ are related to an expansion by the instanton fugacity $u_{SU(2N)}$ of \eqref{inst.fugacity}. Therefore, the sum of the terms with $\sum_{i=1}^{2N}|\nu_i| = k$ gives the contribution of the $k$-instanton of the 5d $SU(2N)$ gauge theory with an antisymmetric hypermultiplet. However, each part of the $k$-instanton contribution is still expanded by the Young diagram summations with respect to $\tilde{\nu}_k, k = 1, \cdots, N-1$ in \eqref{instSU2Nantisym}. Hence only after summing up the Young diagram summations with respect to $\tilde{\nu}_k
,\, k = 1, \cdots, N-1$, we get an exact answer for the $k$-instanton contribution. In fact, we will see in more explicit examples that the summation of the Young diagrams  with respect to $\tilde{\nu}_k,\,  k = 1, \cdots, N-1$ in \eqref{instSU2Nantisym} are simplified due to the division by $Z_{pert}^I$, a phenomenon
similar to the one already observed in \cite{Hayashi:2013qwa, Hayashi:2014wfa}.
Also, note that the expression \eqref{instSU2Nantisym} gives an alternative form of the Nekrasov partition function of the 5d $SU(2N)$ gauge theory with a hypermultiplet in the antisymmetric representation obtained by the localisation technique \cite{Shadchin:2005mx} as the localisation result is written by a contour integral for each instanton sector.

\paragraph{Examples of $N=2, 3$} Here we concentrate on the cases of $N=2, 3$ and see the equivalence of the instanton partition function \eqref{instSU2Nantisym} to the result obtained by the localisation which is written in appendix \ref{sec:NekrasovSUNantisym}. 

For the case of $N=2$ we find that the instanton summations are the following ones
\be	\label{eq:inst2}\begin{split}
Z_{inst}^1 =&
 \sum_{\hat \nu, \,\nu_i}  (-Q_3 Q_m^{-1})^{|\hat \nu|}
\mathcal K (\hat \nu)\left[\prod_{i=1}^4 (-\hat Q_i)^{|
\nu_i|} \mathcal K (\nu_i)\right] f^{3-k}_{\nu_1}(q)f^{2-k}_{\nu_2}(q) f^{-k}_{\nu_3}(q) f^{-k-1}_{\nu_4}(q)\times \\&
\prod_{s \in \hat \nu} \frac{\mathcal L (Q_1 Q_m^{-1},\hat \nu, \nu_1,2)\mathcal L ( Q_m ,\hat \nu, \nu_2,1)\mathcal L (Q_2 Q_m,\hat \nu, \nu_3,1)}{\mathcal L (Q_2 Q_3 Q_m,\hat \nu, \nu_4,1)}\times \\&
\prod_{s \in \nu_1} \frac{\mathcal L (Q_1 Q_m^{-1} ,\nu_1,\hat \nu,1)}{\mathcal L (Q_1, \nu_1,\nu_2,1)^2
\mathcal L (Q_1Q_2, \nu_1,\nu_3,1)^2\mathcal L (Q_1Q_2Q_3, \nu_1,\nu_4,1)}\times\\&
\prod_{s \in \nu_2} \frac{\mathcal L (Q_m,\nu_2,\hat \nu,2)}{\mathcal L (Q_1, \nu_2,\nu_1,2)^2
\mathcal L (Q_2, \nu_2,\nu_3,1)^2\mathcal L (Q_2Q_3, \nu_2,\nu_4,1)}\times\\&
\prod_{s \in \nu_3} \frac {\mathcal L (Q_m Q_2,\nu_3,\hat \nu,2)}{\mathcal L (Q_1Q_2, \nu_3,\nu_1,2)^2
\mathcal L (Q_2, \nu_3,\nu_2,2)^2\mathcal L (Q_3, \nu_3,\nu_4,1)}\times\\&
\prod_{s \in \nu_4} \frac{1}{\mathcal L (Q_m Q_2 Q_3,\nu_4,\hat \nu, 2)\mathcal L (Q_1Q_2Q_3, 
\nu_4,\nu_1,2)\mathcal L (Q_2Q_3, \nu_4,\nu_2,2)\mathcal L (Q_3, \nu_4,\nu_3,2)}\,.
\end{split}\ee
Setting all the $\nu_i$ diagrams to be trivial we find that \eqref{eq:inst2} reduces to \eqref{pert2SU6antisym} and therefore
gives contributions to the perturbative part of the partition function. Since we are interested in computing the 1-instanton part
of the partition function we need to compute \eqref{eq:inst2} setting one of the $\nu_i$ diagrams to be a one box diagram and the 
remaining ones to be trivial and subtract the contributions to the perturbative part. In all cases we are able to make an ansatz for the complete $\hat \nu$ 
summation and check this order
by order in $Q_3$ up to $Q_3^5$. Here we quote the various ans\"atze that we found, calling $\mathcal I_i$ the ansatz for the
summation with $\nu_i = \{1\}$ and $\nu_j = \emptyset$ for $j \neq i$
\be\begin{split}
  \mathcal I_1=&\,\frac{\hat{Q}_1\, q \,(-1)^{\kappa }  Q_1 Q_3  \left(1-Q_m\right) \left(1-Q_2 Q_m\right)}{(1-q)^2
   \left(1-Q_1\right){}^2 \left(1-Q_1 Q_2\right){}^2 \left(1-Q_1 Q_2 Q_3\right){}^2
   Q_m^2}\\&-\frac{\hat{Q}_1 \,q \,(-1)^{\kappa }  \left(Q_1-Q_m\right)}{(q-1)^2
   \left(1-Q_1\right){}^2 \left(1-Q_1 Q_2\right){}^2 \left(1-Q_1 Q_2 Q_3\right) Q_m}
\end{split}\ee
\be\begin{split}
\mathcal I_2=&\,\frac{\hat{Q}_2 \, q\,(-1)^{\kappa+1 }   \left(1-Q_m\right)}{(q-1)^2 \left(1-Q_1\right){}^2
   \left(1-Q_2\right){}^2 \left(1-Q_2 Q_3\right)}+\\&\frac{\hat{Q}_2\, q\, (-1)^{\kappa+1 }  Q_3 
   \left(1-\frac{Q_1}{Q_m}\right) \left(1-Q_2 Q_m\right)}{(1-q)^2 \left(1-Q_1\right){}^2
   \left(1-Q_2\right){}^2 \left(1-Q_2 Q_3\right){}^2}
\end{split}\ee
\be\begin{split}
 \mathcal I_3=&\,\frac{\hat{Q}_3\, q \, (-1)^{\kappa+1 }  Q_2 Q_3  \left(1-\frac{Q_1}{Q_m}\right)
   \left(1-Q_m\right)}{(1-q)^2 \left(1-Q_2\right){}^2 \left(1-Q_1 Q_2\right){}^2
   \left(1-Q_3\right){}^2}+\\&\frac{\hat{Q}_3 \, q\, (-1)^{\kappa +1}  \left(1-Q_2 Q_m\right)}{(1-q)^2
   \left(1-Q_2\right){}^2 \left(1-Q_1 Q_2\right){}^2 \left(1-Q_3\right)}
\end{split}\ee
\be\begin{split}
\mathcal I_4=&\,\frac{\hat{Q}_4\,q\,(-1)^{\kappa } 
   \left(1-\frac{Q_3}{Q_m}\right)}{(1-q)^2 \left(1-Q_3\right)^2 \left(1-Q_2
   Q_3\right)^2 \left(1-Q_1 Q_2 Q_3\right)}+\\&\frac{   \hat{Q}_4\,q\,(-1)^{\kappa}Q_3
\left(1-Q_m\right) \left(1-Q_2 Q_m\right)}{(1-q)^2\left(1-Q_3\right)^2\left(1-Q_2
   Q_3\right)^2 \left(1-Q_1 Q_2 Q_3\right)^2 Q_m}\,.
\end{split}\ee
The 1-instanton result is simply the sum of all these contributions. We compared this result with the one obtained using localisation methods and found
the complete agreement with values of the Chern-Simons level $-2 \leq \kappa \leq 4$.

We followed a similar strategy for the case of $N=3$. Here the instanton summations are
\be\label{eq:Inst3}\begin{split}
Z_{inst}^1 =&
\sum_{\hat \nu_i, \nu_j}  \left(-\frac{Q_2 Q_4}{Q_m^2}\right)^{|\hat \nu_1|}\left(-\frac{Q_4}{Q_m}\right)^{|\hat \nu_2|}
\mathcal K (\hat \nu_1) \, \mathcal K (\hat \nu_2) \left[\prod_{i=1}^6 (-\hat Q_i) \,\mathcal K (\nu_i)\right]\\
&f_{\nu_1}^{4-\kappa}(q) f_{\nu_2}^{3-\kappa}(q) f_{\nu_3}^{2-\kappa}(q) f_{\nu_4}^{-\kappa}(q)f_{\nu_5}^{-1-\kappa}(q)f_{\nu_6}^{-\kappa-2}(q)\\&
\prod_{s \in \hat \nu_1} \frac{\mathcal L \left(\frac{Q_1 Q_2}{Q_m Q_5},\hat \nu_1,\nu_1,2\right) \mathcal L \left(\frac{Q_5 Q_m}{Q_2},\hat \nu_1,\nu_2,1\right) \mathcal L (Q_5Q_m \hat \nu_1 , \nu_3,1) \mathcal L (Q_3 Q_5 Q_m,\hat \nu_1,\nu_4,1)}{\mathcal L (Q_3 Q_4 Q_5^2 Q_m,
\hat\nu_1,\nu_6,1) \mathcal L (Q_5, \hat \nu_1,\hat \nu_2,1)}\\&
\prod_{s \in \hat \nu_2} \frac{\mathcal L \left(\frac{Q_1 Q_2}{Q_m},\hat \nu_2,\nu_1,2\right) \mathcal L \left(\frac{Q_2}{Q_m}, \hat \nu_2,\nu_2,2\right) \mathcal L (Q_m , \hat \nu_2, \nu_3,1) \mathcal L (Q_3 Q_m,\hat \nu_2,\nu_4,1)}{\mathcal L (Q_3 Q_4 Q_5, \hat \nu_2,
\nu_5,1) \mathcal L (Q_5, \hat \nu_2,\hat \nu_1,2)}\\&
\prod_{s \in \nu_1} \frac{\mathcal L \left(\frac{Q_1 Q_2}{Q_m Q_5},\nu_1,\hat \nu_1,1\right)\mathcal L \left(\frac{Q_1 Q_2}{Q_m} ,\nu_1,\hat
\nu_2,1)\right) \mathcal L (Q_1 Q_2 Q_3 Q_4 Q_5,\nu_1,\nu_6,1)^{-1}}{\mathcal L (Q_1,\nu_1,\nu_2,1)^2 \mathcal L (Q_1 Q_2,\nu_1,\nu_3,1)^2 \mathcal L (Q_1 Q_2 Q_3,\nu_1,\nu_4,1)^2
\mathcal L (Q_1 Q_2 Q_3 Q_4,\nu_1,\nu_5,1)}\\&
\prod_{s \in \nu_2} \frac{\mathcal L \left (\frac{Q_5 Q_m}{Q_2}, \nu_2,\hat \nu_1,2\right)\mathcal L \left(\frac{Q_2}{Q_m},\nu_2,\hat \nu_2,1\right)
\mathcal L (Q_2 Q_3 Q_4 Q_5,\nu_2,\nu_6,1)^{-1}}{
\mathcal L (Q_1,\nu_2,\nu_1,2)^2 \mathcal L (Q_2, \nu_2,\nu_3,1)^2 \mathcal L (Q_2 Q_3,\nu_2,\nu_4,1)^2 \mathcal L (Q_2 Q_3 Q_4,\nu_2,\nu_5,1)}
\\&
\prod_{s \in \nu_3} \frac{\mathcal L (Q_m Q_5,\nu_3,\hat \nu_1,2) \mathcal L (Q_m,\nu_3, \hat \nu_2,2)\mathcal L (Q_3 Q_4 Q_5,\nu_3,\nu_6,1)^{-1}}{
\mathcal L (Q_1 Q_2,\nu_3,\nu_1,2)^2 \mathcal L (Q_2,\nu_3,\nu_2,2)^2 \mathcal L (Q_3,\nu_3,\nu_4,1)^2 \mathcal L (Q_3 Q_4,\nu_3,\nu_5,1)}\\&
\prod_{s \in \nu_4} \frac{\mathcal L (Q_3 Q_5 Q_m,\nu_4,\hat \nu_1,2) \mathcal L (Q_3 Q_m,\nu_4,\hat \nu_2,2)\mathcal L (Q_4 Q_5, \nu_4, \nu_6,1)^{-1}}
{\mathcal L (Q_1 Q_2 Q_3,\nu_4,\nu_1,2)^2 \mathcal L (Q_2 Q_3,\nu_4,\nu_2,2)^2 \mathcal L (Q_3,\nu_4,\nu_3,2)^2 \mathcal L(Q_4 ,\nu_4 ,\nu_5,1)
}\\&
\prod_{s \in \nu_5} \frac{\mathcal L (Q_3 Q_4 Q_m,\nu_5,\hat \nu_2,2)^{-1} \mathcal L (Q_1 Q_2 Q_3 Q_4,\nu_5,\nu_1,2)^{-1}}{
\mathcal L (Q_2 Q_3 Q_4 ,\nu_5,\nu_2,2) \mathcal L (Q_3 Q_4,\nu_5,\nu_3,2) \mathcal L (Q_4,\nu_5,\nu_4,2) \mathcal L (Q_5,\nu_5,\nu_6,1)}\\&
\prod_{s \in \nu_6} \frac{\mathcal L (Q_3 Q_4 Q_5^2 Q_m, \nu_6,\hat \nu_2,2)^{-1}\mathcal L (Q_1 Q_2 Q_3 Q_4 Q_5 Q_6,\nu_6,\nu_1,2)^{-1}}
{\mathcal L (Q_2 Q_3 Q_4 Q_5 ,\nu_6,\nu_2,2) \mathcal L (Q_3 Q_4 Q_5,\nu_6,\nu_3,2) \mathcal L (Q_4 Q_5,\nu_6,\nu_4,2)
\mathcal L (Q_5,\nu_6,\nu_5,2)}\,.
\end{split}\ee
Also in this case we were able to to find some ans\"atze for the $\hat \nu$ Young diagrams summations after subtraction of the perturbative part.
Here we quote the results calling $\mathcal I_i$ the summation with $\nu_i =\{1\}$ and all the other diagrams trivial. Equality with \eqref{eq:Inst3}
has been checked up to $Q_4^3$ order in all cases.
\be\begin{split}
\mathcal I_1 =&\frac{(-1)^{\kappa } q \,\hat{Q}_1 }{(1-q)^2 \left(1-Q_1\right){}^2 \left(1-Q_1 Q_2\right){}^2 \left(1-Q_1 Q_2
   Q_3\right){}^2 \left(1-Q_1 Q_2 Q_3 Q_4\right) \left(1-Q_1 Q_2 Q_3 Q_4 Q_5\right)}\\&
   \left[-\frac{Q_1 Q_2 \left(Q_1
   Q_2-Q_m\right)}{Q_5 Q_m^2}+\frac{\left(1-\frac{Q_1 Q_2 Q_4}{Q_m}\right)
   \left(1-\frac{Q_1 Q_2 Q_3 Q_4}{Q_m}\right) \left(1-\frac{Q_1 Q_2^2 Q_3 Q_4}{Q_m}\right) \left(Q_1
   Q_2-Q_m\right)}{\left(1-Q_1 Q_2 Q_3 Q_4\right) \left(1-Q_1 Q_2 Q_3 Q_4 Q_5\right)
   Q_m}\right.\\&\left.
   -\frac{Q_1^2 Q_4 \hat{Q}_1 Q_2^2 \left(Q_2-Q_m\right)
   \left(Q_m-1\right) \left(Q_3 Q_m-1\right)}{\left(1-Q_1 Q_2 Q_3 Q_4\right) Q_5 Q_m^4}+\frac{Q_1 Q_4
   Q_2^2 \left(Q_1 Q_2-Q_m\right)}{Q_5 Q_m^4}\right]\,,
\end{split}\ee
\be\begin{split}
\mathcal I_2 =& \frac{(-1)^{1-\kappa } q \,\hat{Q}_2 }{(1-q)^2 \left(1-Q_1\right){}^2 \left(1-Q_2\right){}^2 Q_2 \left(1-Q_2
   Q_3\right){}^2 \left(1-Q_2 Q_3 Q_4\right) \left(1-Q_2 Q_3 Q_4 Q_5\right)}\\&
   \left[\frac{Q_1 Q_2^2 Q_4 \left(Q_2-Q_m\right)}{Q_m^3}-\frac{Q_2 \left(Q_2-Q_m\right)}{Q_m}-\frac{Q_4
   Q_2^2 \left(Q_1 Q_2-Q_m\right) \left(Q_m-1\right) \left(Q_3 Q_m-1\right)}{\left(Q_2 Q_3
   Q_4-1\right) Q_m^3}+\right.\\&
\left.\frac{Q_5 \left(Q_2-Q_m\right) \left(Q_2 Q_4-Q_m\right) \left(Q_2 Q_3
   Q_4-Q_m\right) \left(Q_1 Q_2^2 Q_3 Q_4-Q_m\right)}{\left(Q_2 Q_3 Q_4-1\right) \left(Q_2 Q_3 Q_4
   Q_5-1\right) Q_m^3}\right]\,,
\end{split}\ee
\be\begin{split}
\mathcal I_3 =& \frac{(-1)^{2-\kappa } q \,\hat{Q}_3}{(1-q)^2 \left(1-Q_2\right){}^2 \left(1-Q_1 Q_2\right){}^2
   \left(1-Q_3\right){}^2 \left(1-Q_3 Q_4\right) \left(1-Q_3 Q_4 Q_5\right)}\\&
  \left[ -\frac{Q_1 Q_4 Q_2^2 \left(Q_m-1\right)}{Q_m^2}+\frac{Q_5 \left(1-\frac{Q_4}{Q_m}\right)
   \left(1-\frac{Q_2 Q_3 Q_4}{Q_m}\right) \left(1-\frac{Q_1 Q_2 Q_3 Q_4}{Q_m}\right)
   \left(1-Q_m\right) Q_m}{\left(1-Q_3 Q_4\right) \left(1-Q_3 Q_4
   Q_5\right)}+\right.\\&
   \left.\frac{Q_m \left(-Q_3 Q_m+Q_m+Q_2 \left(Q_3
   Q_m-1\right)\right)-Q_1 Q_2 \left(Q_2-Q_m\right) \left(Q_3 Q_m-1\right)}{Q_3 Q_m^2}\right.\\&\left.
  - \frac{\left(1-\frac{Q_2}{Q_m}\right) \left(1-\frac{Q_1 Q_2}{Q_m}\right) \left(1-Q_3
   Q_m\right)}{Q_3 \left(1-Q_3 Q_4\right)}\right]\,,
\end{split}\ee
\be\begin{split}
\mathcal I_4 =&\frac{(-1)^{\kappa } q \,\hat Q_4}{(1-q)^2 \left(1-Q_3\right){}^2 \left(1-Q_2 Q_3\right){}^2 \left(1-Q_1 Q_2
   Q_3\right){}^2 \left(1-Q_4\right) \left(1-Q_4 Q_5\right)}\\&
   \left[-\frac{Q_1 Q_3 Q_4 Q_2^2 \left(Q_3 Q_m-1\right)}{Q_m^2}+\frac{Q_3 Q_5
   \left(1-\frac{Q_4}{Q_m}\right) \left(1-\frac{Q_2 Q_4}{Q_m}\right) \left(1-\frac{Q_1 Q_2
   Q_4}{Q_m}\right) Q_m \left(1-Q_3 Q_m\right)}{\left(1-Q_4\right) \left(1-Q_4 Q_5\right)}+\right.\\&
 \frac{Q_m
   \left(Q_2 Q_3 \left(Q_m-1\right)+\left(Q_3-1\right) Q_m\right)-Q_1 Q_2 Q_3 \left(Q_2-Q_m\right)
   \left(Q_m-1\right)}{Q_m^2}\\&
   \left.  -\frac{Q_3 \left(1-\frac{Q_2}{Q_m}\right)
   \left(1-\frac{Q_1 Q_2}{Q_m}\right) \left(1-Q_m\right)}{1-Q_4}\right]\,,
\end{split}\ee
\be\begin{split}
\mathcal I_5 =&\frac{(-1)^{\kappa +1} q \, \hat Q_5}{(1-q)^2 \left(1-Q_4\right) \left(1-Q_3 Q_4\right) \left(1-Q_2 Q_3
   Q_4\right) \left(1-Q_1 Q_2 Q_3 Q_4\right) \left(1-Q_5\right)}\\&
   \left[\frac{Q_3 Q_5 \left(1-Q_m\right) Q_m}{\left(1-Q_4\right) \left(1-Q_5\right)}-\frac{1-Q_5
   Q_m}{1-Q_5}-\frac{\left(Q_m-1\right) \left(Q_m+1\right) \left(Q_3 Q_m-1\right) \left(1-Q_5
   Q_m\right)}{Q_3 \left(1-Q_4\right) \left(1-Q_3 Q_4\right) \left(Q_5-1\right) Q_m^2}\right.\\&\left.
   -\frac{\left(1-Q_m\right) \left(1-Q_3 Q_m\right)
   \left(1-Q_5 \left(Q_3 Q_m+Q_m+1\right)\right)}{Q_3 \left(1-Q_4\right) \left(1-Q_5\right)
   Q_m}-\frac{Q_3 Q_4 \left(Q_4-Q_m\right) \left(Q_5 Q_m-1\right)}{\left(1-Q_4\right)
   \left(1-Q_3 Q_4\right) \left(1-Q_5\right)}\right.\\&
   +\frac{Q_4
   \left(1-Q_m\right) \left(Q_2-Q_m\right) \left(1-Q_3 Q_m\right) \left(1-Q_3 Q_4 Q_5
   Q_m\right)}{\left(1-Q_4\right) \left(1-Q_3 Q_4\right) \left(1-Q_2 Q_3 Q_4\right) \left(1-Q_1 Q_2
   Q_3 Q_4\right) \left(1-Q_5\right) Q_m^2}\\&\left.
   +\frac{\left(Q_m-1\right) \left(Q_3 Q_m-1\right)
   \left(1-Q_3 Q_4 Q_5 Q_m\right)}{Q_3 \left(1-Q_4\right) \left(1-Q_3 Q_4\right) \left(1-Q_2 Q_3
   Q_4\right) \left(Q_5-1\right) Q_m^2}\right]\,,
\end{split}\ee
\be\begin{split}
\mathcal I_6 =& \frac{(-1)^{\kappa } q \, \hat Q_6}{(1-q)^2 \left(1-Q_5\right) \left(1-Q_4 Q_5\right) \left(1-Q_3 Q_4
   Q_5\right) \left(1-Q_2 Q_3 Q_4 Q_5\right) \left(1-Q_1 Q_2 Q_3 Q_4 Q_5\right)}\\&
   \left[\frac{Q_3 Q_4 Q_5^3 \left(Q_m-1\right) \left(Q_m-Q_4\right)}{\left(Q_5-1\right) \left(1-Q_4
   Q_5\right) \left(1-Q_3 Q_4 Q_5\right)}+\frac{Q_5 \left(Q_m-1\right)}{Q_5-1}+\frac{Q_3 Q_5 Q_m
   \left(1-Q_5 Q_m\right)}{\left(Q_5-1\right) \left(1-Q_4 Q_5\right)}+\right.\\&
\frac{Q_4 \left(Q_2-Q_5
   Q_m\right) \left(Q_5 Q_m-1\right) \left(Q_3 Q_5 Q_m-1\right) \left(1-Q_3 Q_4 Q_5
   Q_m\right)}{\left(Q_5-1\right) \left(1-Q_4 Q_5\right) \left(1-Q_3 Q_4 Q_5\right) \left(1-Q_2 Q_3
   Q_4 Q_5\right) \left(1-Q_1 Q_2 Q_3 Q_4 Q_5\right) Q_m^2}\\&
   -\frac{\left(Q_m-1\right) \left(Q_5
   Q_m-1\right) \left(Q_5 Q_m+1\right) \left(Q_3 Q_5 Q_m-1\right)}{Q_3 \left(Q_5-1\right) \left(1-Q_4
   Q_5\right) \left(1-Q_3 Q_4 Q_5\right) Q_5 Q_m^2}\\&
   -\frac{\left(Q_5 Q_m-1\right) \left(Q_3 Q_5
   Q_m-1\right) \left(1-Q_3 Q_4 Q_5 Q_m\right)}{Q_3 \left(Q_5-1\right) \left(1-Q_4 Q_5\right)
   \left(1-Q_3 Q_4 Q_5\right) \left(1-Q_2 Q_3 Q_4 Q_5\right) Q_5 Q_m^2}+\\&\left.\frac{\left(Q_5 Q_m-1\right)
   \left(Q_3 Q_5^2 Q_m \left(Q_3 Q_m+Q_m-1\right)-Q_5 Q_m+Q_5-1\right)}{Q_3 \left(Q_5-1\right)
   \left(1-Q_4 Q_5\right) Q_5 Q_m}\right]\,.
\end{split}\ee
As in the previous case the 1-instanton partition function is simply the sum of all the previous contributions. Again comparison between the topological 
string result and the result obtained via localisation gives the perfect agreement for values of the Chern-Simons level $-3 \leq \kappa \leq 5$.

So far we have limited our discussion to 1-instanton partition functions but a similar procedure may be applied to extract the $n$-instanton partition function
from the topological string partition function. The main difficulty in going beyond 1-instanton is that the summation of the series involving the $\hat \nu_k$
Young diagrams becomes more and more complicated as the instanton number increases, however it is still possible to check the agreement between 
the topological string partition function and the result obtained via localisation checking the expansion in the parameters appearing in the $\hat \nu_k$
summations. We performed this check for the case of 2-instanton level for both $SU(4)$ and $SU(6)$, both with Chern-Simons level $\kappa = 0$.
For the case of $SU(4)$ we checked the agreement between the topological string partition function and the one obtained via localisation up to order $
Q_3^4$ and for the case of $SU(6)$ a similar check was performed up to order $Q_4^2$.



\section{5d pure $USp(2N)$ gauge theory}
\label{sec:USp}

In this section, we move on to the computation of the Nekrasov partition function of the 5d $USp(2N)$ gauge theory without matter. 
The first non-trivial example is the one with $N=2$ for the case of $N=1$ simply corresponds to the $SU(2)$ gauge theory. There are two distinct $USp(2N)$ gauge theories distinguished by a discrete parameter $\theta\in\pi_4(USp(2N)) = \mathbb{Z}_2$. 

\subsection{The 5-brane web}

\begin{figure}
\centering
\includegraphics[width=15cm]{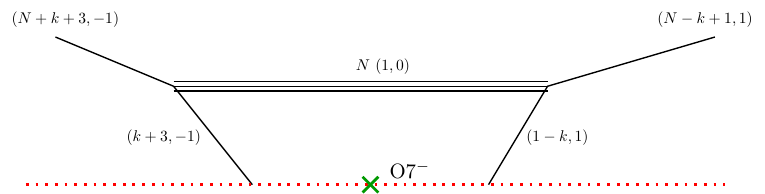}
\caption{The web diagram realising a 5d $USp(2N)$ gauge theory in the presence of an O7$^-$-plane.}
\label{fig:USp2N}
\end{figure}

The 5-brane web realising the 5d $USp(2N)$ gauge theory can be constructed again by using an O7$^-$-plane, and it is depicted in Figure \ref{fig:USp2N}.

The integer $k$ that appears in the definition of the web diagram has an important r\^ole as it determines the discrete $\theta$ angle of the 5d $USp(2N)$
gauge theory.\footnote{Note that we have a different definition of the integer $k$ as opposed to \cite{Bergman:2015dpa}.
Our choice is motivated by having the same definition of $k$ for the diagrams of $SU(2N)$ and $USp(2N)$.} Changing $k$ by one changes the discrete $\theta$ angle and in the case of $N=1$ $k=1\;(\text{mod}\;2)$ corresponds to $\theta = 0$ and $k=2\;(\text{mod}\;2)$ corresponds to $\theta = \pi$ \cite{Bergman:2015dpa}. 
Here we argue that in general the relation between $k$ and the discrete 
$\theta$ angle for $USp(2N)$ is\footnote{A similar result was independently obtained by Oren Bergman and Gabi Zafrir. We thank their correspondence. }
\begin{itemize}
\item If $N=2m$ then odd/even $k$ gives a $USp(4m)$ gauge theory with $\theta = \pi/0$;
\item If $N=2m+1$ then odd/even $k$ gives a $USp(4m+2)$ gauge theory with $\theta = 0/\pi$.
\end{itemize}
One easy consistency check is the following one: if we take the diagram for $USp(2N)$ with a given $k$ we can go to the limit where one of the D5-branes
goes off to infinity yielding the diagram for $USp(2N-2)$ with $k$ unchanged. 
As we explicitly show in Appendix \ref{sec:USp.limit}, 
 the same limit applied to
the instanton partition function of a $USp(2N)$ gauge theory whose discrete angle is $\hat \theta$  yields (after a suitable redefinition of the instanton fugacity) the 
instanton partition function of a $USp(2N-2)$ gauge theory with a new discrete $\theta$ angle equal to $\hat \theta + \pi$. Knowing the relation 
between $k$ and the discrete $\theta$ angle for the case of $USp(2) \simeq SU(2)$ the previous argument shows consistency with the result
stated before.

\begin{figure}
\centering
\includegraphics[width=9cm]{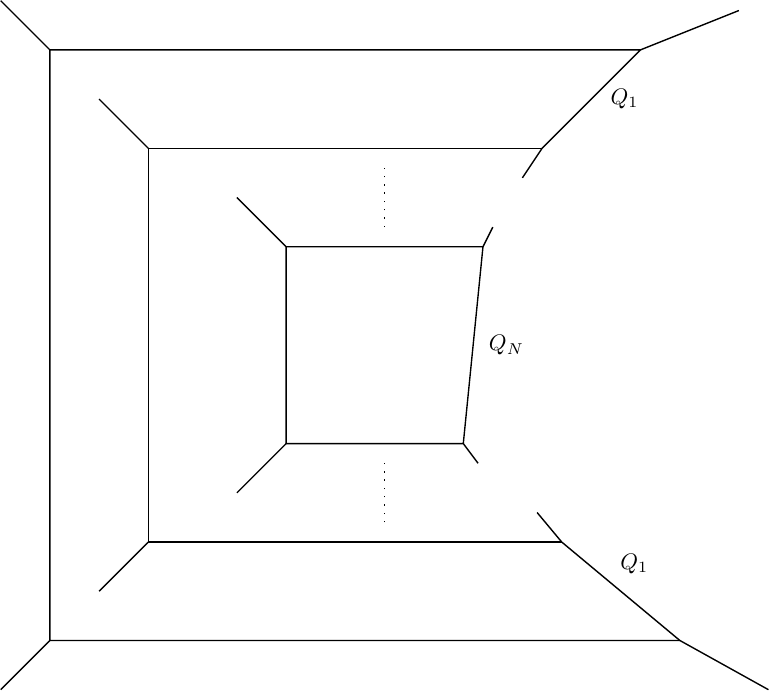}
\caption{The web diagram in Figure \ref{fig:USp2N} after quantum resolution of the orientifold plane and extraction of the 7-branes to infinity.}
\label{fig:USp2N1}
\end{figure}

At the level of the 5-brane web diagram in Figure \ref{fig:USp2N}, we are not able to apply the topological vertex formalism to the diagram. Hence we split 
again the O7$^-$-plane by the quantum resolution. In this case, we can use the usual splitting and the O7$^-$-plane splits into a $(1, -1)$ 7-brane and a $(1, 1)$ 7-brane \cite{Sen:1996vd}. After the arrangement of the branch cuts, we can finally arrive at a 5-brane web given in Figure \ref{fig:USp2N1}\footnote{In Figure \ref{fig:USp2N1}, we have already shrunken some external 5-branes so that we can apply the topological vertex to the diagram. }.
The 5-brane does not have an O7$^-$-plane and we can directly apply the topological vertex by the rules in appendix \ref{sec:top}.

Instead of directly applying the topological vertex to the 5-brane web diagram in Figure \ref{fig:USp2N1}, we can also obtain the same partition function in a different way. The $USp(2N)$ gauge theory can be obtained at far infrared in a Higgs branch of a 5d $SU(2N)$ gauge theory with a hypermultiplet in the antisymmetric representation. When the antisymmetric hypermultiplet is massless, then antisymmetric hypermultiplet can obtain a vev and the gauge group $SU(2N)$ may be broken down to $USp(2N)$. In terms of the 5-brane web in Figure \ref{fig:SU2Nantisym}, we can tune a parameter corresponding the mass of the antisymmetric hypermultiplet so that we have an NS5-brane which stretches from the O7$^-$-plane to the $(0, 1)$ 7-brane. Then, it is possible to remove the NS5-brane from the two-dimensional plane depicted in Figure \ref{fig:SU2Nantisym} to some value in the $(x_7, x_8, x_9)$-plane. These degrees of freedom correspond to the vev of the antisymmetric hypermultiplet. Indeed, after removing the NS5-brane, we can reproduce the 5-brane web in Figure \ref{fig:USp2N}, which gives rise to the 5d $USp(2N)$ gauge theory without matter. The CS level $\kappa = k$ of the $SU(2N)$ gauge theory becomes the $k$ value of the $USp(2N)$ gauge theory and as discussed before it determines the discrete $\theta$-angle of the theory. In terms of the 5-brane web in Figure \ref{fig:SU2Nantisymdiag}, the Higgsing corresponds to combining one external $(1, -1)$ 5-brane with the other $N-1$ $(1, -1)$ 5-brane 
and removing a piece of a 5-brane suspended between the $(1, -1)$ 7-branes.

It has been known that the Higgsing can be applied to the Nekrasov partition function or the topological vertex computation \cite{Hayashi:2013qwa, Hayashi:2014wfa, Hayashi:2015xla}. The argument is based on the Higgsing prescription for the superconformal index of the four-dimensional superconformal field theories \cite{Gaiotto:2012uq, Gaiotto:2012xa}. In this case, the Higgs branch arises when the mass $m$ of the antisymmetric hypermultiplet is zero. In fact, it turns out that the Nekrasov partition function at the far IR in the Higgs branch can be obtained by inserting the condition $m = 0$ 
into the Nekrasov partition function or equivalently the topological string partition function of the UV $SU(2N)$ gauge theory with the antisymmetric hypermultiplet. From the 5-brane web in Figure \ref{fig:SU2Nantisymdiag}, the Higgsing corresponds to combining one external $(1, -1)$ 5-branes with the other $N-1$ $(1, -1)$ 5-branes, 
and hence we obtain a condition
\be\label{higgs1}
\tilde Q_1 \hat {\mathcal Q_1} =e^{N m}= 1\,.
\ee
which is equivalent to $m=0$. It is also important to note that, after imposing this particular tuning condition, there will be 
other legs in the diagram whose lengths shrink to zero size due to the presence of geometric constraints in the 
diagram\footnote{As already noted in \cite{Hayashi:2014wfa} the effect of the geometric constraints can also be interpreted as a propagation of 
the generalised s-rule inside the web diagram.}. More explicitly we find that constraints become
\be\label{higgs2}
\tilde Q_i \hat{\mathcal Q}_i = 1 \,, \quad i = 2,\dots,N-1\,. 
\ee
It is possible to check that the solution to this system has the simple solution $m=0$ together with $\alpha_{2N-i+1} = -\alpha_i$ with $i=1,\dots,N$, which also reduces the number of the Coulomb branch moduli from $2N-1$ to $N$. This is consistent with the fact that the number of the Coulomb branch moduli of the 5d $USp(2N)$ gauge theory is $N$. 

Similarly the instanton fugacity of the $USp(2N)$ gauge theory is obtained by applying \eqref{higgs1} and \eqref{higgs2} to \eqref{inst.fugacity}, namely
\bea
u^2_{USp(2N)} &=& \hat{Q}_1\hat{Q}_{2N}A_1^{-2N-2}\\
&=&\hat{Q}_1^2A_1^{-2N+2k-4}. \label{inst.fugacity.USp}\\
&=&\hat{Q}_1^2\left(\prod_{i=1}^{N-1}Q_i^{-2N+2k-4}\right)Q_N^{-N+k-2} \label{inst.fugacity2.USp}
\eea
Here we used $A_{2N}=A_1^{-1}$ in the first line and $\hat{Q}_{2N}=\hat{Q}_1A_1^{2k-2}$ from the first line to the second line.

In the end, because of these constraints, the actual number of parameters in the web diagram 
is reduced to $N+1$ parameters, namely $N$ Coulomb branch moduli $\alpha_i$ with $i=1, \cdots, N$ and one parameter associated with the instanton fugacity $u_{USp(2N)}$, in agreement with the expectation for a pure $USp(2N)$ gauge theory.

When we insert the condition $m=0$ into \eqref{nek1} together with $\alpha_{2N-i+1} = -\alpha_i$ for $i=1,\dots,N$, we should get the Nekrasov partition function of the 5d $USp(2N)$ gauge theory without matter. Also, this result should agree with the result obtained by directly applying the topological vertex to the 5-brane web diagram given in Figure \ref{fig:USp2N1}, which gives the 5d $USp(2N)$ gauge theory without matter. Therefore, we will make use of \eqref{nek1} with the Higgsing prescription \eqref{higgs1} and \eqref{higgs2} instead of the direct computation of the topological vertex which anyway gives the same result.

\subsection{The partition function}

We can easily apply the Higgsing conditions discussed in the previous section to the partition function of $SU(2N)$ gauge theory 
with one hypermultiplet in the antisymmetric representation and this will give directly
the partition function of the pure $USp(2N)$ gauge theory. The main upshot of the application of the tuning conditions is that all the summations involving
the $\tilde \nu$ diagrams become trivial and therefore the partition function directly factorises into a perturbative part and an instanton part. We will
now discuss separately the two contributions.

\subsubsection{Perturbative part}

As anticipated before the application of the tuning condition trivialises all the summations involving the $\tilde \nu$ Young diagrams which allows
us to extract the perturbative part of the partition function without needing to perform any Young diagram summation. 
Using the parametrisation shown in Figure \ref{fig:USp2N1} we obtain the following result
\be
Z_{pert} ^{II}= \frac{1}{\left[\prod_{i=1}^N \CH (Q_i)^2\right]\left[\prod_{i< j}^N \CH (Q_i \prod_{k>i}^j Q_k)^2 \right]\left[\prod_{i=1}^N
\prod_{k=1}^{N-1} \CH(\prod_{j =i}^N Q_j \prod_{l=1}^k Q_{N-l})^2\right]}\,.
\ee
We find that using the parametrisation
\be\begin{split}\label{coulomb.USp}
Q_i &= e^{-(\alpha_i -\alpha_{i+1})}\,, \quad i = 1,\dots,N-1\,,\\
Q_N &= e^{-2 \alpha_N}\,,
\end{split}\ee
which follows from the application of the Higgsing conditions \eqref{higgs1} and \eqref{higgs2} to \eqref{coulomb.SU2N}, the perturbative part becomes 
\be
Z_{pert}^{II} = \frac{1}{\left[\prod_{i=1}^N \CH (e^{-2\alpha_i})^2\right]\left[\prod_{i<j}^N \CH( e^{-\alpha_i +\alpha_j})^2\CH( e^{-\alpha_i -\alpha_j})^2\right]}.
\ee
This correctly reproduces the perturbative part of the pure $USp(2N)$ gauge theory.
\subsubsection{Instanton part}

Following the same procedure we can write down the instanton part as well for the diagram giving the pure $USp(2N)$ gauge theory. The result is the 
following one\footnote{In writing the instanton partition function we use the Heaviside step function defined as
\begin{displaymath}
H (x) = \left\{\begin{array}{l}1 \quad x\geq 0\\ 0 \quad x <0\end{array}\right.\,.
\end{displaymath}}
\be\begin{split}\label{eq:instUSp}
Z_{inst}^{II} =& \sum_{\nu_i} \left[\prod_{i=1}^{2N} (-\hat Q_i )^{|\nu_i|} \mathcal K (\nu_i)\right] \left[\prod_{i=1}^N f_{\nu_i}^{N+2-\kappa-i}(q)f_{\nu_{N+i}}^{-\kappa-(i-1)}(q)\right]\\&
\prod_{i = 1}^{2N} \prod_{s \in \nu_i} \left[\prod_{k =i+1}^{2N} \mathcal L ( \prod_{j = i}^{k-1} Q_j,\nu_i,\nu_k,1)^{-1}\right]
\left[\prod_{k = 1}^{i-1} \mathcal L (\,\prod_{j=k}^{i-1}Q_j,\nu_i,\nu_k,2)^{-1}\right] \\&
\mathcal L (Q_N \prod_{j=i}^{N-1} Q_j^2 \prod_{i = 2N-i+1}^{N-1}
Q_j^2,\nu_i,\nu_{2N+1-i},1+H(i-1-N))^{-1}\,.
\end{split}\ee
In writing the instanton partition function we used a redundant notation introducing $Q_{2N-i} \equiv Q_i$ for $i = 1,\dots, N-1$.
We shall discuss now the $N=2,3$ cases more in detail.

\paragraph{Examples of $N=2, 3$} For the case of $N=2$ we find that \eqref{eq:instUSp} reduces to
\be\begin{split}
Z_{inst}^{II} =& \sum_{\nu_i} \left[\prod_{i=1}^4 (-\hat Q_i )^{|\nu_i|} \mathcal K (\nu_i)\right] f_{\nu_1}^{3-\kappa} (q) f_{\nu_2}^{2-\kappa}(q)
f_{\nu_3}^{-\kappa} (q)f_{\nu_4}^{-\kappa-1}(q)\\
& \prod_{s \in \nu_1} \left[\mathcal L (Q_1 ,\nu_1,\nu_2,1)\mathcal L (Q_1 Q_2,\nu_1,\nu_3,1)\mathcal L (Q_1^2 Q_2 ,\nu_1,\nu_4,1)^2\right]^{-1}\\
& \prod_{s \in \nu_2} \left[\mathcal L (Q_1 ,\nu_2,\nu_1,2)\mathcal L (Q_2,\nu_2,\nu_3,1)^2\mathcal L (Q_1 Q_2 ,\nu_2,\nu_4,1)\right]^{-1}\\
& \prod_{s \in \nu_3} \left[\mathcal L (Q_1Q_2 ,\nu_3,\nu_1,2)\mathcal L (Q_2,\nu_3,\nu_2,2)^2\mathcal L (Q_1 ,\nu_3,\nu_4,1)\right]^{-1}\\
& \prod_{s \in \nu_3} \left[\mathcal L (Q_1^2Q_2 ,\nu_4,\nu_1,2)^2\mathcal L (Q_1Q_2,\nu_4,\nu_2,2)\mathcal L (Q_1 ,\nu_4,\nu_3,2)\right]^{-1}\,.
\end{split}\ee
To obtain the 1-instanton partition function we need to sum up all possible choices of Young diagrams $\nu_i$ such that $\sum_i |\nu_i|=1$.
This result may be compared with the one obtained using localisation by using the identification of the parameters \eqref{inst.fugacity2.USp} and \eqref{coulomb.USp}. 
Geometric constraints relate the $\hat Q_i$
among themselves leaving only one of them independent. 
The instanton fugacity in the case of $N=2$ is given by $u_{USp(2N)} = \hat{Q}_1Q_1^{-4+k}Q_2^{-2+\frac{k}{2}}$ by using \eqref{inst.fugacity2.USp}. 
With the parametrisation, we find the agreement with the 1-instanton and 2-instanton partition functions of $USp(4)$ with $\theta = k\, \pi \, \,\text{mod} \,\,\pi$.

We may try to follow a similar strategy for the case of $N=3$. Here \eqref{eq:instUSp} becomes
\be\begin{split}
Z_{inst}^{II}=&  \sum_{\nu_i} \left[\prod_{i=1}^6 (-\hat Q_i )^{|\nu_i|} \mathcal K (\nu_i)\right] f_{\nu_1}^{4-\kappa} (q) f_{\nu_2}^{3-\kappa}(q)
f_{\nu_3}^{2-\kappa} (q)f_{\nu_4}^{-\kappa}(q)f_{\nu_5}^{-\kappa-1}(q)f_{\nu_6}^{-\kappa-2}(q)\\&
\prod_{s \in \nu_1}\frac{\mathcal L (Q_1^2 Q_2^2 Q_3,\nu_1,\nu_6,1)^{-2}}{\mathcal L (Q_1, \nu_1,\nu_2,1) \mathcal L (Q_1 Q_2 ,\nu_1,\nu_3 ,1) \mathcal L (Q_1 Q_2 Q_3,\nu_1,\nu_4,1)
\mathcal L (Q_1 Q_2^2 Q_3,\nu_1,\nu_5,1) }\\&
\prod_{s \in \nu_2} \frac{\mathcal L (Q_1 Q_2^2 Q_3,\nu_2,\nu_6,1)^{-1}}{\mathcal L (Q_1,\nu_2,\nu_1,2) \mathcal L (Q_2, \nu_2,\nu_3,1) \mathcal L (Q_2 Q_3,\nu_2,\nu_4,1)
\mathcal L (Q_2^2 Q_3,\nu_2,\nu_5,1)^2 }\\&
\prod_{s \in \nu_3} \frac{\mathcal L (Q_1 Q_2 Q_3,\nu_3,\nu_6,1)^{-1}}{\mathcal L (Q_1 Q_2,\nu_3,\nu_1,2) \mathcal L (Q_2, \nu_3,\nu_2,2) \mathcal L (Q_3,\nu_3,\nu_4,1)^2 
\mathcal L (Q_2 Q_3,\nu_3,\nu_5,1) }\\&
\prod_{s \in \nu_4} \frac{\mathcal L (Q_1 Q_2,\nu_4,\nu_6,1)^{-1}}{\mathcal L (Q_1 Q_2 Q_3,\nu_4,\nu_1,2) \mathcal L (Q_2 Q_3,\nu_4,\nu_2,2) \mathcal L (Q_3,\nu_4,\nu_3,2)^2 
\mathcal L (Q_2,\nu_4,\nu_5,1) }\\&
\prod_{s \in \nu_5} \frac{\mathcal L (Q_1,\nu_5,\nu_6,1)^{-1}}{\mathcal L (Q_1 Q_2^2 Q_3,\nu_5,\nu_1,2) \mathcal L (Q_2^2 Q_3,\nu_5,\nu_2,2)^2 \mathcal L (Q_2 Q_3,\nu_5,\nu_3,2)
\mathcal L (Q_2,\nu_5,\nu_4,2) }\\&
\prod_{s \in \nu_6} \frac{\mathcal L (Q_1,\nu_6,\nu_5,2)^{-1}}{\mathcal L (Q_1^2 Q_2^2 Q_3,\nu_6,\nu_1,2)^2 \mathcal L (Q_1 Q_2^2 Q_3,\nu_6,\nu_2,2) \mathcal L (Q_1 Q_2 Q_3
\nu_6,\nu_3,2) \mathcal L (Q_1 Q_2,\nu_6,\nu_4,2) }\,.
\end{split}\ee
The 1-instanton partition function can be obtained by summing over all Young diagram such that $\sum_i |\nu_i|=1$ with the instanton fugacity $u_{USp(2N)} = \hat{Q}_1Q_1^{-5+k}Q_2^{-5+k}Q_3^{-\frac{5}{2}+\frac{k}{2}}$, which can be determined from \eqref{inst.fugacity2.USp}. 
Using this we compared the topological string result with the 1-instanton and 2-instanton partition functions of $USp(6)$ with $\theta
= (k+1)  \pi \, \, \text{mod} \, \, \pi$, finding the perfect agreement.


\section{5d $SU(2N-1)$ gauge theory with antisymmetric matter}
\label{sec:SU2N-1A}

As for the last example of a 5-brane web involving an O7$^-$-plane, we consider a 5d $SU(2N-1)$ gauge theory with the CS level $\kappa$ and a hypermultiplet in the antisymmetric representation. A non-trivial case starts from $N=3$ as the antisymmetric representation of $SU(3)$ is equivalent to the anti-fundamental representation of $SU(3)$. Note also that the CS level in this case should be always half-integer due to the quantisation condition \cite{Intriligator:1997pq}
\be
\kappa - \frac{2N-5}{2} \in \mathbb{Z}.
\ee

\subsection{The 5-brane web}
\begin{figure}
\centering
\includegraphics[width=15cm]{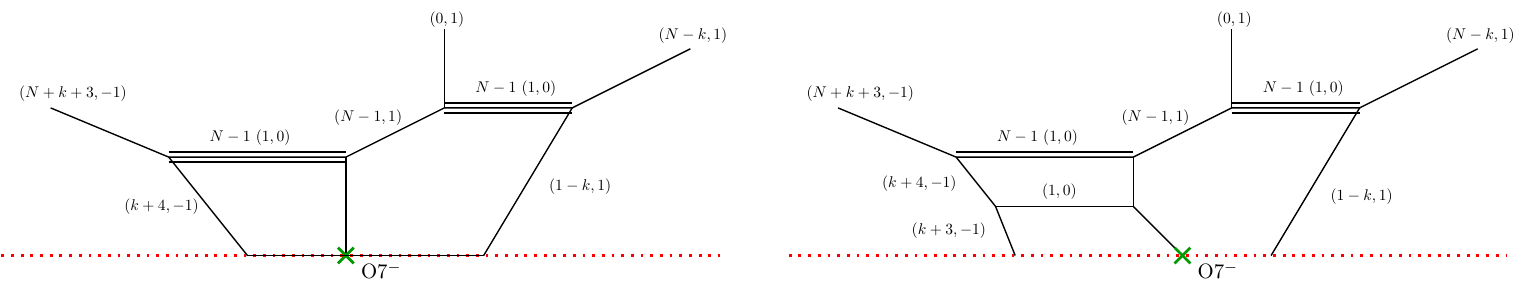}
\caption{The web diagram realising a $SU(2N-1)_{k+\frac{3}{2}}$ gauge theory with antisymmetric matter in the presence of an O7$^-$-plane.}
\label{fig:SU2Nm1antisym}
\end{figure}

The 5-brane web which realises a 5d $SU(2N-1)$ gauge theory with $\kappa = k + \frac{3}{2}$ and a hypermultiplet in the antisymmetric representation is depicted in the left figure in Figure \ref{fig:SU2Nm1antisym}.
One fractional D5-brane is put on the reflection plane of an O7$^-$-plane and a NS5-brane ends on the O7$^-$-plane. A hypermultiplet in the antisymmetric representation comes from strings which cross the middle NS5-brane. We can also obtain an equivalent configuration where one fractional D5-brane is away from the reflection plane, and it appears only on the left-hand side of the NS5-brane as in the right figure in Figure \ref{fig:SU2Nm1antisym}. In this case, a $(1, -1)$ 5-brane ends on the O7$^-$-plane due to the charge conservation. 

\begin{figure}
\centering
\includegraphics[width=12cm]{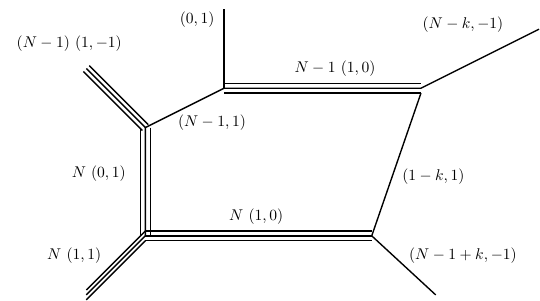}
\caption{The web diagram in Figure \ref{fig:SU2Nm1antisym} after quantum resolution of the orientifold plane and extraction of the 7-branes to infinity. }
\label{fig:SU2Nm1antisym1}
\end{figure}

The O7$^-$-plane again splits into two 7-branes at the quantum level, in this case a $(1, -1)$ 7-brane and a $(1, 1)$ 7-brane. The $(1, 1)$ 5-brane ends on the $(1, 1)$ 7-brane after the quantum resolution of the O7$^-$-plane. The arrangement of the branch cuts of the 7-branes finally gives rise to a 5-brane web diagram given in Figure \ref{fig:SU2Nm1antisym1}. 
This 5-brane web does not involve orientifolds and one can apply the topological vertex to the 5-brane web in Figure \ref{fig:SU2Nm1antisym1}.

\begin{figure}
\centering
\includegraphics[width=12cm]{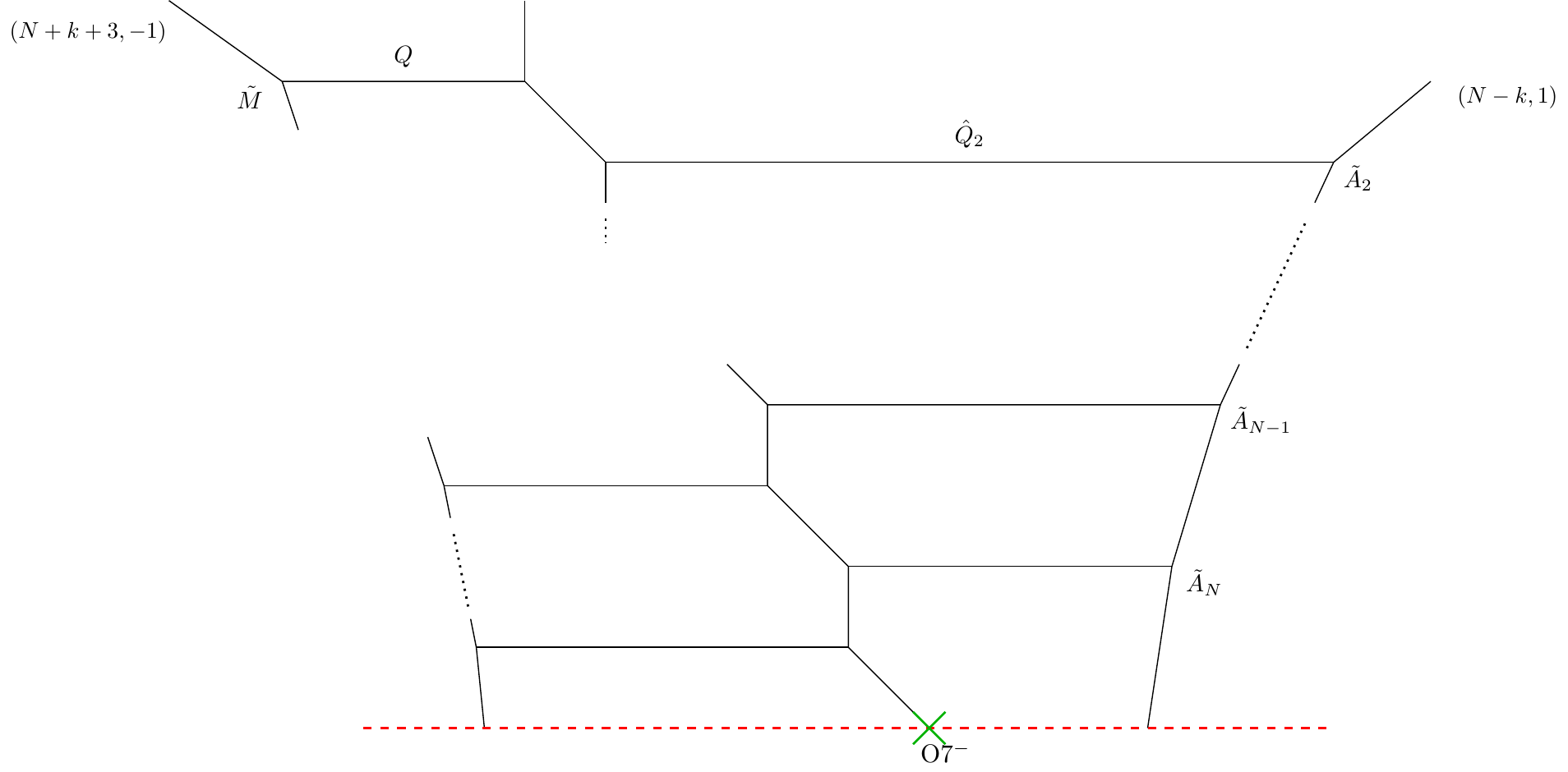}
\caption{The 5-brane web diagram for the 5d $SU(2N-1)$ gauge theory with one antisymmetric hypermultiplet obtained after applying a flop transition to the 5-brane web in Figure \ref{fig:SU2Nm1antisym}.}
\label{fig:SU2Nm1antisymT}
\end{figure}

It is also possible to obtain the equivalent 5-brane web of the right figure in Figure \ref{fig:SU2Nm1antisym} by taking a suitable limit for the 5-brane web of the 5d $SU(2N)_k$ gauge theory with one antisymmetric hypermultiplet given in Figure \ref{fig:SU2NantisymT}. The limit is taking $Q_1 \rightarrow 0$ while keeping $Q_1\tilde Q_1^{-1}$ finite and leaving
untouched all the other parameters. The limit roughly corresponds to sending the top color D5-brane into $+\infty $ reducing therefore the gauge group to $SU(2N-1)$. In terms of the gauge theory parameters, this limit corresponds to 
\bea
\alpha_1 = (2N-1)\gamma, \quad \alpha_i = \beta_{i-1}-\gamma, \;(i= 2, \cdots, 2N), \quad m = m^{\prime} - 2\gamma, \label{limit}
\eea
with $\gamma\rightarrow \infty$. The condition $\sum_{i=1}^{2N}\alpha_i=0$ implies $\sum_{i=1}^{2N-1}\beta_i=0$ and hence $\beta_i, \; (i=1, \cdots, 2N-1)$ can be identified with the Coulomb branch moduli of the $SU(2N-1)$ gauge theory. $m^{\prime}$ is the mass parameter of the hypermultiplet in the antisymmetric representation of the $SU(2N-1)$. The limit decouples some vector multiplets and hypermultiplets and it induces the shift by the CS level by $\frac{3}{2}$, which can be also read off from the 5-brane web in Figure \ref{fig:SU2Nm1antisym}. 
In fact, the parameters $A_i$ and $M$ are divergent under the limit $\gamma \rightarrow \infty$ and hence we shift the origin and define $\tilde{M}$ and $\tilde{A}_i$ as
\be
\tilde{M} = Me^{-\gamma}, \qquad \tilde{A}_i = A_ie^{-\gamma},\quad \text{for}\; i=1, \cdots, 2N.  \label{reparaSU2N-1}
\ee
Then the 5-brane web diagram after applying the limit and the reparameterisation \eqref{reparaSU2N-1} to Figure \ref{fig:SU2NantisymT} is given by Figure \ref{fig:SU2Nm1antisymT}, which is equivalent to the right figure in Figure \ref{fig:SU2Nm1antisym} after flop transitions.

\begin{figure}
\centering
\includegraphics[width=12cm]{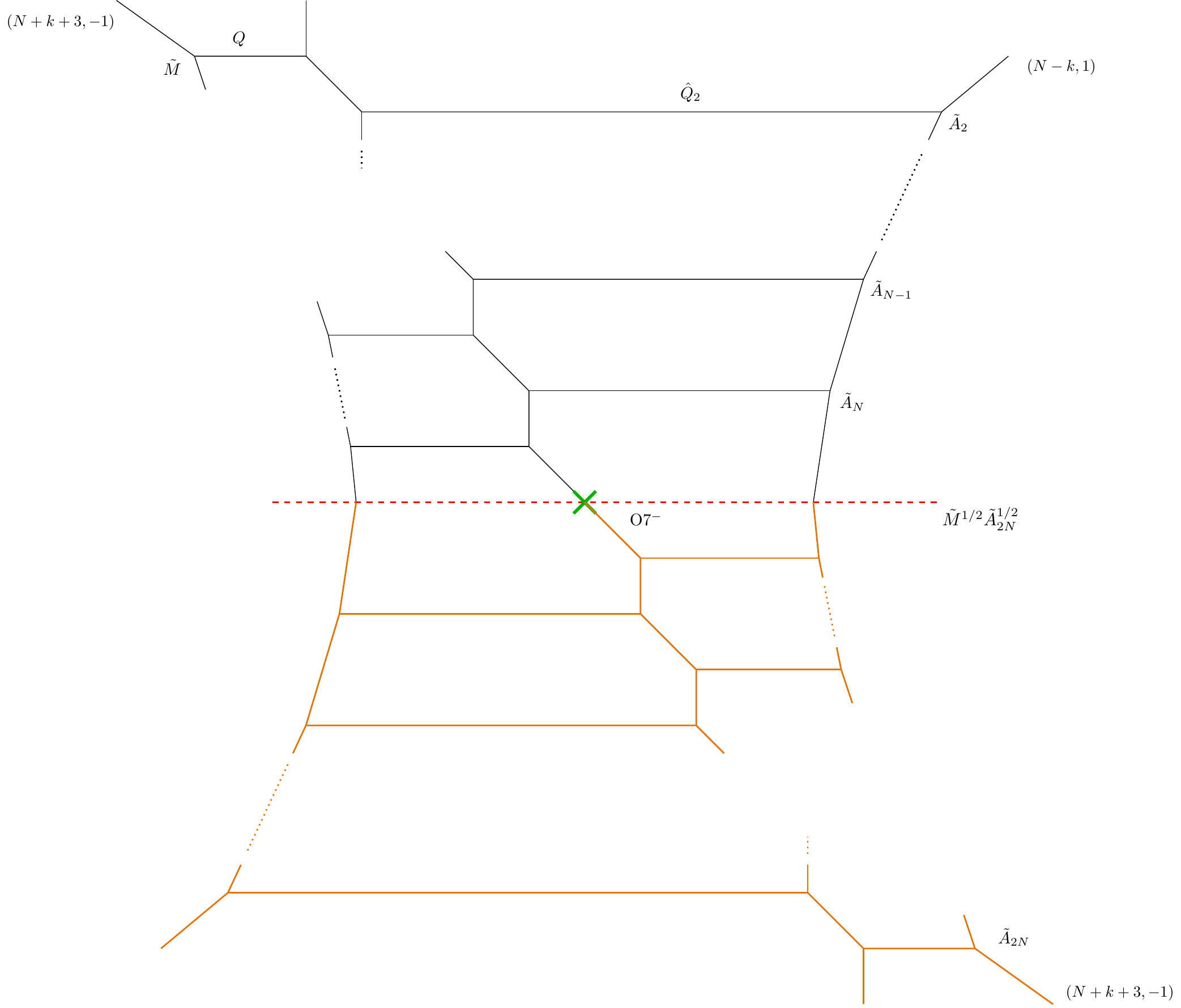}
\caption{The 5-brane web diagram which is a double cover of the one in Figure \ref{fig:SU2Nm1antisymT}.}
\label{fig:SU2Nm1antisymTd}
\end{figure}

\begin{figure}
\centering
\includegraphics[width=12cm]{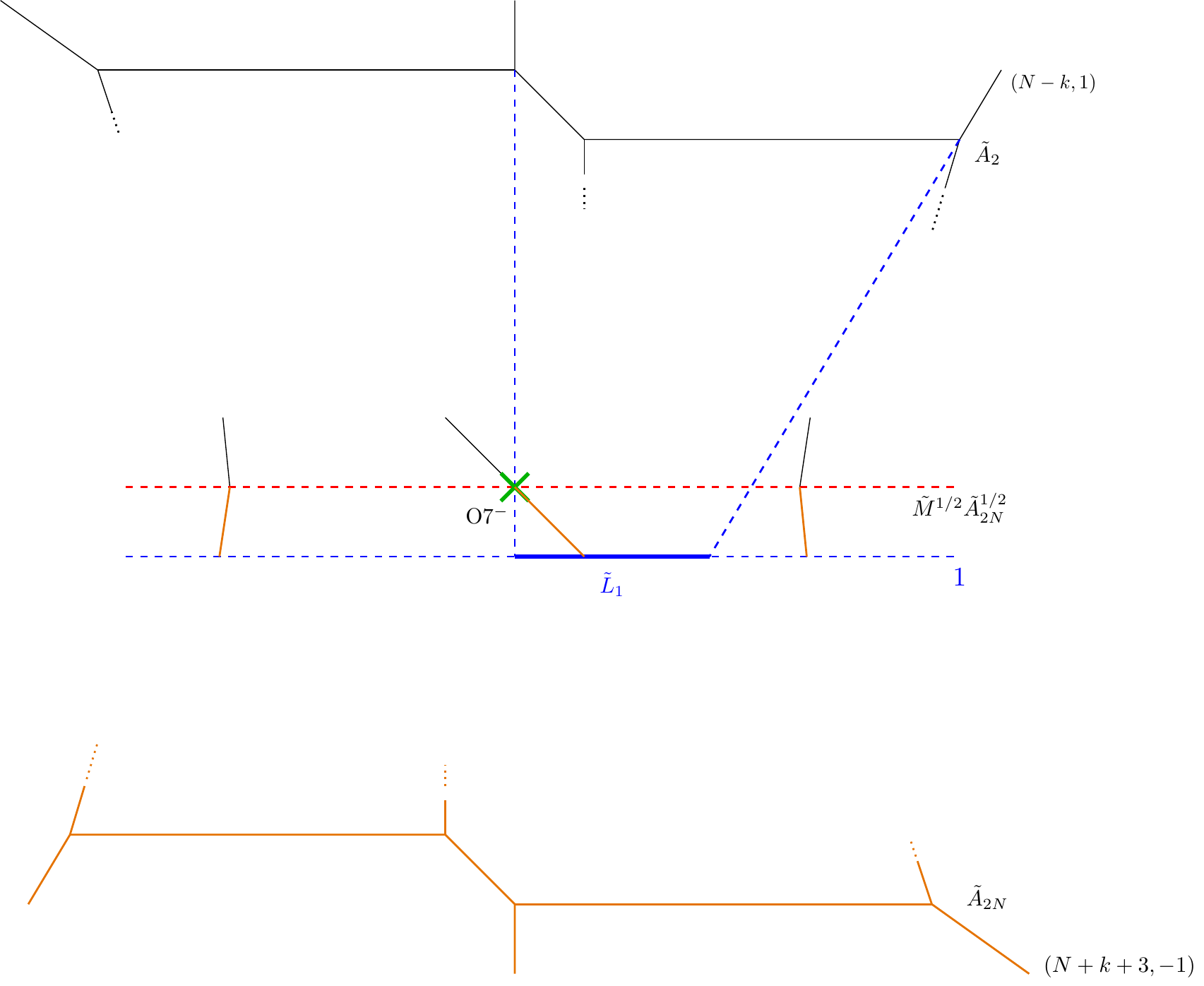}
\caption{The definition of $\tilde{L}_1$.}
\label{fig:SU2Nm1antisymL1}
\end{figure}

\begin{figure}
\centering
\includegraphics[width=12cm]{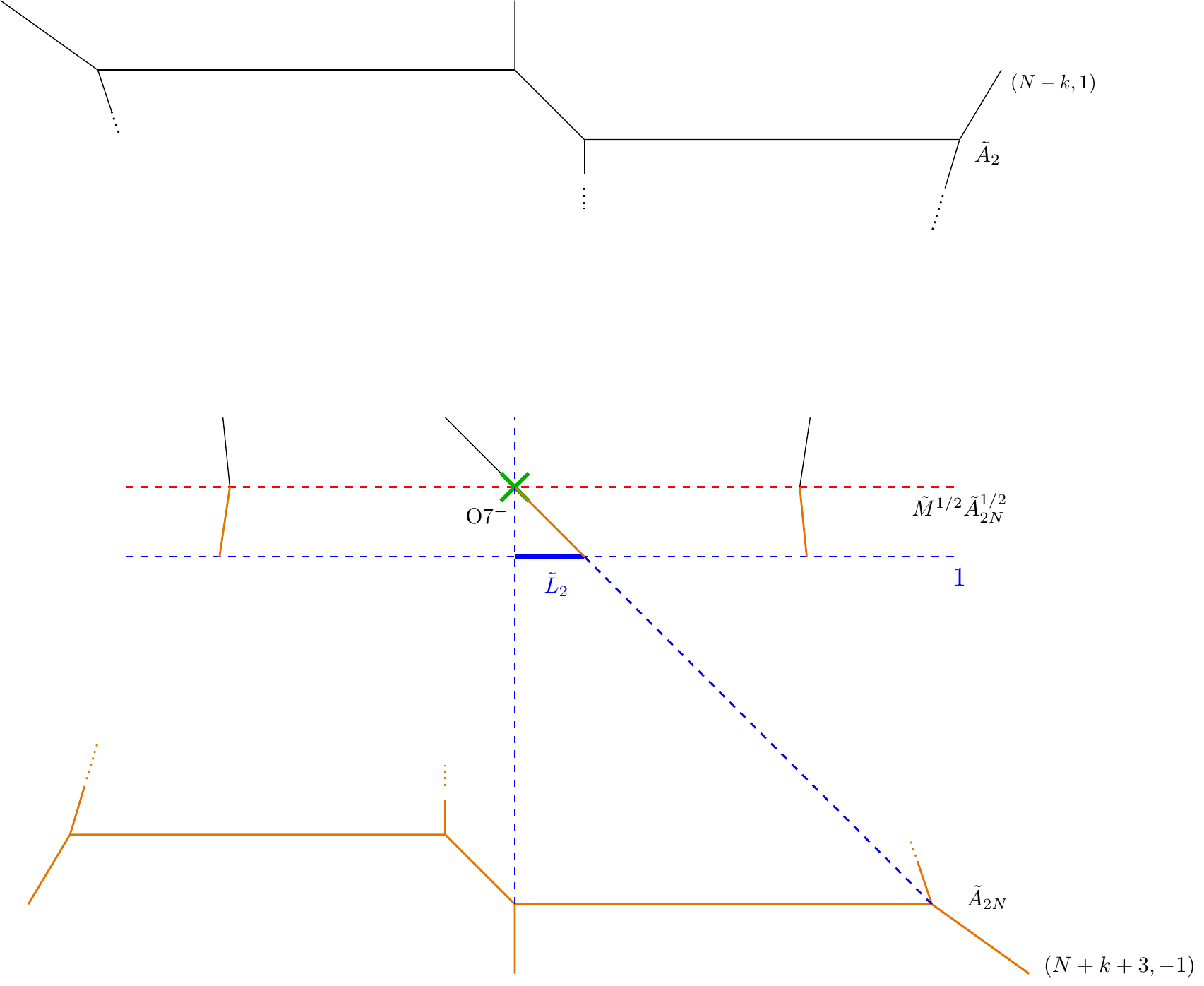}
\caption{The definition of $\tilde{L}_2$.}
\label{fig:SU2Nm1antisymL2}
\end{figure}

Under the reparameterisation \eqref{limit} with the limit $\gamma \rightarrow \infty$, we also need to rescale the instanton fugacity $u_{SU(2N)}$ 
to obtain the instanton fugacity $u_{SU(2N-1)}$ 
for the $SU(2N-1)$ gauge theory. 
In order to see it, we first determine the instanton fugacity of the 5d $SU(2N-1)$ gauge theory from the corresponding 5-brane web diagram. 
The way of how to obtain the instanton fugacity from the 5-brane web in Figure \ref{fig:SU2Nm1antisymT} is essentially the same as the way we performed to obtain the instanton fugacity for the 5d $SU(2N)$ gauge theory with one antisymmetric hypermultiplet in section \ref{sec:SU2Npara}. For that, we first add the mirror image of the 5-brane in the lower half-plane in Figure \ref{fig:SU2Nm1antisymT} as in Figure \ref{fig:SU2Nm1antisymTd}. 
We then extrapolating the two upper-right external 5-branes to the origin as in Figure \ref{fig:SU2Nm1antisymL1} 
and define
\be
\tilde{L}_1 = \hat{Q}_2\tilde{A}_2^{-N+k}\left(\tilde{M}\tilde{A}_2^{-1}\right) = \hat{Q}_2\tilde{M}\tilde{A}_2^{-N+k-1}.
\ee
Similarly, we extrapolate the two lower-right external 5-branes to the origin as in Figure \ref{fig:SU2Nm1antisymL2} 
and define
\be
\tilde{L}_2 = Q\left(\tilde{A}_{2N}^{-1}\right)^{-N-k-3} = Q\tilde{A}_{2N}^{N+k+3}.
\ee
Then the instanton fugacity for the 5d $SU(2N-1)$ gauge theory with one antisymmetric hypermultiplet is given by 
\be
u_{SU(2N-1)}^2 = \tilde{L}_1\tilde{L}_2 = \hat{Q}_2Q\tilde{M}\tilde{A}_2^{-N+k-1}\tilde{A}_{2N}^{N+k+3}. \label{instfugacitySU2Nm1}
\ee

We here compare \eqref{instfugacitySU2Nm1} with the instanton fugacity \eqref{instantonpre2} of the 5d $SU(2N)$ gauge theory with one antisymmetric hypermultiplet. Since \eqref{instantonpre2} is written by using $\hat{Q}_1$, we first rewrite it in terms of $\hat{Q}_2$ and it leads to 
\be
u^2_{SU(2N)}=\hat{Q}_2QA_1^{-1}A_2^{-N+k-1}A_{2N}^{N+k+2}. \label{instSU2Ncomp}
\ee
On the other hand, the instanton fugacity for the 5d $SU(2N-1)$ gauge theory with one antisymmetric hypermultiplet is given by \eqref{instfugacitySU2Nm1} and it can be also written as
\be
u^2_{SU(2N-1)} = \hat{Q}_2QMA_2^{-N+k-1}A_{2N}^{N+k+3}e^{-(2k+3)\gamma}. \label{instSU2Nm1comp}
\ee
By comparing \eqref{instSU2Ncomp} with \eqref{instSU2Nm1comp}, we obtain the following relation
\bea
u^2_{SU(2N)} &=& u^2_{SU(2N-1)}A_1^{-1}M^{-1}A_{2N}^{-1}e^{-(2k+3)\gamma}\\
&=&u^2_{SU(2N-1)}e^{(2N+2k)\gamma + m'}.
\eea
Here we used \eqref{limit} from the first line to the second line. Therefore, the relation between the instanton fugacities is 
\be
u_{SU(2N)} = u_{SU(2N-1)}e^{(N+k)\gamma + \frac{1}{2}m'}. \label{inst.relationSU2NSU2Nm1}
\ee
Namely, we need to redefine the instanton fugacity of the 5d $SU(2N)$ gauge theory as \eqref{inst.relationSU2NSU2Nm1} in order to obtain the instanton fugacity for the 5d $SU(2N-1)$ gauge theory.

Since we determined all the parameters and the moduli of the 5d $SU(2N-1)_{k+\frac{3}{2}}$ gauge theory in terms of the lengths in the 5-brane web, it is possible to compute the partition function of the 5d $SU(2N-1)_{k+\frac{3}{2}}$ gauge theory with antisymmetric matter by applying the topological vertex to the 5-brane web in Figure \ref{fig:SU2Nm1antisym1}. However, instead of directly applying the topological vertex to the 5-brane web in Figure \ref{fig:SU2Nm1antisym1}, we will make use of the result \eqref{nek1} and take the limit \eqref{limit} to obtain the Nekrasov partition function of the 5d $SU(2N-1)_{k+\frac{3}{2}}$ gauge theory with the antisymmetric hypermultiplet. The result computed in this way should give the same result as the one obtained by applying the topological vertex to the 5-brane web in Figure \ref{fig:SU2Nm1antisym1}. Although the CS level of the 5d $SU(2N)$ theory realised on the 5-brane web in Figure \ref{fig:SU2Nantisym1} is restricted in $-N \leq k \leq N+2$, the decoupling does not preserve the shape of the 5-brane web in the cases where $k = N+2, N+1$. In those cases, after the decoupling limit \eqref{limit}, the two external 5-brane web diagrams meet each other and we need to perform further Hanany-Witten transitions. Hence, we will restrict ourselves to the cases where $-N \leq k \leq N$ which gives the 5d $SU(2N-1)_{\kappa}$ gauge theory with a hypermultiplet in the antisymmetric representation and $-N+\frac{3}{2} \leq \kappa \leq N + \frac{3}{2}$. Those cases are indeed consistent with the bound of the CS level $|\kappa| \leq N + \frac{5}{2}$.

\subsection{The partition function}

We then move on to the computation of the partition function of the 5d $SU(2N-1)_{k+\frac{3}{2}}$ gauge theory with a hypermultiplet in the antisymmetric representation by applying the limit \eqref{limit} to the partition function \eqref{nek1} of the 5d $SU(2N)_k$ gauge theory with a hypermultiplet in the antisymmetric representation. We will divide the discussion into the perturbative part and the instanton part and obtain the expressions separately.

\subsubsection{Perturbative part}

We first compute the perturbative partition function of the 5d $SU(2N-1)_{k+\frac{3}{2}}$ gauge theory with antisymmetric matter. The perturbative part of the topological string partition function may be written as\footnote{Note that in the case of $SU(2N-1)$ as opposed to the case of 
$SU(2N)$ there are no contributions due to decoupled factors in the partition function.}
\be
Z^{III}_{pert} = {Z^{III}_{pert1} Z^{III}_{pert2}}
\ee
where 
\be\begin{split}
Z_{pert1}^{III} =& \left[\prod_{i=1}^{2(N-1)} \prod_{j = i}^{2(N-1)} \CH (\prod_{k = i}^j Q_k)\right]^{-1} \left[\prod_{i=1}^{N-2} \prod_{j=i-1}^{N-3} \CH (\tilde  Q_i \prod_{k=i}^j Q_{2N-1-k})\right]\left[
\prod_{j=1}^{N-1}\prod_{i=j}^{N-2} \CH (Q_{j-1}\tilde  Q_{j-1}^{-1} \prod_{k=j}^{i-1} Q_{k})\right]\times \\&
\left[\prod_{i=1}^{N-1} \prod_{j=i+1}^{N} \CH (\prod_{k= i}^{j-1} Q_k)\right]^{-1} \left[\prod_{i=1}^{N-2}\prod_{j=i}^{N-2}
\mathcal H (\prod_{k=i}^{j} Q_{2N-k})  \right]^{-1}\left[\prod_{i=1}^{N-1} \CH (\mathcal Q_i)\right]^{-1}\,,
\end{split}\ee and 
\be
Z_{pert2}^{III}=\sum_{\tilde{\nu}_1, \cdots, \tilde{\nu}_{N-1}}\prod_{i=1}^{N=1}\left(-\hat{\mathcal{Q}}_i\right)^{|\tilde{\nu}_i|}\mathcal{K}\left(\tilde{\nu}_i\right)\mathcal{A}_i\left(\left\{\tilde{\nu}\right\}\right),
\ee
with
\be
\mathcal A_j (\tilde \nu) = \prod_{s \in \tilde \nu_j} \frac{\prod_{i=2}^j \mathcal L (\tilde Q_i \prod_{k=i+1}^j Q_{k-1},\tilde \nu_j,
\emptyset,2)\prod_{i=j+1}^{N+1} \mathcal L (\tilde Q_{j-1}^{-1} Q_{j-1}\prod_{k = j}^{i-2} Q_{k},\tilde \nu_j,\emptyset,1)}{
\mathcal L (\mathcal Q_j,\tilde \nu_j,\emptyset,1)\prod_{i=j+1}^{N-1}
\mathcal L (\prod_{k=j}^{i-1} Q_{2N-k} ,\tilde \nu_j, \tilde \nu_i,1) \prod_{i=1}^{j-1} \mathcal L (\prod_{k=i}^{j-1} Q_{2N-k},\tilde \nu_j,\tilde
\nu_i,2)}\,.
\ee
In writing the perturbative part we introduced the notation  $\tilde Q^{-1}_0 Q_0 \equiv \tilde Q_m$. Similarly to the case of $SU(2N)$ we can conjecture
that the form of $Z_{pert2}^{III}$ is
\be\label{eq:pert2odd}
Z_{pert2}^{III} = \frac{\left[\prod_{k=2}^{N-1} \prod_{\alpha = N-1}^{2N-k-1} \CH(\tilde Q_k \prod_{l=k}^\alpha Q_{2N-1-l})\right]\left[
\prod_{k=1}^{N-1} \prod_{\alpha =N+1}^{2N-k-1} \CH(Q_{k-1} \tilde Q_{k-1}^{-1} \prod_{l=k}^{\alpha-1} Q_{l})\right]\prod_{i=1}^{N-1}\CH(\mathcal Q_i)}{\prod_{i=2}^{N+1}\prod_{j=1}^{N-1}\CH(Q_{N} \prod_{k=i}^{N}Q_{k-1}\prod_{l=1}^{j-1} Q_{N+l})}\,.
\ee
With this result the correct form of the perturbative part of the partition function of $SU(2N-1)$ is reproduced. The form \eqref{eq:pert2odd} for $Z_{pert2}^{III}$
can be checked in the cases $N=2,3$ using the results already obtained for $SU(4)$ and $SU(6)$ finding the perfect agreement.

\subsubsection{Instanton part}

As in the case of the perturbative part we may obtain the instanton part for the $SU(2N-1)_{k+\frac{3}{2}}$ theory by simply applying the aforementioned limit to
the instanton partition function of $SU(2N)_k$. After taking the limit the instanton partition function will involve summations of the $\nu_i$ Young
diagrams with $i=1,\dots,2N-1$ and of the $\tilde \nu_k$ Young diagrams with $k=1,\dots,N-1$. Since the fugacities involved in the $\nu_i$
summations are proportional to the instanton fugacity $u$ of the gauge theory the sum of all the terms involving Young diagrams $\nu_i$ such 
that $\sum |\nu_i| = k$ will give the $k$-instanton contribution of the partition function of $SU(2N-1)$. This differs from the case of the $\hat \nu_k$
summations for terms with different choices of $\hat \nu_k$ Young diagrams will always contribute to the same level of instanton partition function 
and need therefore to be all taken into account. We will discuss now the case of $N=3$ where we are able to perform the computation by simply borrowing
the results already obtained for $SU(6)$ and applying to them the limit in the Coulomb branch moduli \eqref{limit}. 
Note that a similar 
strategy applies to the case of $N=2$, however in this case the result we obtain is not extremely interesting for a hypermultiplet in the antisymmetric
representation of $SU(3)$ is actually equivalent to a hypermultiplet in the anti-fundamental representation.

\paragraph{The case of $N=3$}

For the case of $N=3$ we can obtain the instanton partition function which is
\be\begin{split}
Z_{inst} =&
\sum_{\hat \nu_i, \nu_j}  \left(-\frac{Q_1 Q_3}{Q_m^2}\right)^{|\hat \nu_1|}\left(-\frac{Q_3}{Q_m}\right)^{|\hat \nu_2|}
\mathcal K (\hat \nu_1) \, \mathcal K (\hat \nu_2) \left[\prod_{i=1}^5 (-\hat Q_i) \,\mathcal K (\nu_i)\right]\\
&f_{\nu_1}^{3-\kappa}(q) f_{\nu_2}^{2-\kappa}(q) f_{\nu_3}^{-\kappa}(q)f_{\nu_4}^{-1-\kappa}(q)f_{\nu_5}^{-\kappa-2}(q)\\&
\prod_{s \in \hat \nu_1} \frac{\mathcal L \left(\frac{Q_4 Q_m}{Q_1},\hat \nu_1,\nu_1,1\right) \mathcal L (Q_4Q_m \hat \nu_1 , \nu_2,1) \mathcal L (Q_2 Q_4 Q_m,\hat \nu_1,\nu_3,1)}{\mathcal L (Q_2 Q_3 Q_4^2 Q_m,
\hat\nu_1,\nu_5,1) \mathcal L (Q_4, \hat \nu_1,\hat \nu_2,1)}\\&
\prod_{s \in \hat \nu_2} \frac{\mathcal L \left(\frac{Q_1}{Q_m}, \hat \nu_2,\nu_1,2\right) \mathcal L (Q_m , \hat \nu_2, \nu_2,1) \mathcal L (Q_2 Q_m,\hat \nu_2,\nu_3,1)}{\mathcal L (Q_2 Q_3 Q_4, \hat \nu_2,
\nu_4,1) \mathcal L (Q_4, \hat \nu_2,\hat \nu_1,2)}\\&
\prod_{s \in \nu_1} \frac{\mathcal L \left (\frac{Q_4 Q_m}{Q_1}, \nu_1,\hat \nu_1,2\right)\mathcal L \left(\frac{Q_1}{Q_m},\nu_1,\hat \nu_2,1\right)
}{
\mathcal L (Q_1, \nu_1,\nu_2,1)^2 \mathcal L (Q_1 Q_2,\nu_1,\nu_3,1)^2 \mathcal L (Q_1 Q_2 Q_3,\nu_1,\nu_4,1)\mathcal L (Q_1 Q_2 Q_3 Q_4,\nu_1,\nu_5,1)}
\\&
\prod_{s \in \nu_2} \frac{\mathcal L (Q_m Q_4,\nu_2,\hat \nu_1,2) \mathcal L (Q_m,\nu_2, \hat \nu_2,2)}{
\mathcal L (Q_1,\nu_2,\nu_1,2)^2 \mathcal L (Q_2,\nu_2,\nu_3,1)^2 \mathcal L (Q_2 Q_3,\nu_2,\nu_4,1)\mathcal L (Q_2 Q_3 Q_4,\nu_2,\nu_5,1)}\\&
\prod_{s \in \nu_3} \frac{\mathcal L (Q_2 Q_4 Q_m,\nu_3,\hat \nu_1,2) \mathcal L (Q_2 Q_m,\nu_3,\hat \nu_2,2)}
{\mathcal L (Q_1 Q_2,\nu_3,\nu_1,2)^2 \mathcal L (Q_2,\nu_3,\nu_2,2)^2 \mathcal L(Q_3 ,\nu_3 ,\nu_4,1)
\mathcal L (Q_3 Q_4, \nu_3, \nu_5,1)}\\&
\prod_{s \in \nu_4} \frac{\mathcal L (Q_2 Q_3 Q_m,\nu_4,\hat \nu_2,2)^{-1} }{
\mathcal L (Q_1 Q_2 Q_3 ,\nu_4,\nu_1,2) \mathcal L (Q_2 Q_3,\nu_4,\nu_2,2) \mathcal L (Q_3,\nu_4,\nu_3,2) \mathcal L (Q_4,\nu_4,\nu_5,1)}\\&
\prod_{s \in \nu_5} \frac{\mathcal L (Q_2 Q_3 Q_4^2 Q_m, \nu_5,\hat \nu_2,2)^{-1}}
{\mathcal L (Q_1 Q_2 Q_3 Q_4 ,\nu_5,\nu_1,2) \mathcal L (Q_2 Q_3 Q_4,\nu_5,\nu_2,2) \mathcal L (Q_3 Q_4,\nu_5,\nu_3,2)
\mathcal L (Q_4,\nu_5,\nu_4,2)}\,.
\end{split}\ee
In the cases considered before in order to obtain the 1-instanton result it is necessary to perform the summation over all the $\hat
\nu$'s diagrams when setting one of the $\nu$ diagrams to be a one box diagram. However 
as mentioned before we can simply borrow
the result from the $SU(6)_k$ case and apply the limit to it giving directly the answer for the 1-instanton partition function for 
$SU(5)_{k+\frac{3}{2}}$. This result has been compared with the 1-instanton partition function of $SU(5)_{k+\frac{3}{2}}$
computed using localisation techniques
finding perfect  the agreement for all the values of $k$ in the interval $-3\leq k \leq 3$. In addition to this we have been able to go 
to the 2-instanton level and perform a partial check
of the agreement of the topological string result and the result obtained via localisation techniques. Similarly to the case of $SU(2N)$  discussed in Section \ref{sec:instsu2n} we have checked the
agreement for the case of $SU(5)_{\frac{3}{2}}$ up to order $Q_3^2$ at the 2-instanton.

\section{Conclusions}

In this paper, we have computed the Nekrasov partition function of a 5d $SU(2N)$ gauge theory with one hypermultiplet in the antisymmetric representation by applying the topological vertex to the corresponding 5-brane web. Furthermore, we determined the parameters and the moduli of the gauge theory in terms of the lengths in the 5-brane web. This gives a systematic way to compute the Nekrasov partition function of the 5d $SU(2N)$ gauge theory with one antisymmetric hypermultiplet. The topological string partition function is written by sums of several Young diagrams giving therefore a new expression of the Nekrasov partition function of the 5d $SU(2N)$ gauge theory with antisymmetric matter as opposed to the partition function obtained via localisation techniques which is written in terms of contour integrals. However, the expression has a disadvantage due to extra Young diagram summations not related by the instanton fugacity. In the cases of the 5d $SU(4)$ and $SU(6)$ gauge theory, we guessed the one-instanton expression after summing up the Young diagrams not related to the instanton fugacity and explicitly checked that the resulting expression perfectly agrees with the one-instanton result of the Nekrasov partition function of the 5d $SU(4)$ and $SU(6)$ gauge theory with one antisymmetric hypermultiplet calculated via localisation. This gives another evidence that the 5-brane web diagram indeed realises the 5d $SU(2N)$ gauge theory with a hypermultiplet in the antisymmetric representation. This is a first example of reproducing the Nekrasov partition function of the 5d $SU(2N)$ gauge theory with one antisymmetric hypermultiplet from topological strings to our knowledge. 

We also computed the Nekrasov partition function of a 5d pure $USp(2N)$ gauge theory and of a 5d $SU(2N-1)$ gauge theory with a hypermultiplet in the antisymmetric representation by appropriately taking a limit to the partition function of the 5d $SU(2N)$ gauge theory with a hypermultiplet in the antisymmetric representation. The limit which reduces the 5d $SU(2N)$ gauge theory with one antisymmetric hypermultiplet to the 5d $USp(2N)$ gauge theory corresponds to a Higgsing by the antisymmetric hypermultiplet. The limit which reduces the 5d $SU(2N)$ gauge theory with antisymmetric matter to the 5d $SU(2N-1)$ gauge theory with antisymmetric matter is decoupling one Coulomb branch modulus with a combination of the Coulomb branch modulus and the instanton fugacity kept so that the 5-brane web after the limit reproduces the 5d $SU(2N-1)$ gauge theory with one antisymmetric hypermultiplet. In both cases, we took the limits which precisely reproduce the 5-brane web diagram of the 5d $USp(2N)$ gauge theory and of the 5d $SU(2N-1)$ gauge theory with one antisymmetric hypermultiplet. Therefore, the topological vertex computation applied to the 5-brane web diagrams should reproduce the same results as the results obtained by taking the limits. This also illustrates a power of the topological vertex and 5-brane webs which give rise to the Nekrasov partition function of a 5d $SU(N)$ gauge theory with one antisymmetric hypermultiplet and of a 5d pure $USp(2N)$ gauge theory.

It is straightforward to add flavours to the 5d $SU(N)$ gauge theory with one antisymmetric hypermultiplet or the 5d pure $USp(2N)$ gauge theory. It is known that up to 
$N+6$ flavours may be added in the former case and up to 
$2N+6$ in the latter \cite{Yonekura:2015ksa, Gaiotto:2015una, Bergman:2015dpa, Hayashi:2015zka}. The cases which saturate the bound have UV completion as a 6d SCFT whereas in other cases the UV completion is given by a 5d SCFT. It would be interesting to apply the topological vertex method to those cases with more flavours. When we add more flavours, there is a subtlety regarding the singularity of the ADHM moduli space \cite{Gaiotto:2015una}. Therefore, the standard ADHM quantum mechanics might not be applicable to some cases when the number of flavours is close to the bounds. In those cases the application of topological string results would therefore give a prediction for the field theory result.

\section*{Acknowledgments}
We would like to thank Futoshi Yagi for useful discussions. This work has been partially supported by the grants FPA2012-32828 and FPA2015-65480-P from  MINECO, SEV-2012-0249 of the ``Centro de Excelencia Severo Ochoa" Programme, and the ERC Advanced Grant SPLE under contract ERC-2012-ADG-20120216-320421. 
G.Z. is supported through a grant from ``Campus Excelencia Internacional UAM+CSIC". G.Z. would like to thank GGI in Florence and UW-Madison for hospitality during completion of this work. 

\appendix

\section{The formulae of the topological vertex}
\label{sec:top}

In this appendix we summarise the techniques employed in the computation of the topological string partition function. We collect all the relevant 
definitions and describe how to correctly identify the decoupled factors appearing in the topological string partition function.

\subsection{The topological vertex}

The relevant quantity that we would like to compute using topological strings is the partition function which may be written as an exponential 
of the topological string free energy
\be
Z_{top} = \exp \left(\sum_{g=0}^\infty F_g \,g_s^{2g-2}\right)\,.
\ee
Here $g_s$ is the string coupling and the contribution $F_g$ to the topological string free energy is computed by considering genus $g$ worldsheets.
In turns, when we consider type IIA string theory on a given Calabi--Yau threefold $X$ it is possible to express the genus $g$ free energy in terms
of Gromov--Witten invariants of the Calabi--Yau threefold\footnote{This is the exact result for $g>1$. For $g=0,1$ there are some additional contributions
that may be expressed in terms of topological invariants of the Calabi--Yau threefold $X$ \cite{Bershadsky:1993cx}.}
\be
F_g = \sum_{C \in H_2(X,\mathbb Z)} N_C^g \,Q_C\,,
\ee
where the sum is restricted to genus $g$ two-cycles and the fugacity $Q_C$ is defined as $Q_C := e^{-\int_C J}$ where $J$ is the K\"ahler form of $X$.
The numbers $N_C^g$ are the genus $g$ Gromov--Witten invariants of the Calabi--Yau threefold and they give a way of counting the number of curves in the 
same homology class. 

In general the computation of Gromov-Witten invariants for a given Calabi--Yau threefold is a difficult problem but using a duality
between type IIA string theory and M--theory  it is possible to obtain the complete all genus result \cite{Gopakumar:1998ii,Gopakumar:1998jq}. Moreover if the Calabi--Yau
threefold is toric it is possible to employ the topological vertex \cite{Iqbal:2002we, Aganagic:2003db, Awata:2005fa,Iqbal:2007ii} which gives a way to compute the answer diagrammatically using directly 
the toric diagram. Quite interestingly it has been shown recently \cite{Hayashi:2014wfa} that the same techniques can be still employed for a certain non-toric Calabi--Yau 
threefolds that can be obtained via a complex structure deformation of a toric Calabi--Yau. Here we review briefly how to compute the topological string
partition function by using the topological vertex and also discuss when this may be applied also in more general cases when the Calabi--Yau threefold is
not toric. We shall not discuss the refinement of the topological vertex \cite{Awata:2005fa, Iqbal:2007ii} for this is never employed in the main text.

To compute the topological string partition function for a given toric Calabi--Yau threefold we start by assigning to every leg of the dual of the toric diagram
a Young diagram. For external legs the representation is chosen to be the trivial one. Then to every trivalent vertex in the diagram we associate the factor
\be
C_{\lambda\mu \nu } (q) = q^{\frac{||\mu||^2-||\mu^t||^2 +||\nu||^2}{2}} \tilde Z_{\nu}(q)\sum_\eta s_{\lambda^t/\eta}(q^{-\nu-\rho})
\,s_{\mu/\eta}(q^{-\rho - \nu^t})\,.
\ee
\begin{figure}
\centering
\includegraphics[width=8cm]{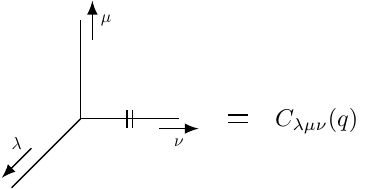}
\caption{The topological vertex.}
\label{fig:vertex}
\end{figure}
The order of the representation is chosen counter-clock wise and in this paper we always take the last representation to be the one associated to a horizontal
leg.  When writing the expression of the topological vertex we defined $||\lambda||^2 = \sum_i \lambda_i^2$ and 
\be
\tilde Z_\nu(q) = \prod_{(i,j) \in \nu} \left(1-q^{l_\nu (i,j) + a_\nu(i,j)+1}\right)^{-1}\,,
\ee
where $l_\nu(i,j) =\nu_i - j$ and $a_\nu(i,j) = \nu^t_j-i$ are the leg and arm lengths of the Young diagram respectively. Finally the functions $s_{\mu/\nu}
(x)$ are skew Schur functions and $\rho = -i +\frac{1}{2} $ with $i=1,2,\dots $.
Moreover in the computation of the topological string partition function it is necessary to assign to every internal leg with
Young diagram $\nu$ an edge factor which has the form
\be
(-Q)^{|\nu|} f^\eta_\nu (q),
\ee
where $|\nu|= \sum_i \nu_i$ is the number of boxes of the Young diagram, $Q= e^{-\int_C J}$ is the fugacity measuring the length of the given leg and $f_
\nu(q)$ is called framing factor and it is defined as
\be
f_\nu (q) = (-1)^{|\nu|}q^{\frac{||\nu^t||^2-||\nu||^2}{2}}\,.
\ee
The exponent $\eta$ is determined in terms of the local embedding of the curve in the Calabi--Yau threefold, and in particular if locally around the curve
the Calabi--Yau threefold is the total space of $\mathcal O (m-1) \oplus \mathcal O(-m-1) \rightarrow \mathbb P^1$ then $\eta = m$.

The full topological string partition function may be obtained by using the rules just described and summing over all possible choices of Young diagrams.

\subsection{Decoupled factors}

An important point in the comparison between the topological string partition function and the Nekrasov partition function is that the former generically
contains contributions from states that carry no charge under any gauge group. Therefore to find perfect agreement between the two results it is necessary
to cancel these contributions from the topological string partition function. Quite remarkably it is possible to give a diagrammatic prescription for the 
identification of these factors and their contribution to the partition function can be easily computed and cancelled. More specifically for toric 
Calabi--Yau threefolds these factors are associated with curves that have zero intersection with any compact divisor in the geometry. Since compact divisors
are associated to gauge symmetry we see directly that the factors associated to these curves indeed carry no gauge charge. In the dual brane picture 
these factors are associated to strings stretching between parallel external legs.
\begin{figure}
\centering
\includegraphics[width=12cm]{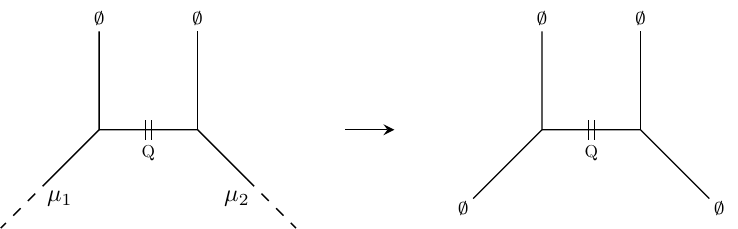}
\caption{The local diagram that allows for the computation of contributions due to decoupled factors.}
\label{fig:strip}
\end{figure}
A simple example of a diagram containing decoupled factors is given in Figure \ref{fig:strip}. Note that this diagram contains two parallel external legs and therefore
we expect the presence of a decoupled factor. To compute the contribution of the decoupled factor we pass to the diagram on the right of Figure \ref{fig:strip}
The Calabi--Yau threefold that has this toric diagram is the total space of $\mathcal O (-2) \oplus \mathcal O \rightarrow \mathbb P^1$  and 
its topological string partition function is 
\be
Z_{dec} = \prod_{i,j=1}^\infty (1- Q q^{i+j-1})^{-1}\,.
\ee
The same logic can be applied in general and gives a practical way for the identification of all the decoupled factors present in a given 
topological string partition function. Knowing the contributions of decoupled factors it is easy to finally obtain the 5d Nekrasov partition function as
\be
Z_{Nek} = \frac{Z_{top}}{Z_{dec}}\,.
\ee

\subsection{Topological vertex for non-toric geometries}

The rules for the topological vertex that we described so far can be applied in all cases in which the Calabi--Yau manifold is toric, but as shown in
\cite{Hayashi:2014wfa} it is possible to apply it to a wider class of geometries suitable to engineer 5d theories. This class includes Calabi--Yau manifolds
that are connected via a complex structure deformation to toric Calabi--Yau manifolds. From the physical point of view the complex structure deformation
amounts to entering the Higgs branch of the theory. Note that for a generic point in the Coulomb branch moduli space it is generically impossible to
access the Higgs branch and it is necessary to go to specific subloci in the Coulomb branch moduli space. The possible Higgs branch deformations 
are identified in the web diagram as deformations that can remove a 5-brane (or part of it) from the diagram. The specific tuning of K\"ahler moduli
necessary to access this deformation may be easily identified as it corresponds to a pole in the 5d superconformal index \cite{Gaiotto:2012uq,Gaiotto:2012xa}. 

\begin{figure}
\centering
\includegraphics[width=15cm]{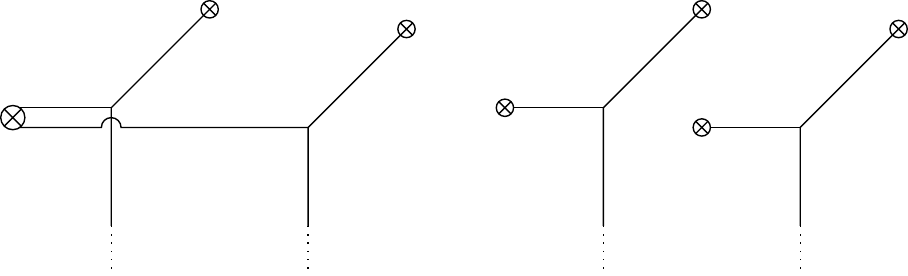}
\caption{On the left: an example of a jumping 5-brane. On the right: the rule for applying the topological vertex directly to the non-toric diagram.
All legs with the crossed circle are external and carry a trivial Young diagram.}
\label{fig:nontoric}
\end{figure}

This class of geometries is in particularly interesting because it was shown in \cite{Hayashi:2014wfa} that it is possible to directly apply the topological vertex
to the non--toric web diagram and correctly obtain the topological string partition function. This gives a great advantage as it makes unnecessary to compute
the topological string partition function of the parent theory and then apply to it the suitable tuning conditions. The rule for jumping branes is to simply
assign to them a Young diagram (trivial if the leg is an external one) and sum over all the possible diagrams. We show in Figure \ref{fig:nontoric}
 an example involving
an external jumping 5-brane and how the representation ought to be assigned to it.
It is possible also to use the non--toric diagram to analyse the contributions due to decoupled factors present in the topological string partition function. 
Quite remarkably the rule that can be used for identifications of decoupled factors in the toric case still holds in the non--toric case, so again it is possible
to simply consider the contributions associated to curves connected to parallel external legs. 

\section{5d Nekrasov partition functions}
\label{sec:Nek}
In this appendix we recapitulate the techniques employed to perform the computations of instanton partition functions for 5d theories with gauge groups $SU(n)$ and $USp(2n)$ via localisation. In both cases the instanton partition function may be written as a Witten index of a suitable ADHM quantum
mechanics, index that can generically be written as
\be
Z_{QM}^k (\epsilon_1,\epsilon_2,\alpha,z) = \text{Tr}\left[(-1)^F e^{-\beta \{Q,Q^\dag\}}e^{-\epsilon_1(J_1 + J_2)}e^{-\epsilon_2 (J_2 + J_R)}
e^{-\alpha_i \Pi_i} e^{-m_a F_a} \right]\,.
\ee
In the definition of the Witten index $Q$ is one of the supercharges that commutes with all the other operators in the trace and $Q^\dag$ its adjoint,
$ F$ is the fermion number, $J_1$ and $J_2$ are the Cartan generators of the spacetime $SO(4)$ symmetry. Moreover $J_R$ is the Cartan generator
of the $SU(2)$ R-symmetry, $\Pi_i$ are the Cartan generators of the gauge group and $F_a$ are the Cartan generators of the flavour symmetry of the theory.
To these symmetries are associated some chemical potentials denoted $\epsilon_1$, $\epsilon_2$, $\alpha_i$ and $m_a$. Their interpretation is as follows:
$\epsilon_1$ and $\epsilon_2$ are the $\Omega$-background deformation parameters, $\alpha_i$ are the Coulomb branch moduli of the theory and
$m_a$ represent the mass parameters. Note that since we wish to compare these results with the ones obtained using topological string computations 
we will always consider the special case $\epsilon_2 = - \epsilon_1 = \epsilon$ and $q = e^{-\epsilon}$. We will now discuss separately the two cases of $SU(N)$ with antisymmetric hypermultiplets and pure $USp(2N)$.

\subsection{$SU(N)$ gauge theory with antisymmetric matter}
\label{sec:NekrasovSUNantisym}

In this case the ADHM quantum mechanics is a gauge theory whose group is $\hat G = U(k)$ for the case of $k$-instantons. Except in the case of $N=2$
we will also introduce a classical Chern-Simons coupling $\kappa$ satisfying the quantisation condition $\kappa -\frac{N-4}{2} \in \mathbb Z$.
The result of localisation is that the partition function may be written as a suitable contour integral over the Coulomb branch moduli of the ADHM quantum
mechanics\footnote{Note that there is a difference in the sign of the Chern-Simons level when comparing with \cite{Gaiotto:2015una}.}
\be
Z_{QM}^k (\epsilon_1,\epsilon_2,\alpha,z) = \frac{1}{k!}\oint \prod_{I=1}^k \frac{d \phi_I}{2\pi i}e^{ \kappa \sum_{J=1}^k \phi_J}
Z_{vec}(\phi,\epsilon_1,\epsilon_2,\alpha)\, Z_A (\phi,\epsilon_1,\epsilon_2,\alpha,z)\,.
\ee
Here $Z_{vec}$ is the contribution due to the vector multiplets of the theory and $Z_A$ is the contribution of the antisymmetric hypermultiplet. Similarly
contributions of additional hypermultiplets in different representations of $SU(N)$ may be added but we did not included them for this is not the case
of interest for the comparison with topological string results. The contribution of vector multiplets takes the form
\be
 Z_{vec}(\phi,\epsilon_1,\epsilon_2,\alpha) = \frac{\prod_{I \neq J} ^k 2 \sinh \frac{\phi_I -\phi_J}{2}\prod_{I,J}^k 2 \sinh \frac{\phi_I -\phi_J+
 2 \epsilon_+}{2}}{\prod_{I,J}^k 2 \sinh \frac{\phi_I - \phi_J +\epsilon_1 }{2}2 \sinh \frac{\phi_I - \phi_J +\epsilon_2 }{2}\prod_{i = 1}^N 
 \prod_{I=1}^k 2 \sinh \frac{\pm(\phi_I -\alpha_i)+\epsilon_+}{2}}\,,
\ee
and the contribution of the antisymmetric hypermultiplet is
\be
 Z_A (\phi,\epsilon_1,\epsilon_2,\alpha,z)= \frac{\prod_{i=1}^N\prod_{I=1}^k 2\sinh \frac{\phi_I +\alpha_i -m}{2}\prod_{I>J}^k
 2 \sinh \frac{\phi_I + \phi_J-m -\epsilon_-}{2}2 \sinh \frac{-\phi_I - \phi_J+m -\epsilon_-}{2}}{\prod_{I>J}^k 2 \sinh \frac{\phi_I + \phi_J -m
 -\epsilon_+}{2}2 \sinh \frac{-\phi_I - \phi_J +m - \epsilon_+}{2}\prod_{I=1}^k 2 \sinh \frac{2 \phi_I -m-\epsilon_+}{2}2 \sinh \frac{-2 \phi_I +m-\epsilon_+}{2}}\,.
\ee

In writing the contributions appearing in the contour integral we introduced the notation $\epsilon_{\pm} = \frac{\epsilon_1 \pm \epsilon_2}{2}$.
The contour integral has to be carefully performed by selecting the appropriate poles in the integrand and the general prescription boils down to the
computation of  Jeffrey-Kirwan (JK) residues \cite{Hwang:2014uwa}. The result for $SU(N)$ turns out to be quite simple and the poles which ought to be
considered are
\be\begin{split}
&\phi_I -\alpha_i +\epsilon_+ = 0\,, \quad \phi_I - \phi_J +\epsilon_1 = 0 \,, \quad \phi_I - \phi_J + \epsilon_2 =0\,, \quad (I>J)\\
& \phi_I = \frac{1}{2} (m + \epsilon_+)\,, \quad \phi_I = \frac{1}{2} (m + \epsilon_+)+i \pi\,, \quad \phi_I + \phi_J  -m -\epsilon_+ =0\,, \quad (I>J)\,.
\end{split}\ee
The poles in the first line come from vector multiplets while the ones in the second line come from the antisymmetric hypermultiplet.
\subsection{$USp(2N)$ gauge theory}

For the case of a $USp(2N)$ gauge theory the dual group appearing in the ADHM quantum mechanics is $\hat G = O(k)$ for $k$ instantons. Since the gauge
group is not connected the instanton partition function will be written as a sum of two contributions that we shall call $Z^k_{\pm}$. It is necessary to take 
into account that a non-vanishing $\theta$-angle will affect the instanton partition function and therefore for the two possible choices of discrete 
$\theta$-angle we will obtain two different partition functions as follows
\be
Z_{QM}^k = \left\{\begin{array}{l l}\frac{1}{2}(Z_+^k+Z_-^k)\,,    &\quad \theta=0\,,\\[2mm]\frac{(-1)^k}{2} (Z^k_+ - Z_-^k)\,,  
 &\quad \theta = \pi\,.\end{array}\right.
\ee
The two distinct contributions to the partition function may be written as 
\be
Z_{\pm}^k (\epsilon_1,\epsilon_2,\alpha)= \frac{1}{|W|_\pm} \oint \prod_{I = 1}^n \frac{d \phi_I}{2\pi i}Z_{\pm}^{vec} (\epsilon_1,\epsilon_2,\alpha)\,.
\ee
While there may be additional contributions when hypermultiplets are present we did not consider them for in all cases we consider pure $USp(2N)$
gauge theories. In writing the form of $Z_\pm^k$ we introduced $n$ via the relation $k = 2n + \chi$ with $\chi = 0,1$ as well as the Weyl group factor
$|W|_\pm$ defined as
\be
|W|_+^{\chi=0}= \frac{1}{2^{n-1}n!}\,, \quad |W|_+^{\chi=1}= \frac{1}{2^{n}n!}\,, \quad |W|_-^{\chi=0}= \frac{1}{2^{n-1}(n-1)!}\,,
\quad |W|_-^{\chi=1}= \frac{1}{2^{n}n!}\,.
\ee
For the $O(k)_+$ the integrand is
\be\begin{split}\label{vecplus}
Z^{vec}_{+} =&\prod_{I < J}^n 2\sinh\frac{\pm\phi_I\pm\phi_J}{2}\left(\frac{\prod_I^n2\sinh\frac{\pm \phi_I}{2}}{2\sinh\frac{\pm\epsilon_-+\epsilon_+}{2}\prod_{i=1}^N2\sinh\frac{\pm\alpha_i+\epsilon_+}{2}}\prod_{I=1}^n\frac{2\sinh\frac{\pm\phi_I+2\epsilon_+}{2}}{2\sinh\frac{\pm\phi_I\pm\epsilon_-+\epsilon_+}{2}}\right)^{\chi}\\
&\prod_{I=1}^n\frac{2\sinh\epsilon_+}{2\sinh\frac{\pm\epsilon_-+\epsilon_+}{2}2\sinh\frac{\pm 2\phi_I\pm\epsilon_-+\epsilon_+}{2}\prod_{i=1}^N2\sinh\frac{\pm\phi_I\pm\alpha_i+\epsilon_+}{2}}\prod_{I<J}^n\frac{2\sinh\frac{\pm\phi_I\pm\phi_j+2\epsilon_+}{2}}{2\sinh\frac{\pm\phi_I\pm\phi_J\pm\epsilon_-+\epsilon_+}{2}}\,.
\end{split}\ee
For $O(k)_-$ the form of the integrand depends on whether $k$ is even or odd. For $k= 2n +1$ the integrand is
\be\begin{split}
Z^{vec}_{-} =&\prod_{I < J}^n 2\sinh\frac{\pm\phi_I\pm\phi_J}{2}\left(\frac{\prod_{I=1}^n2\cosh\frac{\pm \phi_I}{2}}{2\sinh\frac{\pm\epsilon_-+\epsilon_+}{2}\prod_{i=1}^N2\cosh\frac{\pm\alpha_i+\epsilon_+}{2}}\prod_{I=1}^n\frac{2\cosh\frac{\pm\phi_I+2\epsilon_+}{2}}{2\cosh\frac{\pm\phi_I\pm\epsilon_-+\epsilon_+}{2}}\right)\\
&\prod_{I=1}^n\frac{2\sinh\epsilon_+}{2\sinh\frac{\pm\epsilon_-+\epsilon_+}{2}2\sinh\frac{\pm 2\phi_I\pm\epsilon_-+\epsilon_+}{2}\prod_{i=1}^N2\sinh\frac{\pm\phi_I\pm\alpha_i+\epsilon_+}{2}}\prod_{I<J}^n\frac{2\sinh\frac{\pm\phi_I\pm\phi_j+2\epsilon_+}{2}}{2\sinh\frac{\pm\phi_I\pm\phi_J\pm\epsilon_-+\epsilon_+}{2}}\,,
\end{split}\ee
and for $k=2n$ the integrand is
\be\begin{split}\label{vecminus}
Z^{vec}_{-} =&\prod_{I < J}^n 2\sinh\frac{\pm\phi_I\pm\phi_J}{2}\prod_{I=1}^{n-1}2\sinh(\pm \phi_I)\\
&\frac{2\cosh(\epsilon_+)}{2\sinh\frac{\pm\epsilon_-+\epsilon_+}{2}2\sinh(\pm\epsilon_-+\epsilon_+)\prod_{i=1}^N2\sinh(\pm\alpha_i+\epsilon_+)}\prod_{I=1}^{n-1}\frac{2\sinh(\pm\phi_I+2\epsilon_+)}{2\sinh(\pm\phi_I\pm\epsilon_-+\epsilon_+)}\\
&\prod_{I=1}^{n-1}\frac{2\sinh\epsilon_+}{2\sinh\frac{\pm\epsilon_-+\epsilon_+}{2}2\sinh\frac{\pm 2\phi_I\pm\epsilon_-+\epsilon_+}{2}\prod_{i=1}^N2\sinh\frac{\pm\phi_I\pm\alpha_i+\epsilon_+}{2}}\prod_{I<J}^{n-1}\frac{2\sinh\frac{\pm\phi_I\pm\phi_j+2\epsilon_+}{2}}{2\sinh\frac{\pm\phi_I\pm\phi_J\pm\epsilon_-+\epsilon_+}{2}}\,.
\end{split}\ee
Again when writing the integrands we introduced the notation $\epsilon_\pm = \frac{\epsilon_1 \pm \epsilon_2}{2}$.
The contour integral should be again computed using the JK prescription for computing the various residues as explained in \cite{Hwang:2014uwa}. For 
case of $k=1$ no integral is actually necessary and the results for $Z^{k=1}_+$ and $Z^{k=1}_-$ are simply (after taking $\epsilon_+ =0$ for comparison
with the topological string result)
\be
Z_+^{k=1} \rightarrow \sum_{i=1}^N\frac{1}{2 \sinh \frac{\pm \epsilon_-}{2} 2\sinh \frac{\pm \alpha_i}{2}}\,,
\ee
\be
Z_-^{k=1} \rightarrow \sum_{i=1}^N\frac{1}{2 \sinh \frac{\pm \epsilon_-}{2} 2\cosh \frac{\pm \alpha_i}{2}}\,.
\ee
\subsubsection{Relation between $USp(2N)$ and $USp(2N-2)$}
\label{sec:USp.limit}
By decoupling one Coulomb branch modulus from the partition function of the $USp(2N)$ gauge theory, one will obtain the partition function a $USp(2N-2)$ gauge theory. However, we will show that the decoupling actually changes the discrete theta angle of the original $USp(2N)$ gauge theory. Namely the $USp(2N)$ gauge theory has the discrete theta angle $\theta=0$ or $\pi$, then the $USp(2N-2)$ gauge theory obtained by decoupling one Coulomb branch modulus has the discrete theta angle $\theta = \pi$ or $0$ respectively. 

Let us then consider a situation where we decouple one Coulomb branch modulus $\alpha_1$ by sending $\alpha_1 \rightarrow \infty$. From the explicit expressions of the partition function \eqref{vecplus}--\eqref{vecminus}, it is possible to show that
\bea
Z_{+, N}^{vec} &\rightarrow& \left(e^{-\alpha_1}\right)^k (-1)^k Z_{+, N-1}^{vec},\\
Z_{-, N}^{vec} &\rightarrow& \left(e^{-\alpha_1}\right)^k (-1)^{k+1} Z_{-, N-1}^{vec},
\eea
where $k$ is the instanton number and $N$ is the rank of the $USp(2N)$ gauge group. At the moment, we keep the factor $e^{-\alpha_1}$, which will be absorbed in the instanton fugacity. This implies that the $k$-instanton partition function of the ADHM quantum mechanics changes as
\be
\frac{1}{2}\left(Z_{+, N}^k + Z_{-, N}^k\right)u_{USp(2N)}^k \rightarrow \frac{(-1)^k}{2}\left(Z_{+, N-1}^k - Z_{-, N-1}^k\right)\left(e^{-\alpha_1}u_{USp(2N)}\right)^k,
\ee 
for $\theta = 0$ and 
\be
\frac{(-1)^k}{2}\left(Z_{+, N}^k - Z_{-, N}^k\right)u_{USp(2N)}^k \rightarrow \frac{1}{2}\left(Z_{+, N-1}^k + Z_{-, N-1}^k\right)\left(e^{-\alpha_1}u_{USp(2N)}\right)^k,
\ee 
for $\theta = \pi$. By identifying $u_{USp(2N-2)} = e^{-\alpha_1}u_{USp(2N)}$\footnote{Therefore, the precise limit is $\alpha_1 \rightarrow \infty$ and $u_{USp(2N)} \rightarrow \infty$ while $e^{-\alpha_1}u_{USp(2N)}$ kept.}, one obtains
\bea
\frac{1}{2}\left(Z_{+, N}^k + Z_{-, N}^k\right)u_{USp(2N)}^k &\rightarrow& \frac{(-1)^k}{2}\left(Z_{+, N-1}^k - Z_{-, N-1}^k\right)u_{USp(2N-2)}^k,\\
\frac{(-1)^k}{2}\left(Z_{+, N}^k - Z_{-, N}^k\right)u_{USp(2N)}^k &\rightarrow& \frac{1}{2}\left(Z_{+, N-1}^k + Z_{-, N-1}^k\right)u_{USp(2N-2)}^k.
\eea
Therefore, when the $USp(2N)$ gauge theory has the discrete angle $\theta = 0$ or $\pi$ then the $USp(2N-2)$ gauge theory obtained after the limit has the discrete theta angle $\theta = \pi$ or $0$ respectively. Namely the limit indeed changes the discrete theta angle. 

Under the limit we defined the instanton fugacity of the $USp(2N-2)$ gauge theory as $u_{USp(2N-2)} = e^{-\alpha_1}u_{USp(2N)}$. This is in fact completely agrees with the expectation from the 5-brane web in section \ref{sec:USp}. The instanton fugacity of the original the $USp(2N)$ gauge theory is given by \eqref{inst.fugacity.USp}, namely
\be
u_{USp(2N)}^2 = \hat{Q}_1^2A_1^{-2N+2k-4}. \label{USp2N.inst.app}
\ee
On the other hand, after decoupling $\alpha_1$, we obtain a 5-brane web with the top and bottom color D5-branes removed. In particular, the instanton fugacity of the $USp(2N-2)$ gauge theory should be given by
\be
u_{USp(2N-2)}^2 = \hat{Q}_2A_2^{-2(N-1)+2k-4}.\label{USp2N-2.inst.app}
\ee
From the 5-brane web in Figure \ref{fig:USp2N1}, the relation between $\hat{Q}_1$ and $\hat{Q}_2$ is $\hat{Q_2} = \hat{Q}_1\left(A_1A_2^{-1}\right)^{k-N-1}$. Combining this relation with \eqref{USp2N.inst.app} and \eqref{USp2N-2.inst.app}, one obtains
\be
u_{USp(2N)}A_1 = u_{USp(2N-2)},
\ee
which exactly agrees with the redefinition of the instanton fugacity obtained from the analysis of the partition function.

\subsection{Perturbative partition functions}

For a full comparison between field theory and topological string computations it is necessary to compare also the perturbative part of the partition functions.
Here we recall the result for $SU(N)$ with antisymmetric and pure $USp(2N)$.
\subsubsection{$SU(N)$ with antisymmetric}

For the case of $SU(N)$ with a hypermultiplet in the antisymmetric representation we have that the perturbative part of the partition function
is
\be
Z_{pert} =\prod_{i,j=1}^\infty\left[\frac{1}{(1-q^{i+j-1})^{N-1}}\prod_{I>J}^N \frac{(1-e^{-\alpha_I-\alpha_J - m} q^{i+j-1})(1-e^{\alpha_I+\alpha_J - m} q^{i+j-1})}{ (1-e^{-\alpha_I + \alpha_J} q^{i+j-1})
(1-e^{\alpha_I - \alpha_J} q^{i+j-1})}\right]\,.
\ee

\subsubsection{$USp(2N)$}

For the case of pure $USp(2N)$ the perturbative part is
\be\begin{split}
Z_{pert} =& \prod_{i,j=1}^\infty \left[\frac{1}{(1-q^{i+j-1})^{N}}\prod_{I>J}^N\frac{1}{ (1-e^{-\alpha_I +\alpha_J}q^{i+j-1})(1-e^{\alpha_I -\alpha_J}q^{i+j-1})}\right.\\&\left.\prod_{I,J}^N\frac{1}{(1-e^{-\alpha_I -\alpha_J}q^{i+j-1})(1-e^{\alpha_I +\alpha_J}q^{i+j-1})}\right]\,.
\end{split}\ee


\bibliographystyle{JHEP}
\bibliography{refs}

\end{document}